# Low-temperature formation of pyridine and (iso)quinoline via neutral–neutral reactions

Zhenghai Yang,[1] Chao He,[1] Shane J. Goettl,[1] Alexander M. Mebel,[2]* Paulo F. G. Velloso,[3] Márcio O. Alves,[3] Breno R. L. Galvão,[3]* Jean-Christophe Loison,[4]* Kevin M. Hickson,[4] Michel Dobrijevic,[5] Xiaohu Li,[6,7]* Ralf I. Kaiser[1]*

[1] Department of Chemistry, University of Hawaii at Manoa, 2545 McCarthy Mall, Honolulu, HI 96822 (USA)

[2] Department of Chemistry and Biochemistry, Florida International University, Miami, Florida 33199, USA

[3] Centro Federal de Educação Tecnológica de Minas Gerais, CEFET-MG, Av. Amazonas 5253, 30421-169 Belo Horizonte, Minas Gerais, Brazil

[4] Institut des Sciences Moléculaires, CNRS, Univ. Bordeaux, 351 Cours de la Libération, 33400 Talence, France

[5] Laboratoire d'Astrophysique de Bordeaux, Univ. Bordeaux, CNRS, B18N, Allée Geoffroy Saint-Hilaire, 33615 Pessac, France

[6] Xinjiang Astronomical Observatory, Chinese Academy of Sciences, Urumqi, Xinjiang 830011, P. R. China

[7] Key Laboratory of Radio Astronomy, Chinese Academy of Sciences, Urumqi, Xinjiang 830011, P. R. China

* Correspondence to:
ralfk@hawaii.edu, mebela@fiu.edu, brenogalvao@gmail.com, jean-christophe.loison@cnrs.fr, xiaohu.li@xao.ac.cn

The file includes:
Main text: 4880 words
Methods: 840 words
Legends: 465 words
Number of main text references: 76
Number of methods references: 8
Number of figures: 6

**Aromatic molecules represent fundamental building blocks in prebiotic chemistry and are contemplated as vital precursors to DNA and RNA nitrogen bases. However, despite the identification of some 300 molecules in extraterrestrial environments, the pathways to pyridine ($C_5H_5N$), pyridinyl ($C_5H_4N^{\bullet}$), and (iso)quinoline ($C_9H_7N$) – the simplest representative of mono and bicyclic aromatic molecule carrying nitrogen – are elusive. Here, we afford compelling evidence on the gas-phase formation of methylene amidogen ($H_2CN^{\bullet}$) and cyanomethyl ($H_2CCN^{\bullet}$) radicals via molecular beam studies and electronic structure calculations. The modeling of the chemistries of Taurus Molecular Cloud (TMC-1) and Titan's atmosphere contemplates a complex chain of reactions synthesizing pyridine, pyridinyl, and (iso)quinoline from $H_2CN^{\bullet}$ and $H_2CCN^{\bullet}$ at levels of up to 75%. This study affords unique entry points to precursors of DNA and RNA nitrogen bases in hydrocarbon-rich extraterrestrial environments thus changing the way we think about the origin of prebiotic molecules in our Galaxy.**

Since the very first discovery of biorelevant, heteroaromatic molecules such as vitamin B3 (niacin)[1,2] and nucleobases (pyrimidines, purines)[3] in carbonaceous chondrites including Murchison[3,4], critical questions have arisen on their formation routes in extraterrestrial environments. The identification of a series of terrestrially rare nucleobases such as 6-diaminopurine along with the $^{15}N/^{14}N$ isotope enrichment suggests an interstellar origin[4] thus providing a vital link between cold molecular clouds as their origin and their identification in our solar system. However, well defined formation routes of these molecules are still lacking. Their stem compounds - polycyclic aromatic hydrocarbons (PAHs) along with their cations and (partially) hydrogenated counterparts[5-7] - have been proposed to be associated with the synthesis of these biorelevant molecules in the interstellar medium (ISM), though not having unraveled how a stable C-H moiety in PAHs can be replaced by an isoelectronic nitrogen atom (N) in NPAHs. The 6.2 μm (1613 $cm^{-1}$) infrared emission band in deep space has been linked to protonated PAHs[8], but has also been discussed as the result of NPAHs[9] with PAHs and NPAHs accounting for up to 30 % of the cosmic carbon budget[10]. Whereas well-defined low-temperature (cold molecular clouds; TMC-1) and high-temperature routes (circumstellar envelopes; IRC +10216) to PAH formation in interstellar and circumstellar environments have begun to emerge[11], surprisingly little is known on the gas-phase synthesis of their nitrogen-substituted counterparts (NPAHs). This lack of knowledge is rather staggering considering that these aromatics carry the cyclic nitrogen-carbon skeletons of a key class of astrobiologically important molecules: nitrogen bases of deoxyribonucleic acid (DNA) and ribonucleic acid (RNA)[9,12].

Recent astrochemical models advocated that the carbon – nitrogen chemistries in molecular clouds can be linked with complex reaction networks[13] of gas phase ion–molecule[14] and neutral–neutral reactions[15] of aromatic (AR) and resonantly stabilized free radicals (RSFR) such as phenyl ($C_6H_5^\bullet$) and propargyl ($C_3H_3^\bullet$) along with their nitrogen counterparts pyridinyl ($C_5H_4N^\bullet$) and cyanomethyl ($H_2CCN^\bullet$)[14,16-18]. Further, the synthesis of pyridine ($C_5H_5N$) has been proposed to be driven by radical mediated reactions of hydrogen cyanide (HCN)[19] and via de-facto methylidyne radical (CH) insertion into pyrrole ($C_4H_5N$)[20]. These reaction networks have been 'borrowed' from the planetary science community attempting to rationalize the existence of both stratospheric PAHs and NPAHs in Titan's atmosphere determined from Cassini's Visual and Infrared Mapping Spectrometer (VIMS) measurements at 3.28 μm (3,049 $cm^{-1}$)[21] and Composite Infrared

Spectrometer (CIRS) measurements at 71.43 μm (140 $cm^{-1}$)[22,23], and through Cassini's Plasma Spectrometer (CAPS)[24]. The latter detected positively and negatively charged particles with molecular weights less than 8,000 amu containing (N)PAHs along with their fundamental building blocks benzene ($C_6H_6$; m/z = 78) and pyridine ($C_5H_5N$; m/z = 79)[14,24,25]. Overall, to date, an understanding of the synthesis of benzene along with aromatics carrying up to six rings such as corannulene ($C_{20}H_{10}$)[26] and helicenes ($C_{26}H_{16}$)[27] is beginning to emerge[11]. However, the underlying elementary processes even leading to the simplest representative of mono- and bicyclic aromatic molecule carrying nitrogen, i.e. pyridine ($C_5H_5N$; **1**) and (iso)quinoline ($C_9H_7N$; **2/3**) – together with their cyanomethyl ($H_2CCN^\bullet$; **4**) and methylene amidogen ($H_2CN^\bullet$; **5**) precursors is still in its infancy (Figure 1). The understanding of these gas phase reactions and the formation of the first carbon – nitrogen bonds from the 'bottom up' is fundamental to our knowledge of how nitrogen containing aromatics can be produced abiotically in low temperature interstellar and solar system environments from the simple closed shell nitrogen containing hydride (ammonia; $NH_3$) and reactive carbon-based reactants in form of atomic carbon (C) and dicarbon ($C_2$).

Here, we report on the gas phase preparation of the methylene amidogen radical ($H_2CN^\bullet$, $X^2B_2$) and of the resonantly stabilized cyanomethyl radical ($H_2CCN^\bullet$, $X^2B_1$) via bimolecular reactions of atomic carbon (C, $^3P$) and of dicarbon ($C_2$, $X^1\Sigma_g^+/a^3\Pi_u$) with ammonia ($NH_3$, $X^1A_1$) exploiting crossed molecular beams experiments. The role of the methylene amidogen and the cyanomethyl radicals in the formation of pyridine ($C_5H_5N$; **1**), pyridinyl ($C_5H_4N$; **9-11**), and (iso)quinoline ($C_9H_7N$; **2/3**) are also elucidated. These data are combined with electronic structure calculations and modeling of the chemistries of hydrocarbon rich environments of cold molecular clouds and atmospheres of planets of their moons exploiting the Taurus Molecular Cloud (TMC-1) and Titan as benchmarks. And a complex chain of exoergic, barrierless routes is contemplated with $H_2CN^\bullet$ and $H_2CCN^\bullet$ radicals representing fundamental molecular building blocks of pyridine and pyridinyl radicals ($C_5H_4N^\bullet$, **9-11**) synthesized through successive barrierless reactions involving propargyl ($C_3H_3^\bullet$, $X^2B_1$, **6**) and 1-butene-3-yne-2-yl/1-butene-3-yne-1-yl (*i/n*-$C_4H_3^\bullet$, $X^2A'$, **7-8**) under low temperature conditions of molecular clouds (10 K) and Titan's atmosphere (70-180 K) (Figure 1)[28]. Since pyridinyl radicals are isoelectronic to the phenyl radical ($C_6H_5^\bullet$), pyridinyl may play a critical role in the gas phase formation of (iso)quinoline upon reaction with vinylacetylene ($C_4H_4$) via the low-temperature hydrogen abstraction – vinylacetylene addition (HAVA) pathway[29]. These results thus offer fundamental knowledge on the previously elusive reaction routes to prototype nitrogen heteroaromatics in low-temperature extraterrestrial environments from cold molecular clouds such as TMC-1 to atmospheres of planets and their moons like Titan. These mechanisms are not constrained to Titan, but may present a versatile strategy for the synthesis of nitrogen heteroaromatics in low temperature, hydrocarbon and nitrogen-rich atmospheres of outer Solar System bodies such as Triton[30] and Pluto[31]. Hence, the present work sheds light on sensible processes coupling the carbon and nitrogen chemistries eventually leading to the formation of molecular nitrogen – carbon motives of astrobiological relevance as found in, e.g., nucleobases[29] in our Universe.

## Results

### Carbon – D3-Ammonia and Dicarbon – Ammonia Systems: Laboratory Frame

Exploiting a crossed molecular beams machine (Figure S1)[32], reactive scattering signal of the reaction of ground state atomic carbon (C, $^3P$) with D3-ammonia ($ND_3$, $X^1A_1$) was observed at *m/z* = 30 ($D_2CN^+$) after electron impact ionization of the neutral reaction products. Within our signal-to-noise, no signal was monitored at 32 ($D_3CN^+$) indicating that no $D_3CN$ adducts were formed. These data alone indicate the existence of the atomic deuterium loss channel (reaction (1)). Note that for technical reasons, the reaction was conducted with D3-ammonia, but not with ammonia ($NH_3$, $X^1A_1$) since this would have shifted reactive scattering signal to *m/z* = 28 ($H_2CN^+$). Signal at m/z = 28 is obscured by significant background counts from carbon monoxide, which outgasses from stainless steel even under our ultra-high vacuum (UHV) conditions of $6\times10^{-12}$ Torr. Therefore, angular resolved TOF spectra were recorded at 30 ($D_2CN^+$) in 5° intervals within the scattering plane, integrated, and scaled with respect to the TOF recorded at the center-of-mass (CM) angle of 35.9 ± 0.5° leading to the laboratory angular distribution (LAD) (Table S1). This distribution holds a maximum around the CM angle and is nearly forward-backward symmetric (Figure 2, A-B) implying that the carbon − D3-ammonia reaction proceeds through indirect reaction dynamics involving the formation of $D_3CN$ collision complex(es).

Reactive scattering signal of the reaction of dicarbon ($C_2$, $X^1\Sigma_g^+/a^3\Pi_u$) with ammonia ($NH_3$, $X^1A_1$) was detected at *m/z* = 40 ($C_2NH_2^+$) and 39 ($C_2NH^+$) (reaction (2)) with signal at 39 collected at level of 66 ± 5% compared to signal at 40. The TOF spectra at 40 and 39 are identical after scaling (Figure S2) suggesting that signal at 39 originates from dissociative ionization of the nascent product (40 amu) in the electron impact ionizer. Further, signal at 40 indicated the existence of a dicarbon versus atomic hydrogen loss and inherent formation of a molecule with the molecular formula $C_2NH_2$ via reaction (2). Consequently, data were collected at 40 in 2.5° steps (Figure 2, C-D). The derived LAD is also nearly forward-backward symmetric revealing the existence of $C_2H_3N$ intermediate(s) and indirect scattering dynamics.

$$(1)\ C\ (12\ amu) + ND_3\ (20\ amu) \rightarrow D_2CN\ (30\ amu) + D\ (2\ amu)$$

$$(2)\ C_2\ (24\ amu) + NH_3\ (17\ amu) \rightarrow H_2C_2N\ (40\ amu) + H\ (1\ amu)$$

**Carbon – D3-Ammonia and Dicarbon – Ammonia Systems: Center-of-Mass Frame**

Having provided compelling evidence on the formation of $D_2CN$ and $H_2C_2N$ isomers in reactions (1) and (2), respectively, we are turning attention to elucidating the nature of the structural isomer(s) and the underlying reaction mechanism(s). This information is obtained by transforming the laboratory data (TOFs, LAD) into the CM reference frame[32]. This forward convolution approach yields the CM translational energy ($P(E_T)$) and angular ($T(\theta)$) flux distributions as detailed in Figure 3 (A-C). For C ($^3P$) − $ND_3$ ($X^1A_1$) system, signal at 30 could be replicated with a single reaction channel of an atomic deuterium loss channel (reaction (1)) with a product mass combination of 30 amu ($D_2CN$) plus 2 amu (D). A detailed inspection of the $P(E_T)$ reveals a high-energy cutoff of 237 ± 25 kJ mol$^{-1}$, which denotes the sum of the collision energy $E_c$ (28.1 ± 0.9 kJ mol$^{-1}$) plus the reaction exoergicity for molecules generated without internal excitation. Therefore, reaction (1) is suggested to be exoergic by 209 ± 26 kJ mol$^{-1}$. Further, the $P(E_T)$ display a distribution maximum at 26 ± 3 kJ mol$^{-1}$ suggesting a tight exit transition state upon unimolecular decomposition of the $D_3CN$ intermediate to the separated products and a significant electron density reorganization. The $T(\theta)$ function depicts flux across the complete angular range together with a forward-backward symmetric scattering pattern. These findings reveal indirect scattering dynamics through long-lived $D_3CN$ complex(es) holding lifetime longer than the(ir) rotational periods[32].

Considering the $C_2$ ($X^1\Sigma_g^+/a^3\Pi_u$) − $NH_3$ ($X^1A_1$) system, the laboratory data can be replicated with a single atomic hydrogen loss channel with a mass combination of 40 amu ($H_2C_2N$) and 1 amu (H) (reaction (2)) (Figure 3, D-F). The $P(E_T)$ as depicted in Figure 3D shows an $E_{max}$ of 327

± 26 kJ mol$^{-1}$; this results in an exoergicity of 310 ± 26 kJ mol$^{-1}$ or 318 ± 26 kJ mol$^{-1}$ considering the $E_c$ of 17.0 ± 0.3 kJ mol$^{-1}$ for dicarbon reactants in ground $X^1\Sigma_g^+$ or excited $a^3\Pi_u$ state. It should be noted that the $P(E_T)$ peaks only slightly away from zero translational energy at 9 ± 1 kJ mol$^{-1}$ indicating a loose exit transition state with only a minor rearrangement of the electron density, which contrasts to the results in C-$ND_3$ system. Finally, the $T(\theta)$ function exhibits a forward-backward symmetry with sideway scattering and hence a distribution maximum at 90°. These results propose indirect scattering dynamics via the formation of long-lived $C_2NH_3$ intermediates[33].

**Electronic Structure Calculations and Reaction Mechanisms**

We are merging now our experimental data with electronic structure calculations (Figures 4 and 5). *First*, our computations on the triplet $CND_3$ potential energy surface (PES) reveal four reaction intermediates (**i1-i4**), four product channels (**p1-p4**), and ten transition states. These high-level calculations predict relative energies of the transition states, local minima, and products within 8 kJ mol$^{-1}$.[34,35] Four deuterium atom (D) loss pathways lead to D2-methylene amidogen (**p1**, $D_2CN^\bullet$, $C_{2v}$, $X^2B_2$), trans-D2-iminomethyl (**p2**, DCND, $C_s$, $X^2A'$), cis-D2-iminomethyl (**p3**, DCND, $C_s$, $X^2A'$), and D2-aminomethylidyne (**p4**, $D_2NC$, $C_{2v}$, $X^2B_2$). The relative energies of these isomers are within 3 kJ mol$^{-1}$ compared to two previous studies[36,37]. A comparison of the theoretically predicted reaction energies (Figure 4A) with our experimentally derived exoergicity of 209 ± 26 kJ mol$^{-1}$ reveals that at least the thermodynamically most stable product **p1** is formed. Contributions from high energy isomers **p2-p4** might be masked in the low-energy section of the $P(E_T)$ and cannot be eliminated. How can **p1** be formed? Our calculations predict that the reaction is initiated via a barrierless addition of the carbon atom to the non-bonding electron pair at the nitrogen atom of ammonia. This reaction can be seen as an acid-base reation and leads to **i1** ($CND_3$, D3-ammoniamethylidyne) bound by 103 kJ mol$^{-1}$. This collision complex may eliminate a deuterium atom forming **p4** by passing a transition state located only 2 kJ mol$^{-1}$ below the separated reactants, or isomerizes to **i2** ($DCND_2$, D3-aminomethylidene) via deuterium migration from nitrogen to carbon. The transition state resides 11 kJ mol$^{-1}$ below the separated reactants and hence can be accessed preferentially compared to the unimolecular decomposition of **i1** to **p4** plus D. The calculations suggest that **i2** can fragment via D loss to **p2** or **p3** through tight transition states residing 34 and 27 kJ mol$^{-1}$ above the separated products; **p4** may be formed barrierlessly, too. Alternatively, **i2** may isomerize to **i3** ($D_2CND$, D3-methanimine) via a deuterium shift. This intermediate can undergo three distinct decomposition pathways through deuterium loss leading to **p3, p2,** and/or the thermodynamically most stable isomer **p1** involving tight exit transition states (**i3→p3** 20 kJ $mol^{-1}$; **i3→p2** 22 kJ $mol^{-1}$; **i3→p1** 29 kJ $mol^{-1}$; the numbers indicate barrier with respect to the products). Finally, **i3** may undergo a deuterium shift to **i4** ($D_3CN$, D3-methylnitrene), the global minimum of the $D_3CN$ PES, followed by unimolecualr decomposition to **p1** through a tight exit transition state located 13 kJ $mol^{-1}$ above the separated products. These considerations reveal that the experimentally detected D2-methylene amidogen (**p1**) can be formed via **i1→i2→i3→p1**+D and/or **i1→i2→i3→i4→p1**+D.

Statistical Rice–Ramsperger–Kassel–Marcus (RRKM) calculations were also conducted to predict the branching ratios of **p1-p4** theoretically[34]. Within the limit of a complete energy randomization, **p1**, **p2**, **p3**, and **p4** contribute 7 %, 33 %, 29 %, and 31%, respectively, at the experimental collision energy of 28.1 $kJmol^{-1}$ (Table S1). At the low temperature conditions of TMC-1 and Titan, fractions of 8 %, 41 %, 33 %, and 18 % are predicted (Table S2-3). Under the atmospheric conditions of Titan, one peculiarity exists. Suprathermal hydrogen atoms with excess kinetic energies of a few electron volts can be formed via photodissociation of hydrides like methane ($CH_4$) by solar photons, via neutral-neutral reactions of, e.g., methylidyne (CH) with methane, and/or electron recombination reactions of with abundant ions ($CH_5^+$, $CH_3^+$, and $C_6H_7^+$)[14,38,39]. These suprathermal hydrogen atoms are not in thermal equilibrium with the surrounding low-temperature atmosphere and their kinetic energies can easily exceed 100 kJ $mol^{-1}$ (1.04 eV)[38,40]. Thus, rapid suprathermal hydrogen-atom assisted isomerization processes can efficiently convert the high-energy isomers **p3** and **p4** to **p1**[41,42]. These processes are very efficient due to absence of a barrier of **p4 + H → i2** and a small barrier of only 20 $kJmol^{-1}$ for **p3 + H → i3**. As depicted in Figures S3-4, the elevated rate constants of **p3** + H of a few $10^{-10}$ $cm^3s^{-1}$ can be reached with suprathermal hydrogen atoms possessing high kinetic energies.

*Second*, for the $C_2$-$NH_3$ system, the primary reactant beam contains dicarbon in its electronic ground state ($X^1\Sigma_g^+$), but also in its first electronically excited state ($a^3\Pi_u$). The calculations reveal that $C_2$ ($a^3\Pi_u$) does not form any bound intermediates upon collision with ammonia, but rather reacts in a direct fashion via hydrogen abstraction forming the ethynyl radical ($C_2H^\bullet$, $X^2\Sigma^+$) plus the amino radical ($NH_2^\bullet$, $X^2B_1$) in an exoergic reaction (-36 $kJmol^{-1}$) through a barrier of 22 $kJmol^{-1}$. However, on the singlet surface, dicarbon adds barrierlessly to the non-bonding electron pair of the nitrogen atom of ammonia to **i1'** ($CCNH_3$, ammoniaethynyl). Overall, nine intermediates (**i1'-i9'**), seven products (**p1'-p7'**), and nine transition states were identified. Among the product isomers, **p1'** ($H_2CCN^•$; cyanomethyl; $C_{2v}$; $^2B_1$) represents the thermodynamically most stable isomer followed by **p2'** ($H_2CNC$; isocyanomethyl; $C_{2v}$; $^2B_1$) and **p3'** (HCCNH; imidogenacetylene; $C_s$; $^2A''$). The calculated relative energies of the $C_2H_2N$ isomers agree well with previous calculations[43]. A comparison of these energies with the experimentally extracted exoergicity of 310 ± 26 kJ mol$^{-1}$ proposed that at least **p1'** is formed under single collision conditions. Contributions from high energy isomers **p2'-p5'** might be hidden in the low-energy section of the *P*($E_T$) and cannot be eliminated. **i1'** can isomerize to an exotic cyclic intermediate **i2'** ($HC(NH_2)C$) via ring closure along with a hydrogen shift from the nitrogen to the carbon or to **i4'** ($HCCNH_2$, aminoacetylene) through hydrogen migration. Our calculations also identify a loose H loss channel to **p7'** ($CCNH_2$; aminoethynyl; $C_{2v}$; $^2B_1$) from **i1'**. Extensive hydrogen migration and ring opening pathway access intermediates **i3'** to **i9'**, with all transition state to isomerization residing well below the energy of the separated reactants. The experimentally detected product **p1'** can eventually be accessed via H loss from **i6'** ($H_2CCNH$, aminovinyl) and/or **i7'** **(**$CH_3CN$, acetonitrile) through loose transition states. Note that **i6'** and **i7'** are connected through a hydrogen migration barrier of 262 kJmol$^{-1}$. Overall, the experimentally predicted loose exit transition state agrees well with the computational predictions of two open channels to **p1'** ($H_2CCN^•$; cyanomethyl) via simple bond rupture processes on the singlet surface (Figure 4B).

## From the Laboratory to Hydrocarbon-Rich Atmospheres of Planets and their Moons

We are now conveying our findings to the atmosphere of Saturn's moon Titan. *First*, the carbonaceous reactants (atomic carbon, dicarbon) originate from photolysis of methane ($CH_4$) and acetylene ($C_2H_2$), respectively[14,44]. Carbon can also be generated from dissociative electron – ion ($CH_2^+$, $CH_3^+$) recombination in atmospheric layers above 1,200 km[18]. The mole fraction of ammonia ($NH_3$) in Titan's atmosphere is still a tricky question. Ammonia abundances have been inferred indirectly via the Cassini's Ion and Neutral Mass Spectrometer (INMS) detection of the ammonium cation ($NH_4^+$) predicting high mole fraction of ammonia of (3-4)×$10^{-5}$ at around 1000 km[45]. However, photochemical models cannot replicate these findings and underestimate mole fractions by up to two orders of magnitude[46,47]. In the stratosphere, the calculated ammonia mole fraction is consistent with upper limits derived from Composite Infrared Spectrometer (CIRS) and Herschel[47,48]. *Second*, the absence of entrance barriers in the exoergic bimolecular reactions signifies the crucial prerequisite for reaction operating in Titan's low temperature atmosphere (70–180 K), which would prohibit reactions with significant entrance barrier. Therefore, chemical reactions relevant to Titan's atmospheric chemistry must be exoergic, proceed without an entrance barrier, and involve transition states with lower energy than the separated reactants. All these benchmarks are fulfilled in the formation of the $H_2CN^{\bullet}$ and the $H_2CCN^{\bullet}$ radical holding rate constants of, e.g., a few $10^{-10}$ $cm^3$ $s^{-1}$ for the carbon – ammonia system at 50 K[49]. *Third*, the aforementioned radicals are isoelectronic with the vinyl ($C_2H_3^{\bullet}$, $X^2A'$) and propargyl ($C_3H_3^{\bullet}$, $X^2B_1$) radical. Therefore, both the methylene amidogen and the cyanomethyl radical are involved in fundamental molecular mass growth processes in Titan's atmosphere. These processes are in strong analogy to the $C_2H_3^{\bullet}$-$C_4H_3^{\bullet}$[16,50] and $C_3H_3^{\bullet}$-$C_3H_3^{\bullet}$[51] systems, respectively, which access the phenyl radical ($C_6H_5^{\bullet}$) under single collision conditions and benzene along with its isomers, if the collision complexes can be stabilized through collision with a third body. This conclusion is verified through electronic structure calculation for the $H_2CN^{\bullet}$-$C_4H_3^{\bullet}$, and $H_2CCN^{\bullet}$-$C_3H_3^{\bullet}$ systems synthesizing three distinct pyridinyl radicals (o, m, p; $C_5H_4N^{\bullet}$) under single collision conditions (Figure 5A). Three distinct entrance channels lead barrierlessly to **i1''** to **i3''**. Extensive hydrogen shifts, and cyclization accompanied by aromatization to pyridine ($C_5H_5N$, **i12''**), which undergoes three barrierless hydrogen loss pathways to distinct pyridinyl radicals. It is interesting to point out that analogous barrierless pathways are identified in the cis-HCNH-$C_4H_3^{\bullet}$ system, in which two distinct entrance channels lead barrierlessly to **i14''** and **i19''** followed by hydrogen shifts and cyclization to pyridine and pyridinyl radicals (Figure 5B). In the presence of a dense atmosphere such as of Titan, pyridine can be stabilized by a third body with the bath molecule such as molecular nitrogen. Once stabilized, pyridine can be photolyzed to pyridinyl radicals (o, m, p; $C_5H_4N^{\bullet}$) (Figure 1) followed by barrierless reactions with vinylacetylene ($C_4H_4$) to (iso)quinoline ($C_9H_7N$)[29]. *Finally*, previous photochemical models suggest that $H_2CN^{\bullet}$ is produced via the reaction of atomic nitrogen with the methyl radical ($CH_3$)[52,53]. However, this modeling study did not include reactions of atomic carbon and dicarbon with ammonia due to the foregoing lack of reliable laboratory and computational data of the $C/NH_3$ and $C_2/NH_3$ systems, whereas extensive computational and experimental data exist for the $N/CH_3$ system[37,54,55]. Here, we prove that both $H_2CN^{\bullet}$ and cis-HCNH radicals can react barrierlessly with i/n-$C_4H_3$ isomers forming pyridine and pyridinyl radicals indicating the potential significant role of $C/NH_3$ in the prebiotic chemistry of Titan. Overall, we depict evidence that in Titan's atmosphere, where abundant suprathermal hydrogen atoms exist, **p3** can efficiently undergo suprathermal hydrogen atom − assisted isomerization to **p1**, the most stable isomer, followed by the reactions with i/n-$C_4H_3$ isomers to pyridine and pyridinyl radicals. Even **p3** itself can itself react with i/n-$C_4H_3$ isomers barrierlessly leading to pyridine and pyridinyl radicals.

The aforementioned findings are implemented into a one-dimensional photochemical atmospheric model of Titan to evaluate the eventual formation of pyridine ($C_5H_5N$) and (iso)quinoline ($C_9H_7N$) (Supplementary Information). This model incorporates an unbiassed chemistry of neutrals and cations along with the coupling between them from the lower atmosphere to the ionosphere[56,57]. The chemical scheme operated in the present model has been enhanced with the new reactions studied included[57-59]. To evaluate the uncertainties of the nominal model profiles, a Monte Carlo simulation was performed according to the method described in Benne et al.[58] (Figure 6A, Supplementary Information). These photochemical models yield exciting results. First, these studies reveal that two $C_5H_5N$ isomers, pyridine ($C_5H_5N$) and ethylcyanoacetylene ($C_2H_5CCCN$), display significant mole fractions of $1.4 \times 10^{-7}$ and $2.3 \times 10^{-7}$ in the ionosphere of Titan, respectively. The maximum mole fraction for (iso)quinoline is predicted to be $1.7 \times 10^{-11}$ around 1,100 km, which should be observable spectroscopically. These models also predict maximum mole fractions of $C_5H_5NH^+$ and $C_2H_5C_3NH^+$ of $1.5 \times 10^{-10}$ and $3.9 \times 10^{-10}$ at around 1,100 km, respectively. Accounting for the uncertainties, the mole fraction of $(1.5 \pm 0.3) \times 10^{-9}$ at m/z = 80 ($C_5H_5NH^+$) derived from the Cassini INMS data[18,60] agrees within the error limits well with the sum of mole fractions of $C_5H_5NH^+$ and $C_2H_5C_3NH^+$ of the atmospheric models ranging between $2.5 \times 10^{-9}$ to $5.4 \times 10^{-10}$. Although the atmospheric models provide compelling constrains on the abundances of pyridine and (iso)quinoline, we have to concede that the uncertain abundances of ammonia in Titan's atmosphere, which may vary over at least two orders of magnitude, make it difficult to quantify the contributions of distinct pathways to the $H_2CN^\bullet$ radical. With a low predicted ammonia mole fraction of a few $10^{-7}$, the carbon – ammonia reaction hardly competes with the nitrogen-methyl system providing small fractions of up to one percent at most. However, considering the bimolecular nature of the carbon – ammonia system, an increase of the fractional abundance of ammonia will lead to an enhancement of $H_2CN^\bullet$ radicals. Only future direct spectroscopic measurements of ammonia can resolve this issue. In fact, based on the nominal model results, it can be determined that at least 10% of $H_2CN^\bullet$ radicals are produced by the carbon

– ammonia reaction under conditions of high ammonia mole fraction derived from the INMS data of Cassini mission[45]. Therefore, fundamental bimolecular reactions including atomic carbon and dicarbon with ammonia may initiate a chain of barrierless reactions ultimately to pyridine and (iso)quinoline, i.e. the two simplest mono- and bicyclic aromatic molecules. These processes are not limited to Titan, but represent versatile pathways eventually leading to NPAHs in hydrocarbon and nitrogen containing atmospheres of planets and their moons in the outer Solar systems such as on Triton[30] and Pluto[31] with organic haze layers recently detected by the *New Horizons* mission[61]. Only recently, NPAH product quinolizinium$^+$ ($C_9H_8N^+$) was formed via reaction of the pyridine cation with acetylene in low-temperature pathways, highlighting the role of ion-molecule reactions in the NPAH formation in Titan's atmosphere[62]. Here, the low-temperature neutral-neutral formation pathways to pyridine and (iso)quinoline are revealed. The combined ion-molecule and neutral-neutral reaction network may finally reproduce the astronomical detected abundances of PAHs which are drastically underestimated in the current modeling[63]. Thus our understanding of fundamental low-temperature molecular mass growth processes to nitrogen substituted aromatics and their radicals is deepened.

**From the Laboratory to Cold Molecular Clouds: TMC-1**

The low temperature chemical mass growth processes in cold molecular clouds such as TMC-1 is fundamentally distinct from those in atmospheres of planets and their moons with both interstellar (TMC-1) and solar system (Titan) low-temperature environments require the absence of any entrance barriers to overall exoergic reactions[64]. However, the low number densities of molecules in molecular clouds ranging from $10^4$ to $10^6$ $cm^{-3}$ necessitate bimolecular reactions; third body collisions, in which collisions of the reaction intermediate with a bath molecule divert the internal energy of the intermediate and hence stabilize the latter, are absent. This requires changes to the reaction network from Titan's atmosphere (Figure 1) to TMC-1 thus implementing a reaction network of barrierless and exoergic bimolecular reactions (Figure S5, Table S4) such as reactions (1) and (2). To explore the implications of our findings to the chemistry leading eventually to pyridine ($C_5H_5N$), pyridinyl ($C_5H_4N^\bullet$), and (iso)quinoline ($C_9H_7N$), we untangled the role of the cyanomethyl ($H_2CCN^\bullet$) and methylene amidogen ($H_2CN^\bullet$) radicals in the formation of nitrogen heteroaromatics (pyridine ($C_5H_5N$), pyridinyl ($C_5H_4N^\bullet$), (iso)quinoline ($C_9H_7N$)) using the *University of Manchester Institute for Science and Technology* Database (RATE2012)[65] operated with the single-point time-dependent astrochemical models[27]. Physical parameters were modernized according to Markwick et al.[66], McElroy et al.[65], and Yang et al.[33] with a temperature of 10 K, a cosmic ray ionization rate of $1.3 \times 10^{-17}$ s$^{-1}$, a visual extinction of 10 Mag, and a number density of molecular hydrogen of $10^4$ cm$^{-3}$. The predictive capabilities of the model are verified by comparing the relevant species observed with modeled fractional abundances.

These models revealed fascinating findings (Figure 6B). First, the remarkable performance of the astrochemical model for TMC-1 can be benchmarked for the methylene amidogen radical ($H_2CN^\bullet$)[67], the cyanomethyl radical ($H_2CCN^\bullet$)[68], vinyl cyanide ($C_2H_3CN$)[69], and methyl cyanide ($CH_3CN$)[70] with astronomically observed fractional abundances of $(1.1 \pm 0.9) \times 10^{-10}$, $(3.5 \pm 1.5) \times 10^{-9}$, $(7.0 \pm 1.0) \times 10^{-10}$, and $(6.0 \pm 3.0) \times 10^{-10}$ relative to molecular hydrogen. Predicted peak abundances of methylene amidogen ($H_2CN^\bullet$) and of the cyanomethyl radical ($H_2CCN^\bullet$) of $(3.3 \pm 0.3) \times 10^{-10}$ at $1.3 \times 10^5$ years and $(6.0 \pm 0.4) \times 10^{-9}$ at $2.0 \times 10^5$ years replicate the astronomical observations nicely. Here, the carbon – ammonia system can account for 30 % to 75 % of the observed methylene amidogen radicals; generally spoken, as the initial abundances of nitrogen increase or carbon decreases, the fraction of methylene amidogen rises. For example, the contribution from the carbon – ammonia system rises to 50 % when the fraction of nitrogen increases to $10^{-2}$ and even to 75 % with at a fraction of $10^{-1}$; these cases operate in those regions of TMC-1 where nitrogen-rich species are injected into the gas phase from the icy grains[71,72]. Even for standard abundances of carbon versus nitrogen in TMC-1 without grain ejection, the reaction of dicarbon with ammonia contributes up to 75 % to the peak abundance of the cyanomethyl radical ($H_2CCN^\bullet$). Model outputs of the closed shell nitriles vinyl cyanide ($C_2H_3CN$) and methyl cyanide ($CH_3CN$) are reported to be $(8.6 \pm 0.6) \times 10^{-10}$ after $3.2 \times 10^5$ year and $(1.0 \pm 0.2) \times 10^{-9}$ after $2.5 \times 10^5$ years also close to the observed data with important routes of barrierless reactions of the cyano radical (CN) with ethylene ($C_2H_4$)[73] and of the vinyl radical ($C_2H_3$) with the imidogen radical (NH)[74]. Second, a complex chain of reactions initiated by barrierless reactions of the cyanomethyl ($H_2CCN^\bullet$) and methylene amidogen ($H_2CN^\bullet$) radicals (Figure S5) drive molecular mass growth processes to pyridine ($C_5H_5N$), pyridinyl ($C_5H_4N^\bullet$), and (iso)quinoline ($C_9H_7N$) with predicted peak fractional abundances of $(6.0 \pm 0.3) \times 10^{-9}$ ($6.3 \times 10^5$ years), $(1.2 \pm 0.1) \times 10^{-10}$ ($3.2 \times 10^5$ years), and $(6.0 \pm 0.4) \times 10^{-12}$ ($6.3 \times 10^5$ years), respectively. The $H_2CN^\bullet/C_4H_3^\bullet$ and $H_2CCN^\bullet/C_3H_3^\bullet$ reactions produce 46 % and 54 % of the predicted abundances of pyridinyl ($C_5H_4N^\bullet$), respectively. These results suggest that at least pyridine ($C_5H_5N$) and pyridinyl ($C_5H_4N^\bullet$) might be detectable by radio telescopes such as the Green Bank Observatory and Yebes Radio Telescope in TMC-1.

**Conclusion**

To conclude, our combined crossed molecular beam and electronic structure studies provided persuasive evidence on the formation of the methylene amidogen radical ($H_2CN^•$, $X^2B_2$) and of the resonantly stabilized cyanomethyl radical ($H_2CCN^•$, $X^2B_1$) via bimolecular reactions of atomic carbon (C; $^3P$) and of dicarbon ($C_2$; $X^1\Sigma_g^+/a^3\Pi_u$) with ammonia ($NH_3$; $X^1A_1$) in low temperature extraterrestrial environments such as in the cold molecular cloud TMC-1. Combined with modeling, these findings reveal further that both the methylene amidogen and the cyanomethyl radicals can initiate a complex chain of reactions leading to pyridine (**1**) and pyridinyl radicals (**9-11**) and eventually to (iso)quinoline (**2/3**) as the simplest prototype NPAHs and potential feedstock for more complex nitrogen-based aromatics in deep space. Whereas the elementary reactions of carbon and dicarbon with ammonia can account for up to 75 % of the methylene amidogen and of the cyanomethyl radical in TMC-1, their contributions in Titan's atmosphere are less constrained; this is predominantly based on the uncertain abundances of atmospheric ammonia diverging by at least two orders of magnitude. This can only be solved through future in situ observations by, e.g., the prospective *Dragon Fly* mission. However, pyridine (**1**) and (iso)quinoline (**2/3**) – the most primitive nitrogen aromatics – have been detected in the Murchison (CM2) carbonaceous chondrite with abundances of up to 0.5 µg $g^{-1}$ (ppm)[75,76] thus providing a critical link between the low temperature chemistry in cold molecular clouds and their delivery to our solar system in form of meteorites. Overall, the present study provides a template for a systematic investigation of elementary reactions so that a comprehensive picture of the low temperature chemistry leading to biorelevant molecules in extraterrestrial environments emerges.

**Correspondence and requests for materials** should be addressed to R.I.K., A.M.M., B.R.L.G., J.-C.L., and X.L.

**Acknowledgements**

This work was supported by the U.S. Department of Energy, Basic Energy Sciences, by Grants No. DE-FG02-03ER15411 to the University of Hawaii at Manoa. The support of Conselho Nacional de Desenvolvimento Científico e Tecnológico (CNPq), Grant 311508/2021-9 and 405524/2021-8, is also acknowledged. We would like to acknowledge fruitful discussions on the fractional abundances of ammonia with Drs. Conor A. Nixon (NASA Goddard) and Karen Willacy (JPL).

**Author Contributions**

R.I.K. designed the experiments; Z.Y., C.H., and S.J.G. preformed the experiments; A.M.M., P.F.G.V., M.O.A., and B.R.L.G. conducted the electronic structure calculations; J.-C.L, K.M.H., and M.D. conducted the atmospheric modeling of Titan; X.L. performed the astrochemical modeling of TMC-1; Z.Y. and R.I.K. analyzed data and wrote the manuscript. All authors discussed the data.

**Competing interests**

The authors declare no competing financial interests.

## Methods

## Crossed Molecular Beams

### C-$ND_3$ system

The bimolecular reaction of ground-state atomic carbon (C; $^3P$) with D3-ammonia ($ND_3$; $X^1A_1$) was explored under single collision conditions employing a crossed molecular beams machine[32]. A supersonic beam of atomic carbon was produced by ablating a rotating graphite rod at 266 nm (Nd:YAG, 10-12 mJ pulse$^{-1}$, 30 Hz). The ablated carbon species were seeded in helium gas (99.9999%, 4 atm, 60 Hz). The primary carbon beam was velocity-selected by a chopper wheel (120 Hz) after passed through a skimmer revealing a well-defined $v_P$ (peak velocity) of 2512 ± 49 m s$^{-1}$ and S (speed ratio) of 2.9 ± 0.3 (Table S1). Carbon atoms in the primary beam are only in the ground state ($^3P$) under these conditions. Operation conditions were optimized that dicarbon in the primary beam was reduced to levels of less than 5%, which does not interfere with the scattering signal. The secondary beam of D3-ammonia ($ND_3$; Sigma-Aldrich; 99% D) was released with a backing pressure of 550 Torr and 60 Hz, characterized by a $v_P$ of 1091 ± 25 m s$^{-1}$ and S of 10.1 ± 1.3. Finally, the carbon beam crossed the $ND_3$ beam perpendicularly in the interaction region resulting in a collision energy ($E_C$) of 28.1 ± 0.9 kJ mol$^{-1}$ and a CM angle ($\Theta_{CM}$) of 35.9 ± 0.5°.

### $C_2$-$NH_3$ system

A pulsed supersonic dicarbon beam [$C_2$ ($X^1\Sigma_g^+/a^3\Pi_u$)] was produced exploiting the same ablation source described above. Briefly, the graphite rod was ablated by focusing the 266 nm laser output at 30 Hz and energy of 8-10 mJ pulse$^{-1}$. The ablated species were seeded in Neon (Ne, 4 atm, 99.9999%). Operation conditions and delay times were optimized to maximize dicarbon concentrations in the primary beam. Laser-induced fluorescence (LIF) of dicarbon revealed both singlet ground state ($X^1\Sigma_g^+$) and the lowest lying triplet state ($a^3\Pi_u$) along with the ro-vibrational distribution. The rotational temperature ($T_{rot}$) for the vibrational levels of ν = 0, 1 of the $a^3\Pi_u$ state were measured to be 240 ± 30 and 190 ± 30 K with fractions of 0.67 ± 0.05 and 0.33 ± 0.05, respectively, via the Swan band transition ($d^3\Pi_g$-$a^3\Pi_u$). The singlet state was detected via the Mulliken excitation ($D^1\Sigma_u^+$-$X^1\Sigma_g^+$) and the bimodal rotational distributions of both ν = 0, 1 were revealed at fractions of 0.83 ± 0.10 and 0.17 ± 0.04, respectively. Rotational temperature of the first and second vibrational levels was derived to be 200 K with population fraction of 0.44 ± 0.05 and 0.06 ± 0.02 together with 1000 K with fraction of 0.39 ± 0.05 and 0.11 ± 0.02, respectively[77].

The secondary ammonia ($NH_3$, Matheson, 99.99%) beam was released with 550 Torr backing pressure. The peak velocities, speed ratios of the primary and secondary beam along with the derived collision energy and CM angles of the $C_2$-$NH_3$ system are tabulated in Table S1.

For both reactions, the products were detected by a rotatable detection system at ultrahigh-vacuum conditions ($6\times10^{-12}$ Torr). In detail, the neutral species were ionized with an electron impact ionizer (80 eV; 2 mA) before they are mass-selected by a quadrupole mass spectrometer (QMS) in the time-of-flight (TOF) mode. The ions at a selected *m/z* will eventually lead to the signal detected and filtered by a photomultiplier tube (PMT; model 8850; -1.35 kV) and a discriminator (1.6 mV). Finally, a multichannel scaler is used to collect the TOF spectra at different angles. These laboratory data are converted into the CM frame with a forward-convolution method yielding the CM translational energy ($P(E_T)$) and angular ($T(\theta)$) flux distributions, with which the information of the reaction dynamics can be extracted[32]. The reactive differential cross section, $I(u, \theta) \sim P(u) \times T(\theta)$, which reports the product intensity ($I$) as a function of the center-of-mass angle $\theta$ and the velocity $u$, represents an overall image of the reaction and contains all the information of the scattering process[32].

**Electronic Structure Calculations**

The electronic structure calculations reported in this work for the carbon and dicarbon reactions with ammonia (Figure 4, Table S5-6) were performed with the MOLPRO[78] software. The geometry optimizations and harmonic frequencies calculations employed the coupled-cluster singles and doubles plus perturbative triples – CCSD(T) – method[79]. For such optimizations and frequencies, the augmented correlation consistent family basis set aug-cc-pV*X*Z[80] was employed, with a quadruple-zeta ($X=Q$) basis being used for the $CNH_3$ system, and a triple-zeta ($X=T$) for $CCNH_3$. For both systems, a final single point energy calculation was performed at the optimized geometries with the explicitly correlated CCSD(T)-F12 method[35] with the quadruple-zeta basis set cc-pVQZ-F12[81]. Using the conventional notation, the reported energies for the carbon-ammonia system are therefore CCSD(T)-F12/cc-pVQZ-F12//CCSD(T)/aug-cc-pVQZ+ZPE(CCSD(T)/aug-cc-pVQZ) and for dicarbon-ammonia CCSD(T)-F12/cc-pVQZ-F12//CCSD(T)/aug-cc-pVTZ+ZPE(CCSD(T)/aug-cc-pVTZ). For the larger $C_5H_5N$ system where the potential energy surface is accessed by the $C_3H_3$ + $H_2CCN$, $C_4H_3$ + $H_2CN$, and $C_4H_3$ + cis-HCNH reactions (Figure 5, Table S7), geometry optimizations and harmonic frequencies calculations used the density functional (DFT) B3LYP method[82,83] with the cc-pVTZ basis and single-point energies refined at the CCSD(T)-F12/cc-pVTZ-F12 level. Thus, the reported energies for $C_5H_5N$ species are obtained at CCSD(T)-F12/cc-pVTZ-F12//B3LYP/cc-pVTZ+ZPE(B3LYP/cc-pVTZ) employing the Gaussian 16[84] and MOLPRO[78] software packages.

## Data availability

The data that support the plots within this paper and other finding of this study are available from the corresponding author upon reasonable request.

## Supplementary information

Astrochemical modeling details, Figure S1 to S5, Table S1 to S7.

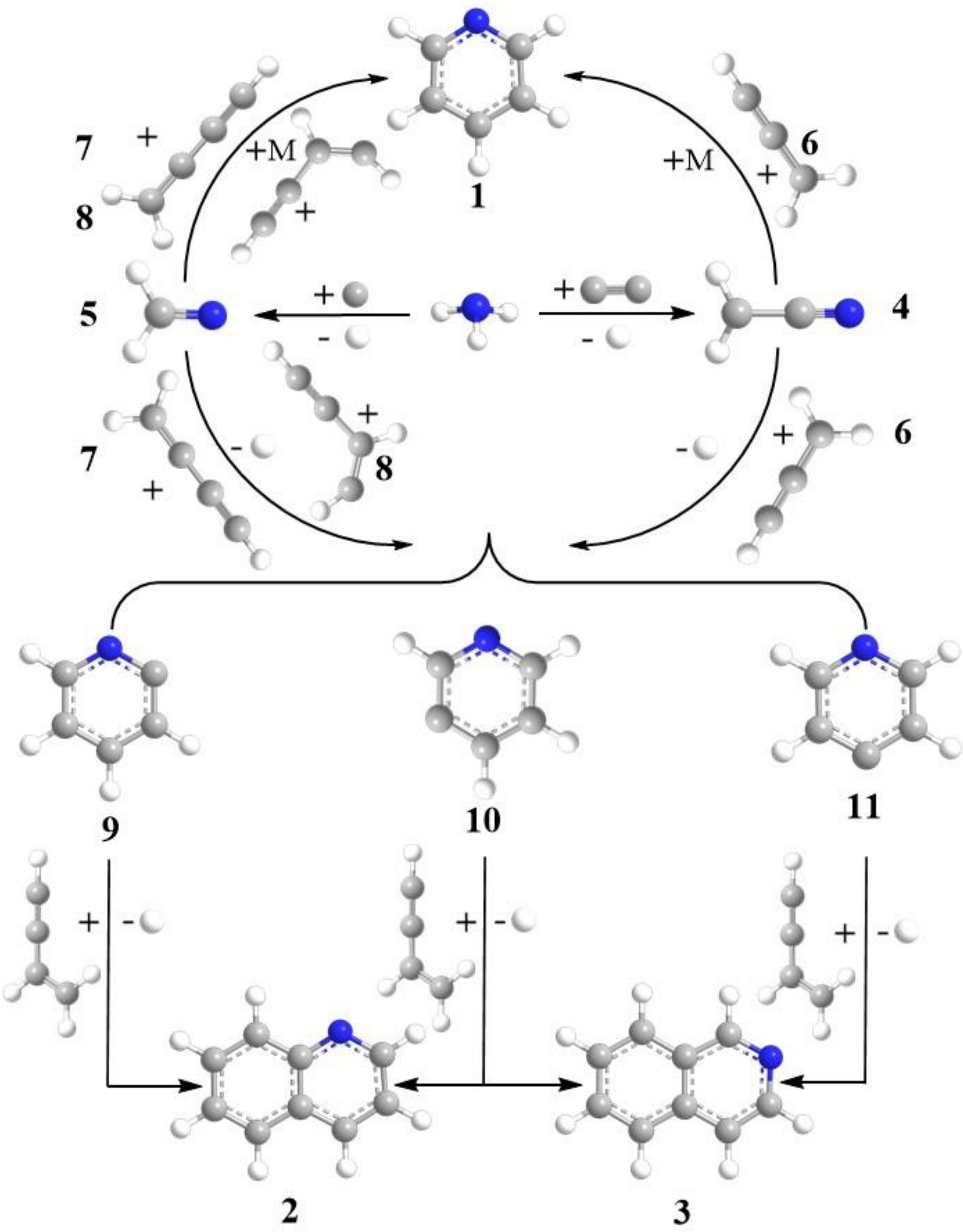

**Figure 1**. **Pathways to pyridine and (iso)quinoline.** A chain of reactions initiated through the formation of methylene amidogen radical ($H_2CN^{\bullet}$) and cyanomethyl radical ($H_2CCN^{\bullet}$) lead to the simplest representative of mono and bicyclic aromatic molecules carrying nitrogen. The reactions of atomic carbon and dicarbon with ammonia leading to methylene amidogen radical (**5**) and cyanomethyl radical (**4**) are investigated via our crossed molecular beam machine; our calculations also predict the formation of pyridine (**1**) and pyridinyl radicals (**9-11**) through the reactions of methylene amidogen with i/n-$C_4H_3$ and of cyanomethyl with propargyl. The reactions of pyridinyl radicals (**9-11**) with vinylacetylene ($C_4H_4$) forming (iso)quinoline are depicted in Ref. 29.

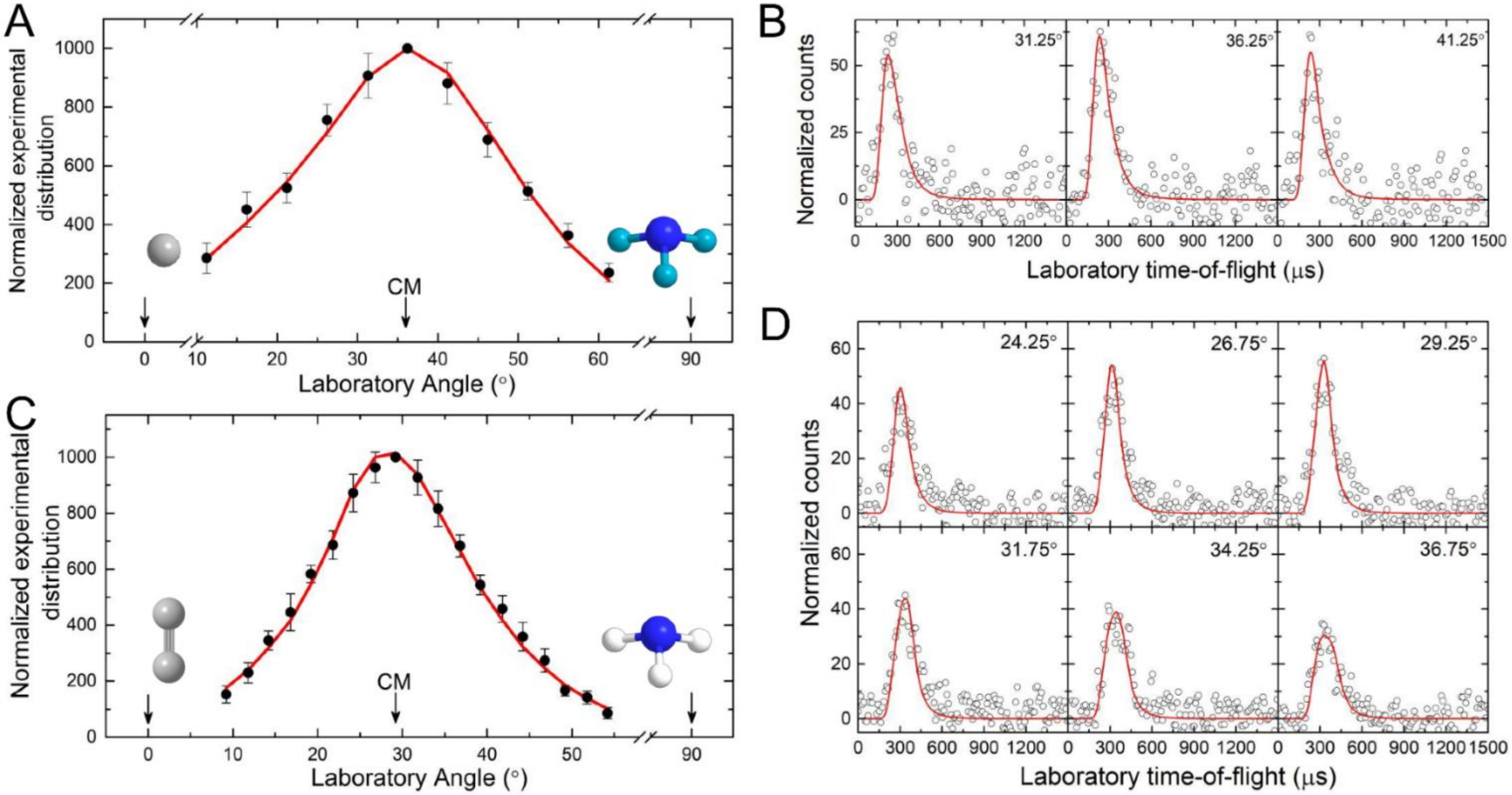

**Figure 2**. **Laboratory data of the C-ND3 and C2-NH3 reactions**. Laboratory angular distributions (A, C), and time-of-flights (B, D) for the carbon – D3-ammonia (A, B) and dicarbon – ammonia (C, D) reactions. The solid circles with their error bars represent the normalized experimental distribution; the open circles indicate the experimental data. The red lines represent the best fits obtained. Atoms are color coded as follows: carbon, gray; nitrogen, blue; deuterium, light blue; and hydrogen, white.

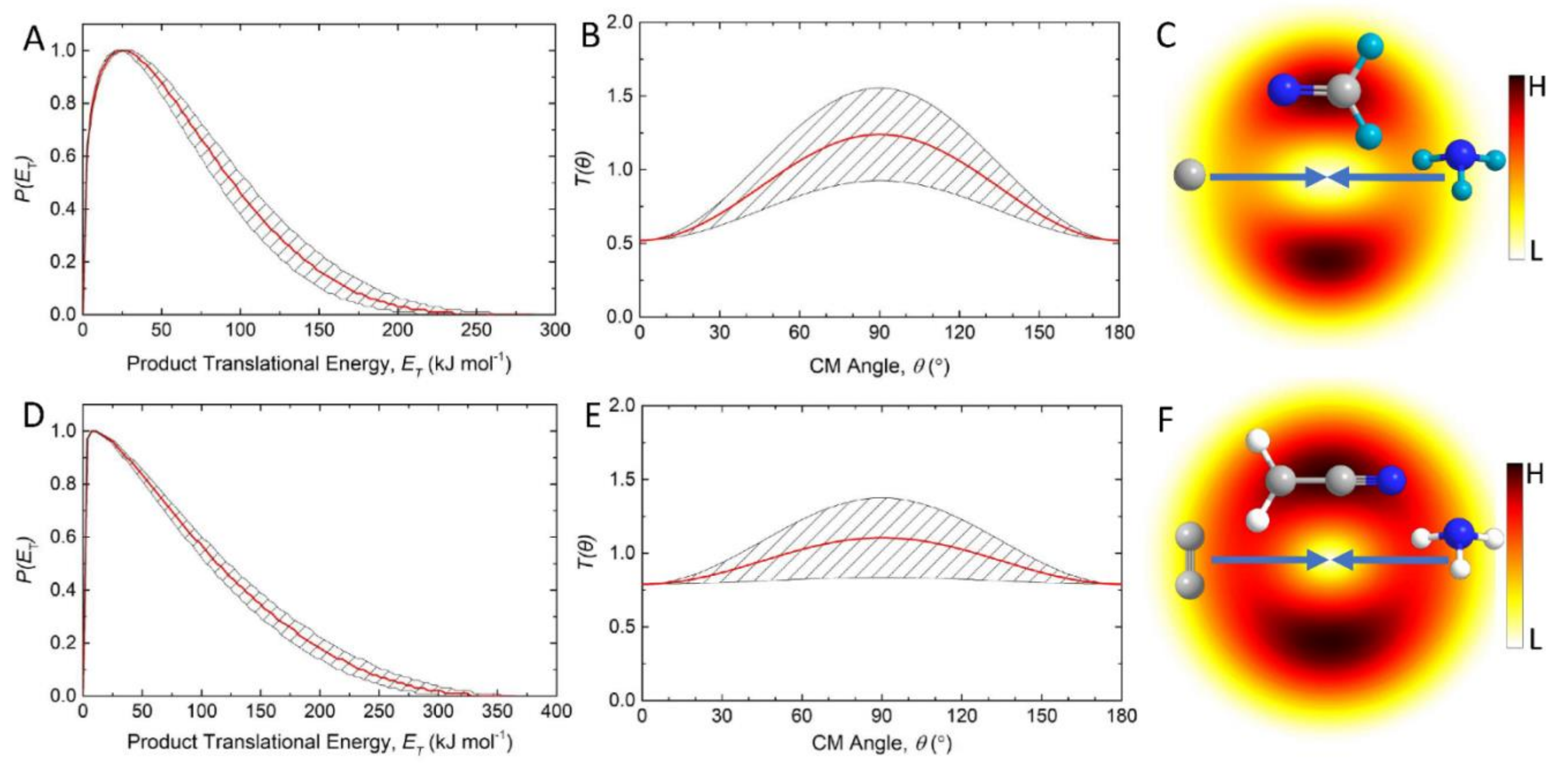

**Figure 3**. **CM functions of the C-$ND_3$ and $C_2$-$NH_3$ reactions**. Center-of-mass translational energy distributions (*$P(E_T)$*; A, D), angular flux distributions (*$T(\theta)$*; B, E), and the corresponding flux contour map (C, F) for the carbon – D3-ammonia (A, B, C) and dicarbon – ammonia (D, E, F) reactions. The red lines represent the best-fit; shaded areas depict the error limits of the best fits.

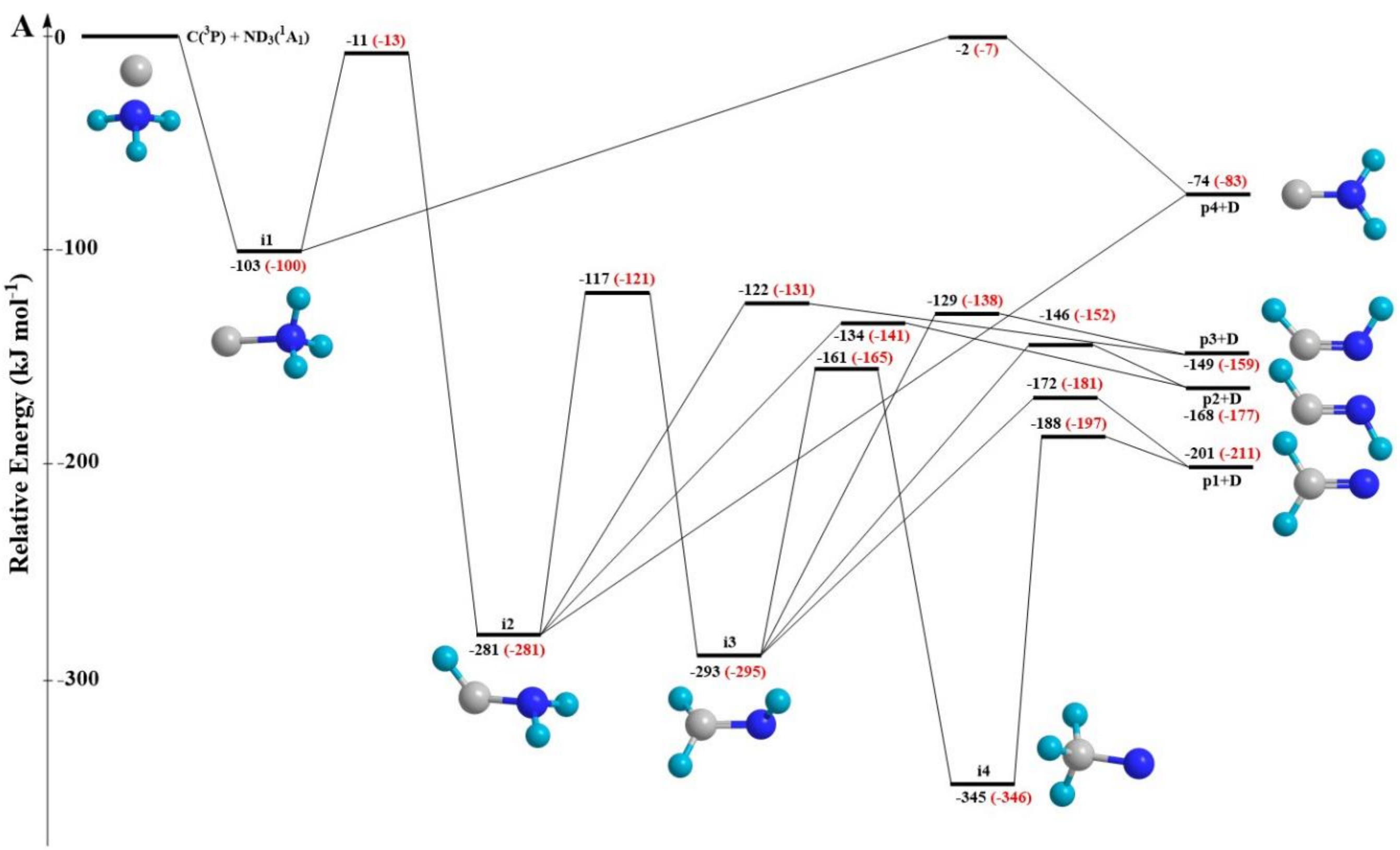

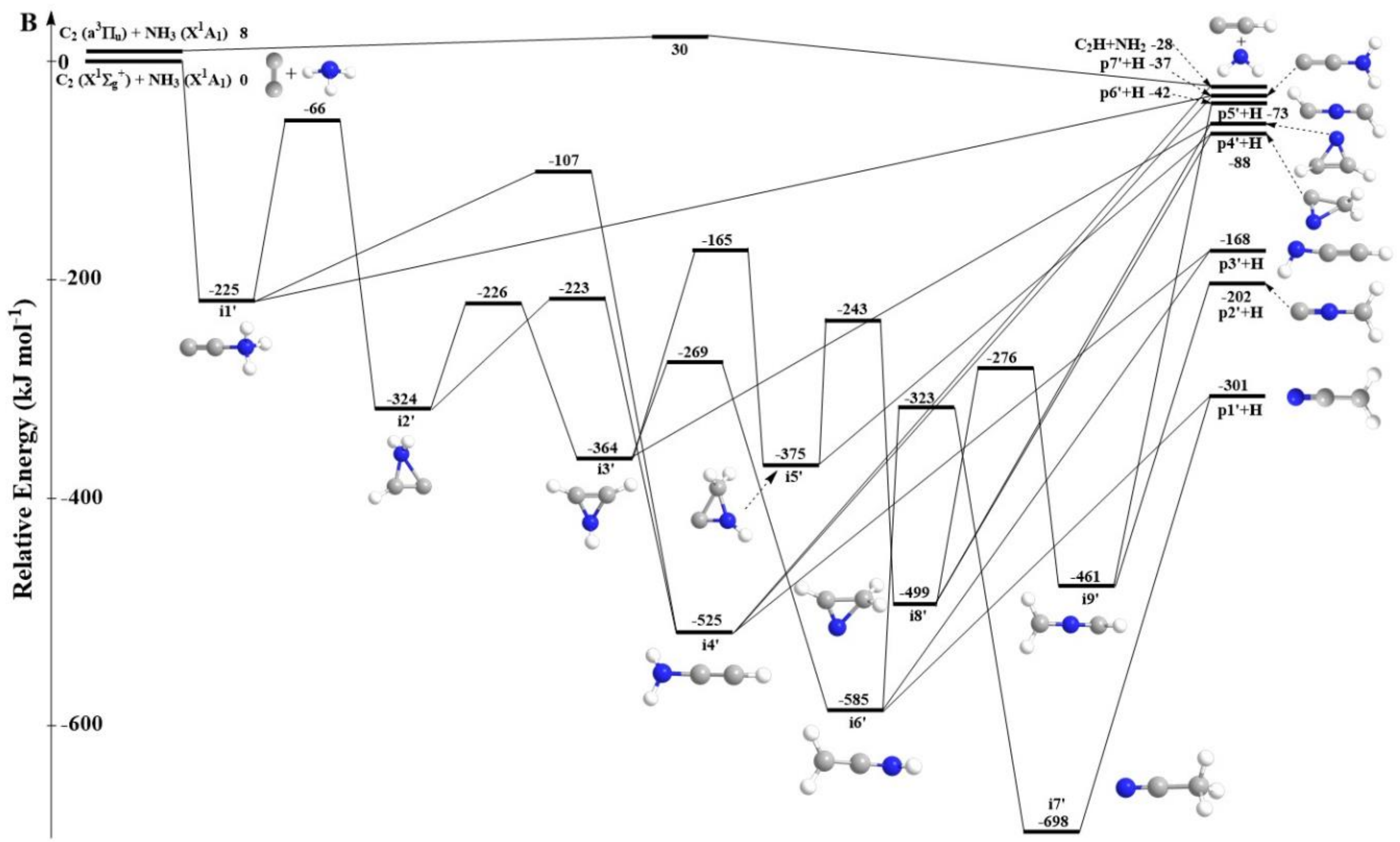

**Figure 4**. **Potential energy surfaces of the reactions of carbon – D3-ammonia (A) and dicarbon – ammonia (B).** For the carbon – D3-ammonia reaction, energies provided in black are relative energies for the deuterated reactants, whereas the energies in red refer to the hydrogenated reactants. Corresponding Cartesian coordinates and vibrational modes are compiled in the Supplementary Information.

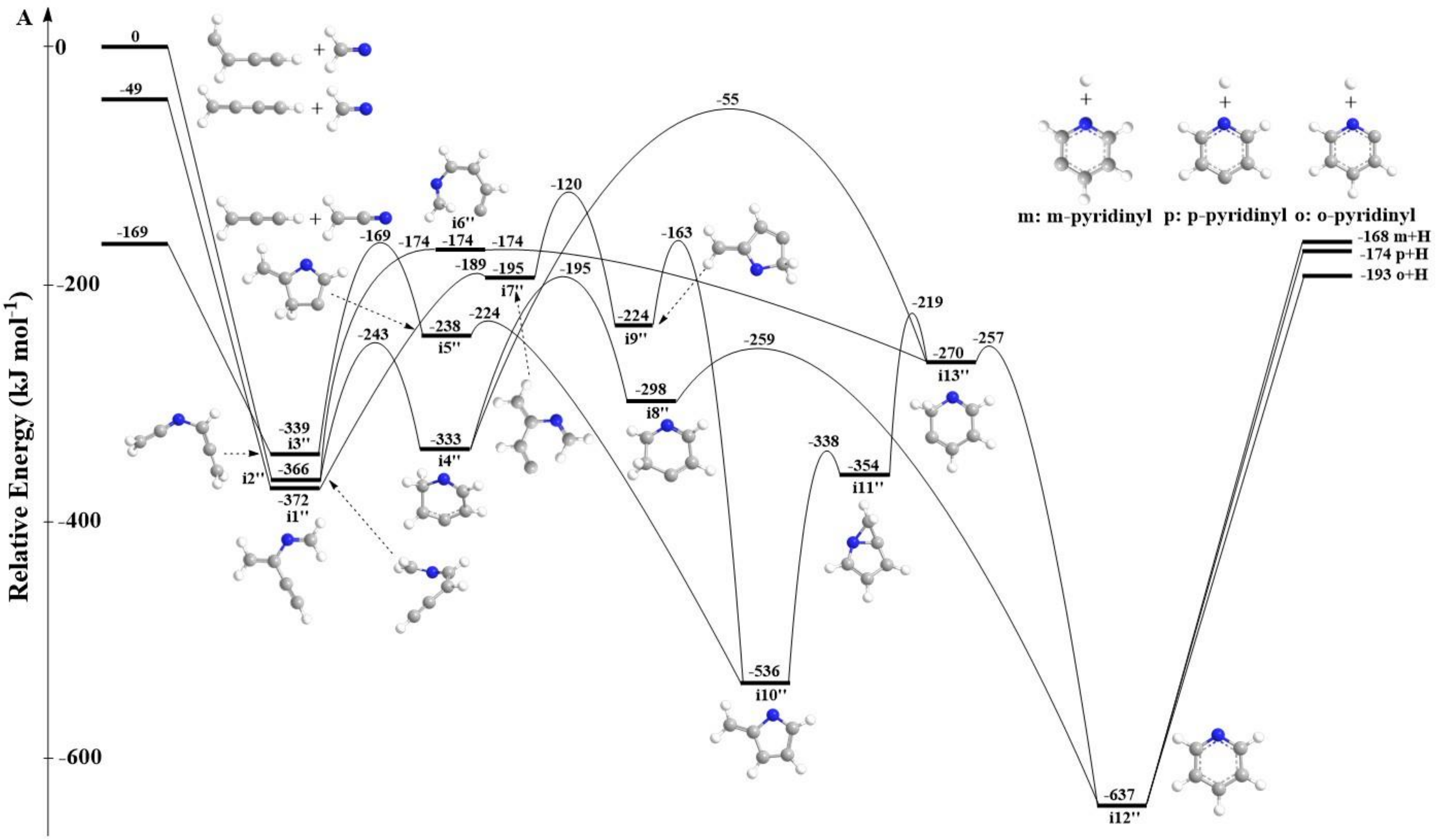

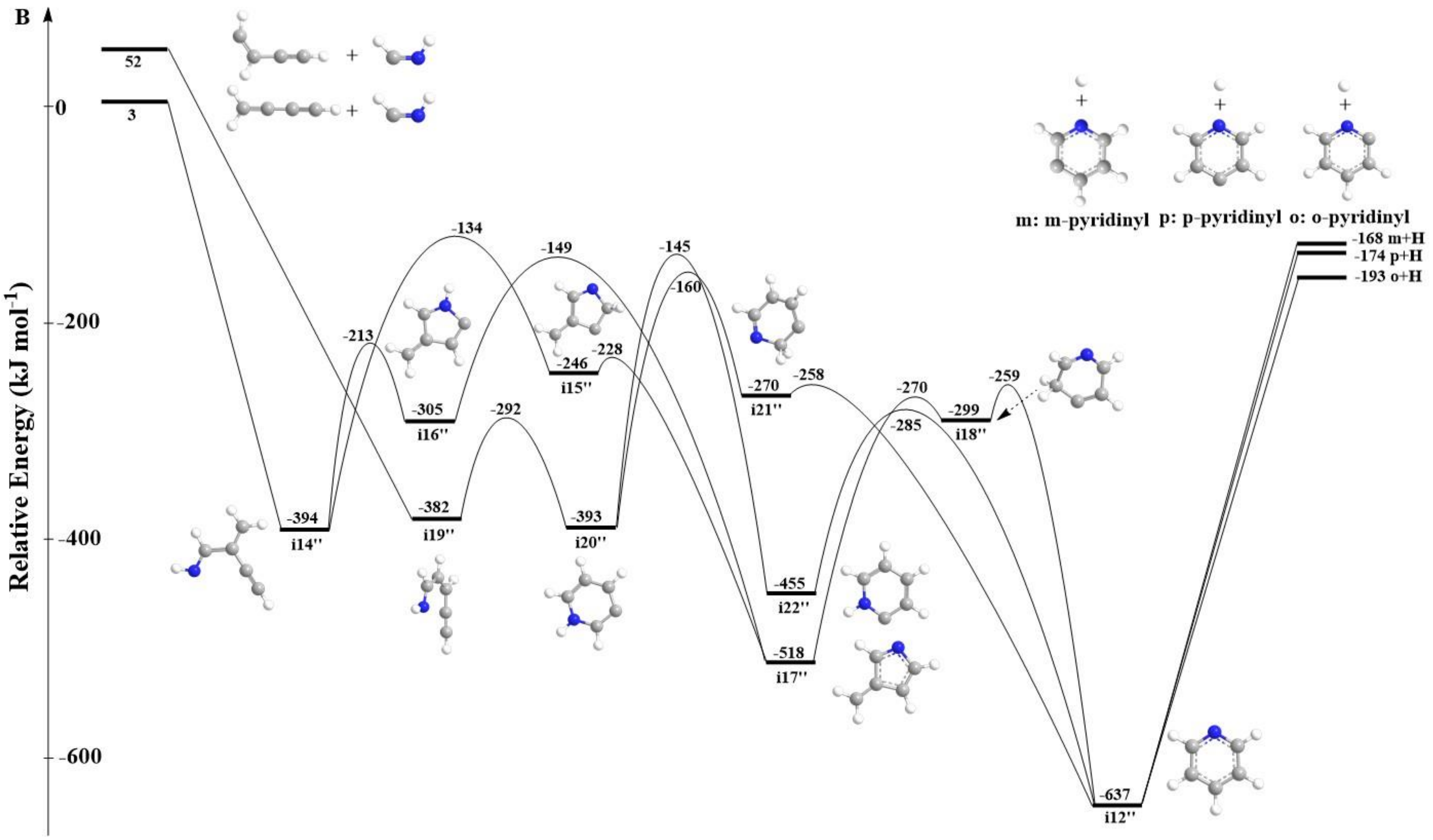

**Figure 5. Formation pathways to pyridinyl radicals and the pyridine intermediate.** Distinct pyridinyl radicals and pyridine can be formed from reactions of methylene amidogen ($H_2CN$) with i/n-$C_4H_3$ isomers and the cyanomethyl ($H_2CCN$) with propargyl (A) and cis-iminomethyl (HCNH) with i/n-$C_4H_3$ isomers (B).

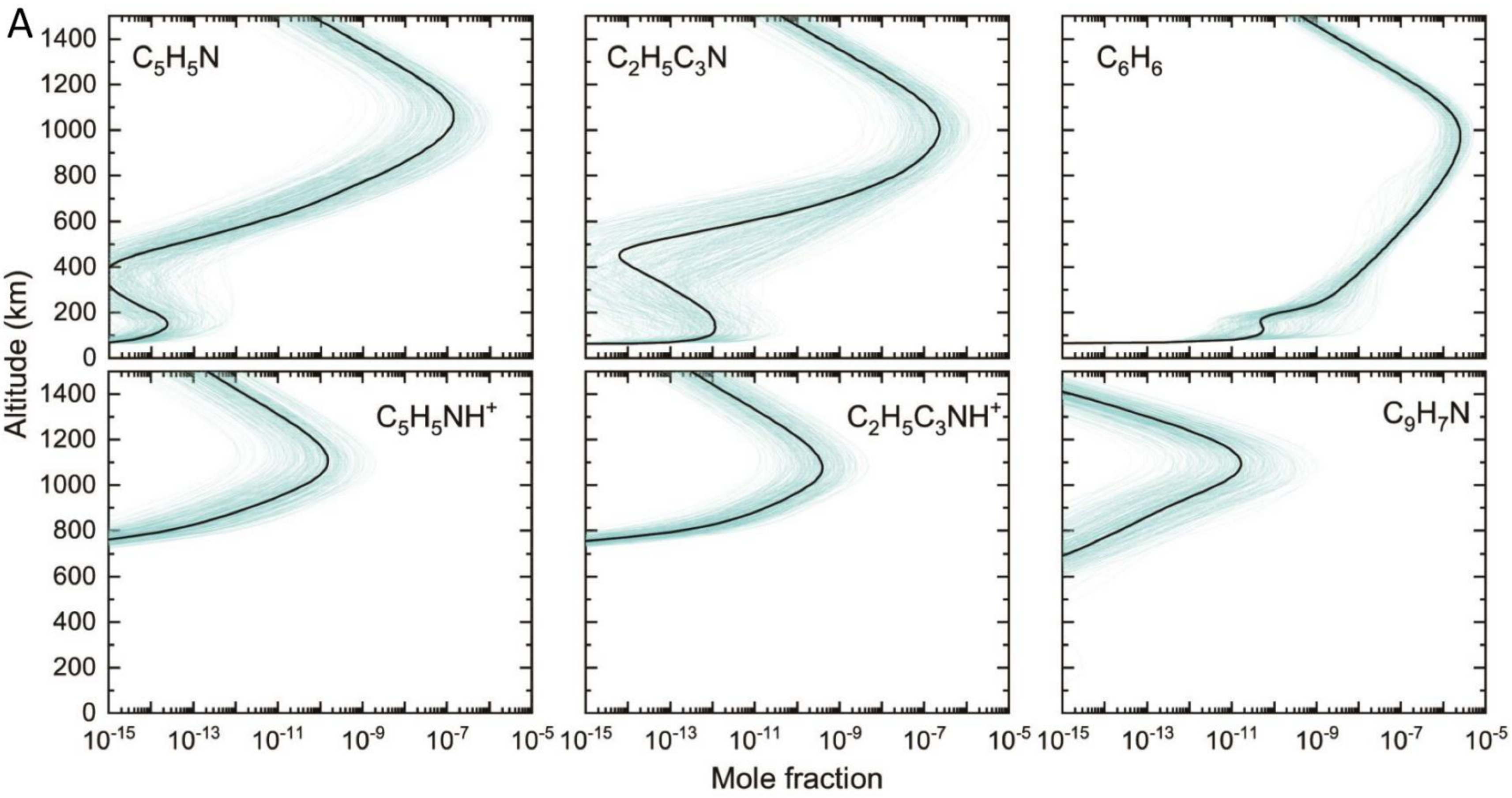

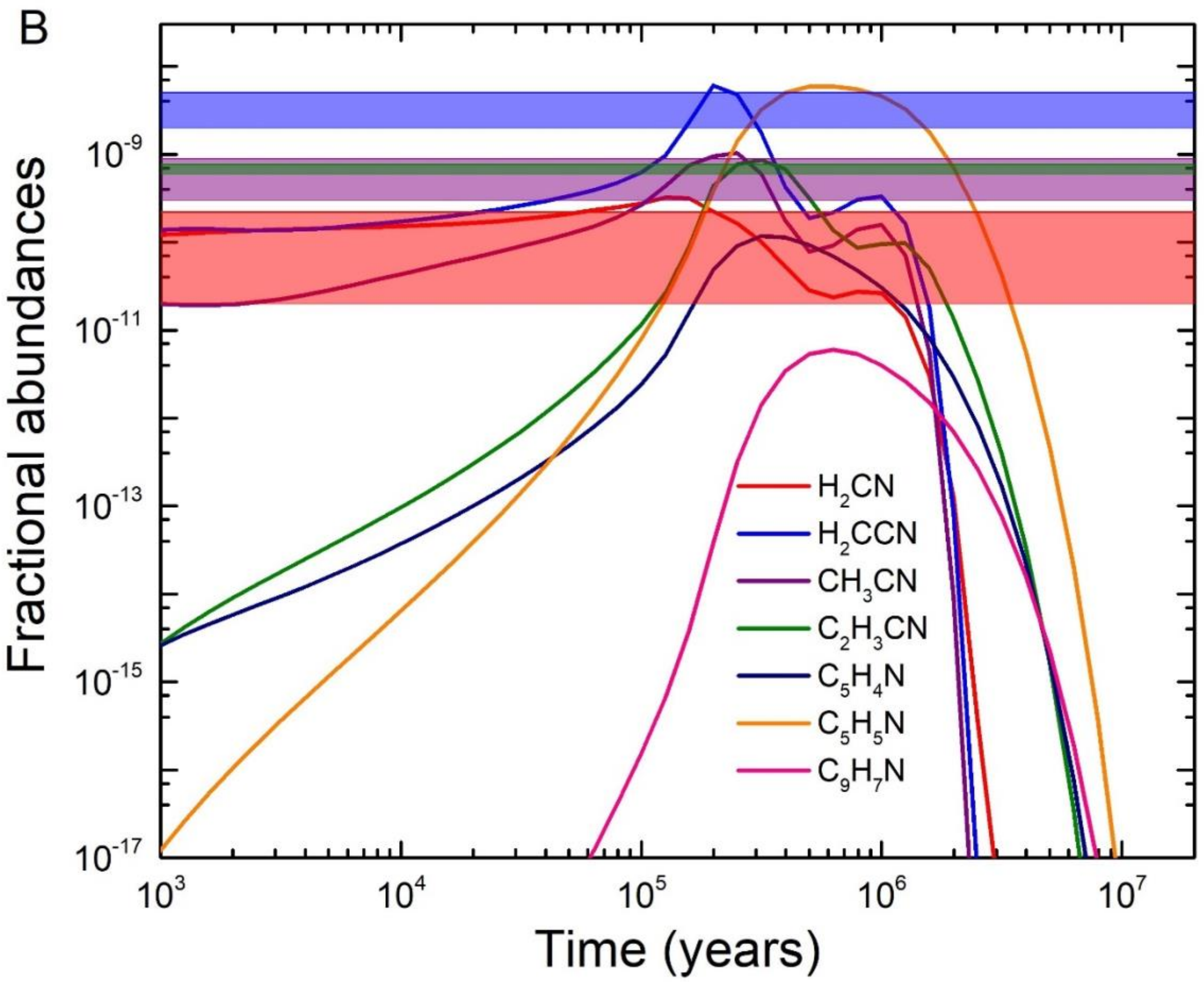

**Figure 6**. **Results of the astrochemical model for Titan's atmosphere and TMC-1.** Mole fraction profiles for six species obtained from a one-dimensional photochemical model of Titan's atmosphere (A) and the fractional abundance of the gas-phase methylene amidogen ($H_2CN$, red), cyanomethyl ($H_2CCN$, blue), methyl cyanide ($CH_3CN$, purple), vinyl cyanide ($C_2H_3CN$, olive), pyridinyl ($C_5H_4N$, navy), pyridine ($C_5H_5N$, orange), and (iso)quinoline ($C_9H_7N$, pink) plotted as a function of time (B). The solid black lines in Figure 6A represent the nominal model results, with 280 runs of the Monte Carlo analysis displayed as cyan lines. Please confer to the text for details on the error analysis and the assignment of the species. Astronomically observed fractional abundances of the four species are visualized with the colored horizontal bars in Figure 6B.

# Supplementary information

**This file includes:**

The file includes:
Text: 853 words
Number of references: 14
Number of figures: 5
Number of tables: 7

**Astrochemical modeling**

The main aspects of the chemistry of the formation of pyridine ($C_5H_5N$) and ethylcyanoacetylene ($C_2H_5C_3N$) (the most abundant linear $C_5H_5N$ isomer deduced from production fluxes) and (iso)quinoline ($C_9H_7N$) are presented below[1-5]:

| | | |
|---|---|---|
| $N + CH_3$ | $\rightarrow H_2CN + H$ | $\Delta H_R$ = -160 kJ/mol |
| $N + C_2H_3$ | $\rightarrow CH_2CN + H$ | $\Delta H_R$ = -292 kJ/mol |
| $C + NH_3$ | $\rightarrow H_2CN + H$ | $\Delta H_R$ = -217 kJ/mol |
| $C + CH_2NH$ | $\rightarrow CH_2CN + H$ | $\Delta H_R$ = -334 kJ/mol |
| | $\rightarrow CH_2$ + HCN/HNC | $\Delta H_R$ = -293/-241 kJ/mol |
| $C_2 + NH_3$ | $\rightarrow CH_2CN + H$ | $\Delta H_R$ = -332 kJ/mol |
| $CN + CH_3CH_2CCH$ | $\rightarrow C_2H_5C_3N + H$ | $\Delta H_R$ = -99 kJ/mol |
| | $\rightarrow C_2H_5 + HC_3N$ | $\Delta H_R$ = -118 kJ/mol |
| $CN + CH_2CHCHCH_2$ | $\rightarrow C_5H_5N + H$ | $\Delta H_R$ = -212 kJ/mol |
| | $\rightarrow CH_2CHCHCHCN + H$ | $\Delta H_R$ = -96 kJ/mol |
| $H_2CN + C_4H_3$ | $\rightarrow C_5H_5N$ | $\Delta H_R$ = -585 kJ/mol |
| | $\rightarrow$ o-$C_5H_4N + H$ | $\Delta H_R$ = -147 kJ/mol (-122, -129 for m-, p-) |
| | $\rightarrow HCN + C_4H_4$ | $\Delta H_R$ = -301 kJ/mol |
| | $\rightarrow CH_2NH + C_4H_2$ | $\Delta H_R$ = -177 kJ/mol |
| $CH_2CN + C_3H_3$ | $\rightarrow C_5H_5N$ | $\Delta H_R$ = -464 kJ/mol |
| | $\rightarrow$ o-$C_5H_4N$ + H | $\Delta H_R$ = -26 kJ/mol |
| | $\rightarrow HCCCH_2CH_2CN$ | $\Delta H_R$ = -292 kJ/mol |
| | $\rightarrow CH_2CCHCH_2CN$ | $\Delta H_R$ = -297 kJ/mol |
| $CH_2NCH + C_3H_3$ | $\rightarrow C_5H_5N + H$ | $\Delta H_R$ = -299 kJ/mol |
| H + o-$C_5H_4N$ | $\rightarrow C_5H_5N$ | $\Delta H_R$ = -438 kJ/mol |
| $CH_3 + CH_2C_3N$ | $\rightarrow C_2H_5C_3N$ | $\Delta H_R$ = -318 kJ/mol |
| $CH_3$ + o-$C_5H_4N$ | $\rightarrow$ o-$C_5H_4CH_3N$ | $\Delta H_R$ = -410 kJ/mol |
| | $\rightarrow$ o-$C_5H_4CH_2N$ + H | $\Delta H_R$ = -34 kJ/mol |
| $C_5H_4N + C_4H_4$ | $\rightarrow C_9H_7N + H$ | $\Delta H_R$ = -241 kJ/mol |

The rates of these barrier-free reactions are incorporated from capture theory when no experimental measurements exist, while the rates of the three-body reactions are taken to be equal to similar reactions of benzene derivatives. For some reactions such as CN + $CH_2CHCHCH_2$ (and also $CH_2CN + C_3H_3$) the formation of pyridine is in competition with the formation of nitriles, see for example[1,4]. The destruction of pyridine, $C_2H_5C_3N$ and quinoline occurs mainly through photodissociation, through reactions with N($^2$D)[6] and $C_2H$, and through reactions with ions such as $HCNH^+$, $C_2H_5^+$, $NH_4^+$. Unlike benzene where ionic chemistry plays an important role in its formation through the $C_4H_3^+ + C_2H_2$ and $C_4H_3^+ + C_2H_4$ reactions[7,8], ionic reactions do not appear to be effective in producing pyridine (nor likely quinoline). Indeed the pyridine forming reaction $C_4H_3^+$ + HCN is slow, in contrast to the benzene forming reaction $C_4H_3^+ + C_2H_2$[9], and even if the reaction $C_4H_3^+ + CH_2NH$ is fast (this reaction has never been studied to the best of our knowledge) the flux will be much smaller than the reaction $C_4H_3^+ + C_2H_4$ because $CH_2NH$ is much less abundant than $C_2H_4$ ($CH_2NH$ has not yet been detected in Titan's atmosphere). Moreover, the proton affinity of pyridine (937 kJ mol$^{-1}$) is very high which will induce a proton transfer from ions ($HCNH^+$, $C_2H_5^+$, $NH_4^+$ and so on) to $C_5H_5N$. If the electronic recombination reaction of $C_5H_5NH^+$ does not lead to a 100% yield of $C_5H_5N$ + H, not only will the ionic chemistry not induce the formation of pyridine, but it will promote its destruction instead.

The absorption spectrum of pyridine and quinoline and the products for the photodissociation of pyridine has been studied. The products for the photodissociation of quinoline are deduced from photodissociation of similar species[10-13]. For pyridine (and quinoline) photodissociation, the rate we use is in fact the maximum limit. Indeed, as for benzene, the lifetime of $C_5H_5N^{**}$ produced after photon absorption is quite long so some of these molecules will stabilize before dissociating[13]. On the other hand, the lifetime of pyridine is shorter than benzene (the pyridine lifetime is 0.1 μs following excitation at 193 nm compared with 10 μs for benzene) so photodissociation will be more efficient for pyridine than for benzene. The absorption of $C_2H_5C_3N$ has not been studied to the best of our knowledge but the photodissociation of this species will be dominated by the absorption of the chromophore -$C_3N$, so its absorption spectrum is taken to be the same as $CH_3C_3N$. The photodissociation products of $C_2H_5C_3N$ are assumed to be $C_2H_4$ + $HC_3N$ and $CH_3$ + $CH_2C_3N$.

To evaluate the uncertainties on the nominal model profiles, obtained using the recommended rate constants, a Monte Carlo simulation was performed according to the method described in Benne et al.[14]. Briefly, the rate constants for all reactions were recalculated using their associated uncertainty factors, $F_i$, the temperature-dependent uncertainty factor, and $g_i$, a coefficient used to extrapolate $F_i$ to lower temperature, allowing $F_i(T)$ to be determined. Each rate constant was considered to be a random variable, $k_i$, with a log-normal distribution centered on the nominal value, $k_{0_i}$, with a standard deviation, log $F_i$. For bimolecular reactions, $k_i$ was given by

$$\log(k_i) = \log(k_{0_i}) + \varepsilon_i \log[F_i(T)]$$

As $\varepsilon_i$ is a random number with a normal distribution centered on zero and a standard deviation of one, this led to a 68.3 % probability of finding a $k_i$ value within the interval $\left[\frac{k_{0_i}}{F_i}, k_{0_i} \times F_i\right]$. 280 simulations were performed using this procedure to establish the possible dispersion from the nominal mole fraction profiles displayed in Figure 6.

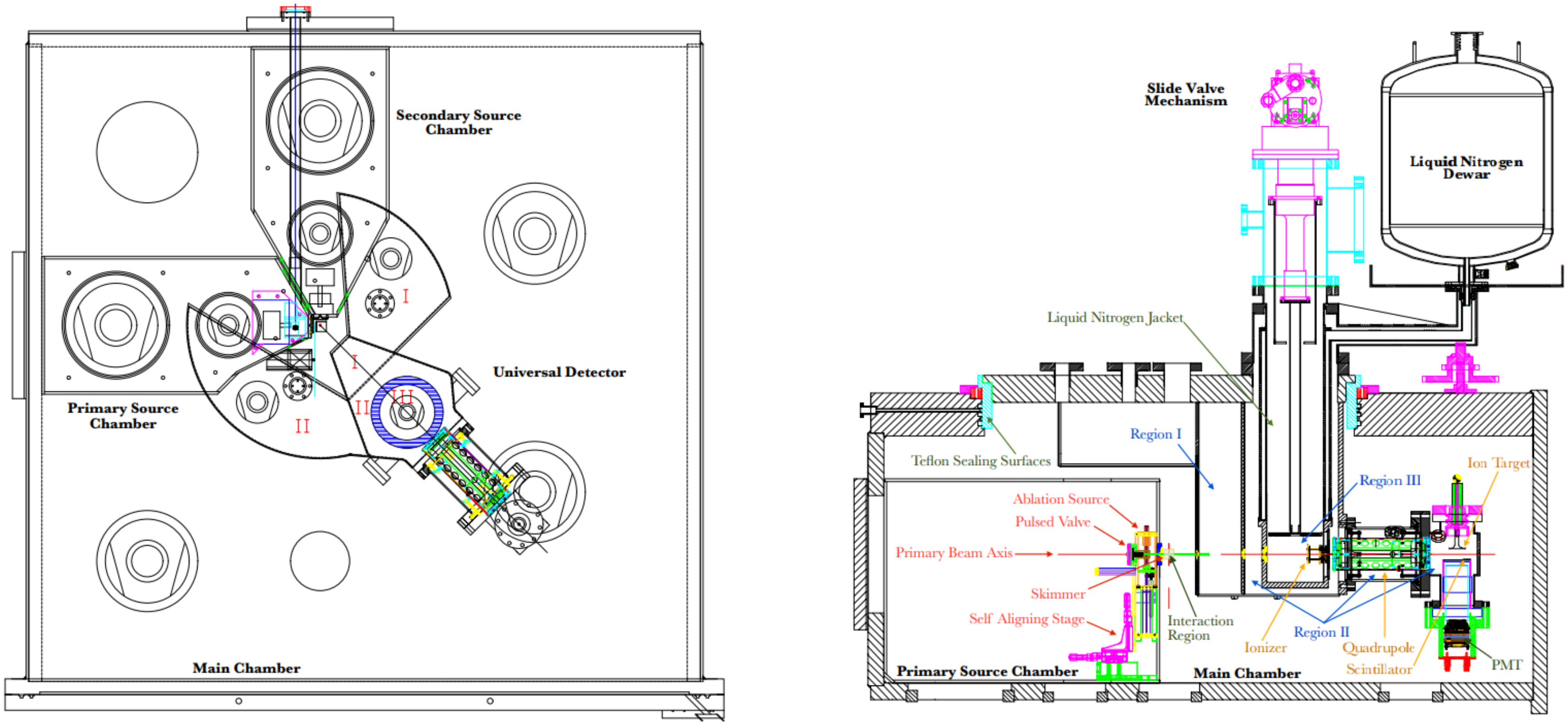

**Figure S1**. Schematic of the crossed molecular beam machine.

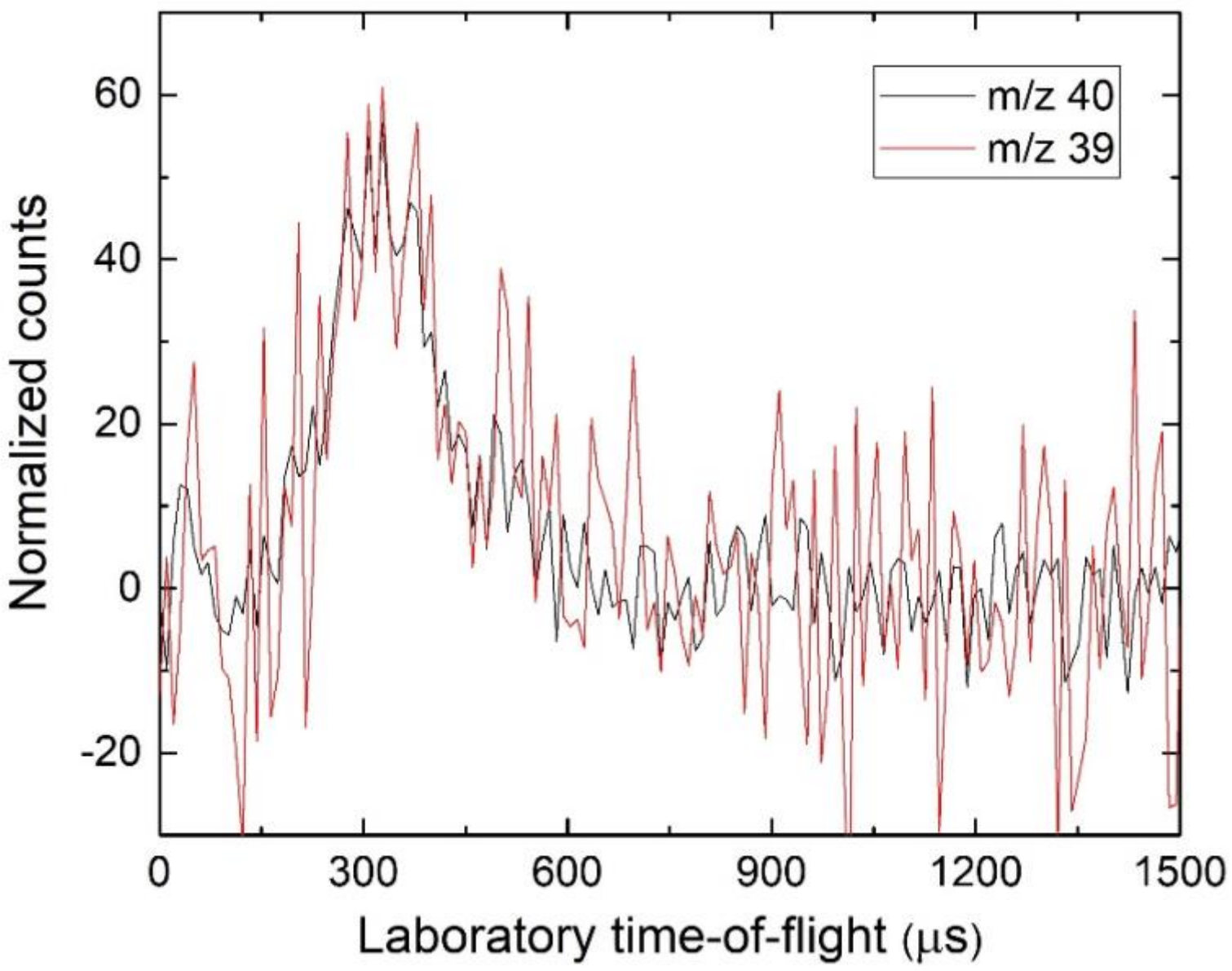

**Figure S2**. Scaled time-of-flight spectra recorded at the CM angles for m/z = 40 and 39 for the $C_2$-$NH_3$ reaction.

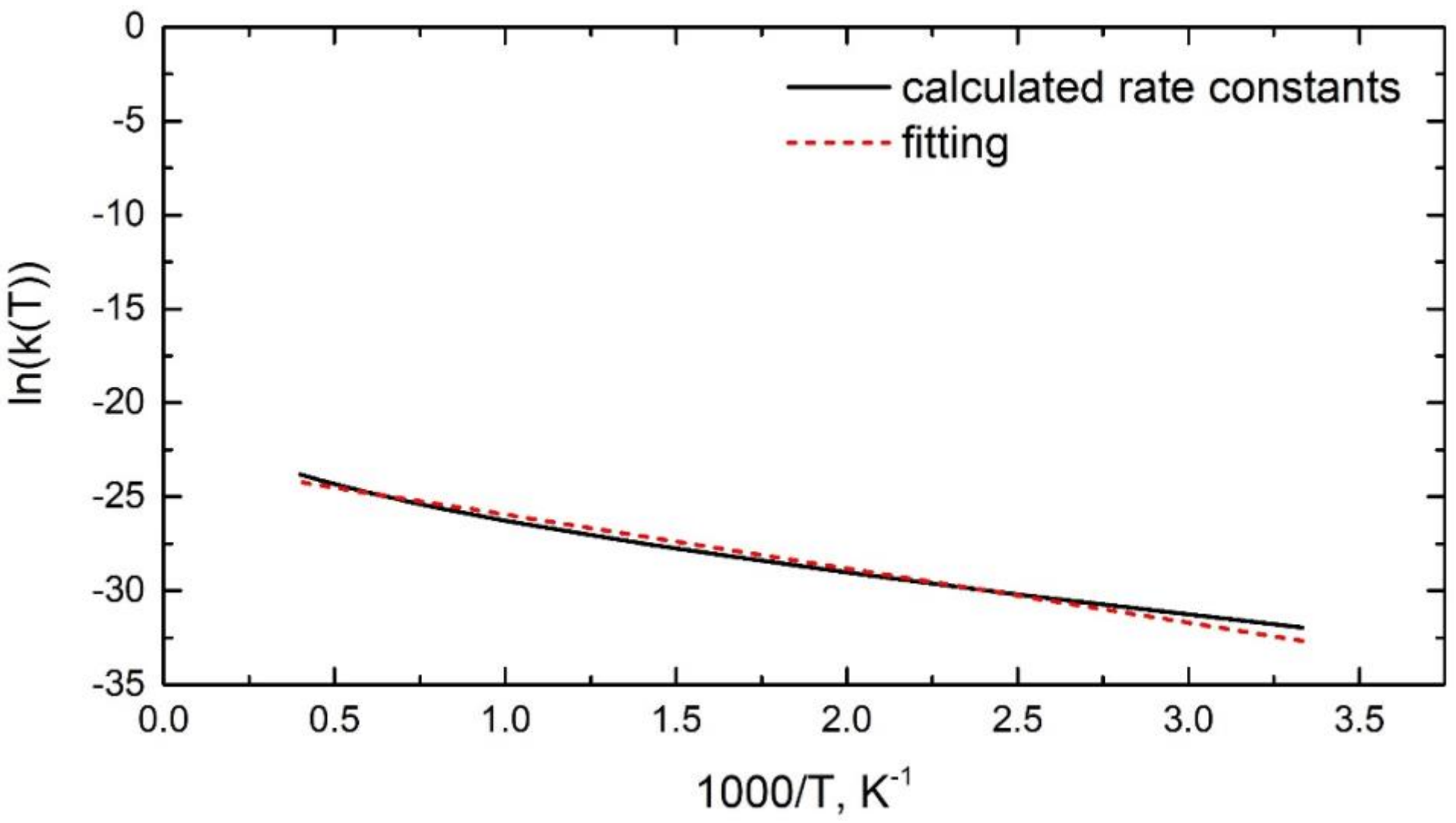

**Figure S3**. Temperature dependence of the thermal rate constant k(T) for the **p3** + H in the C + $NH_3$ reaction. The thermal rate constants are obtained using TST calculations.

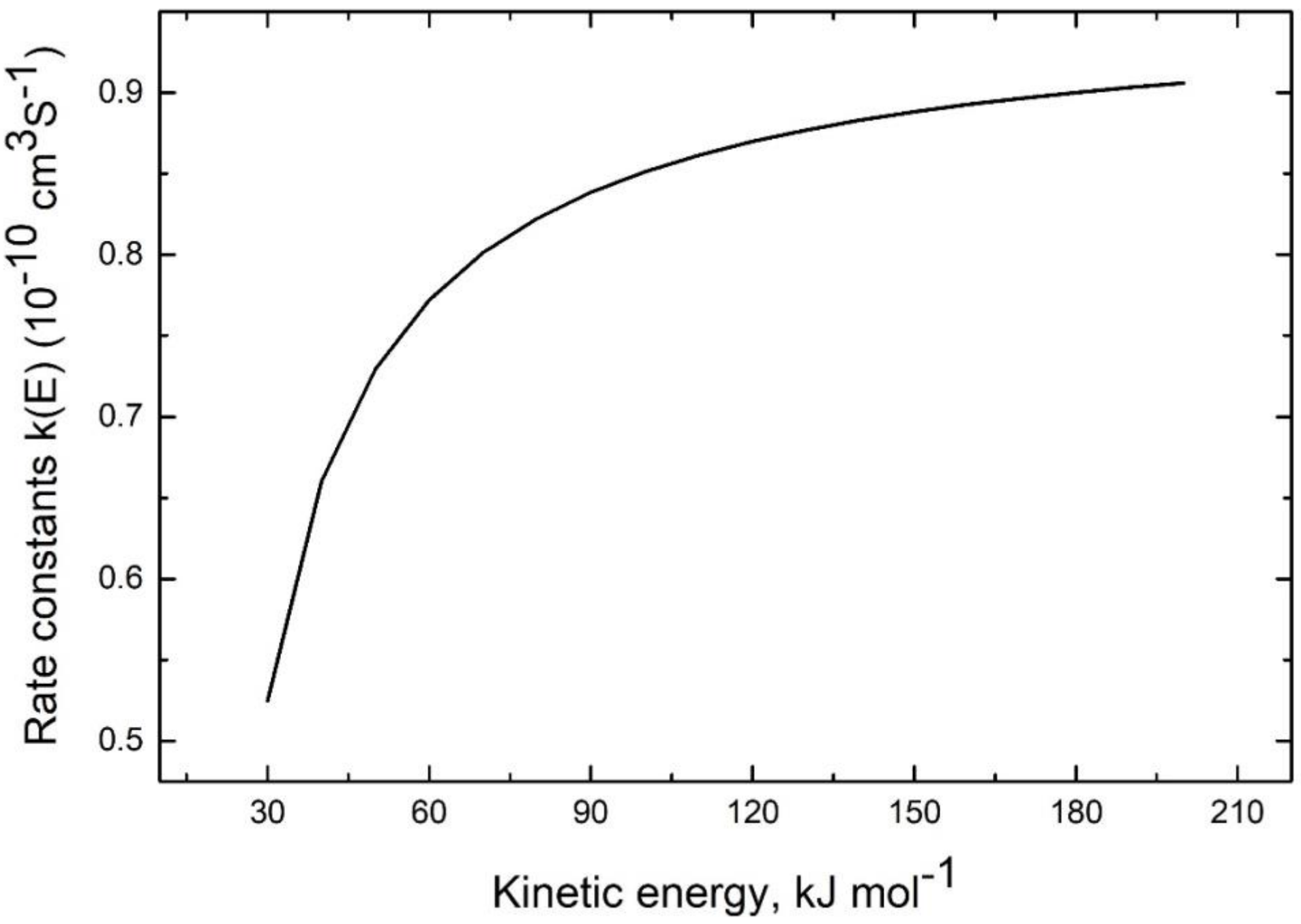

**Figure S4**. Energy dependence of the suprathermal rate constants k(E) for the **p3** + H in the C + $NH_3$ reaction. The calculation details are described in Ref. 38.

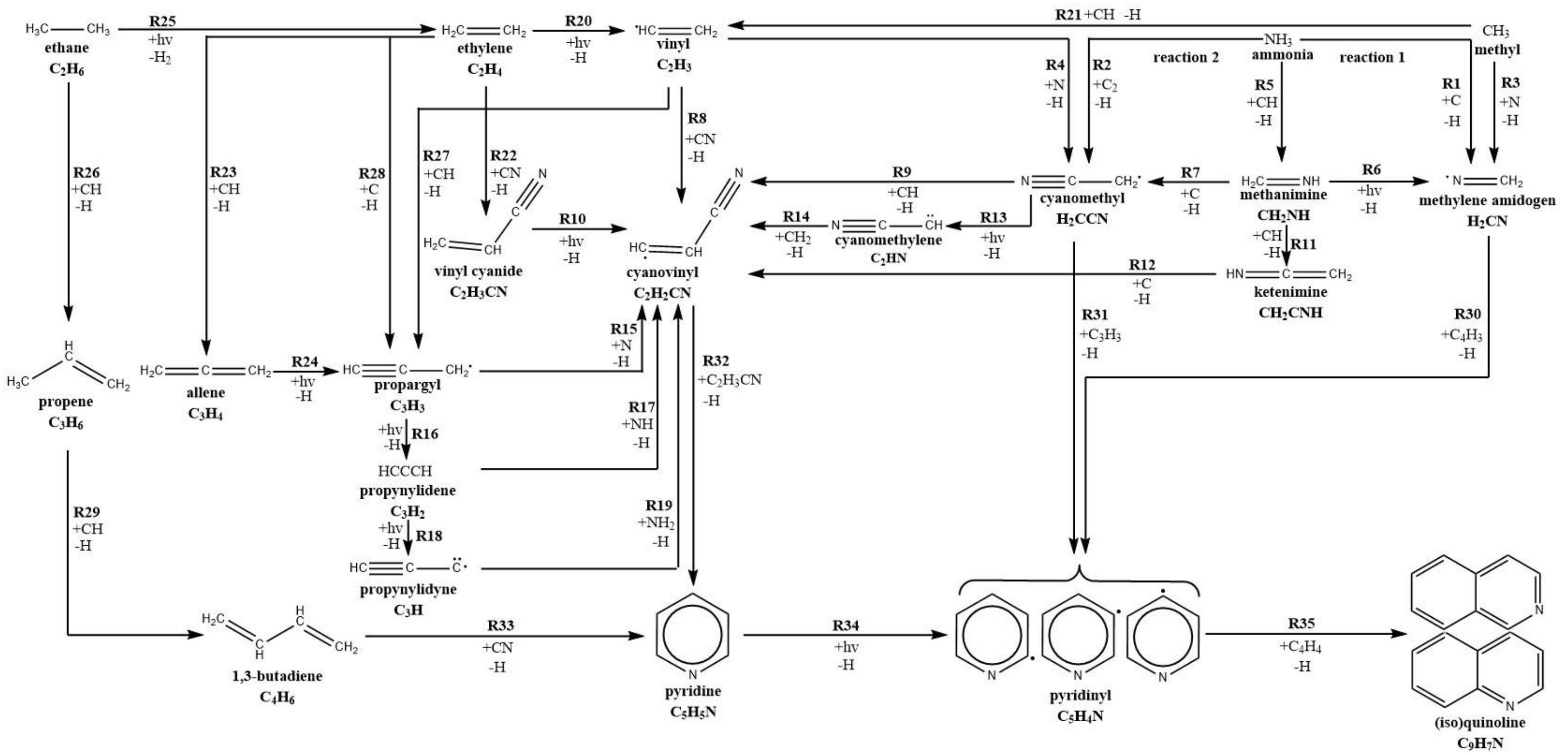

**Figure S5**. Compilation of key bimolecular reactions and photodissociation processes newly introduced into the astrochemical model for TMC-1 leading to pyridine, pyridinyl, and (iso)quinoline.

**Table S1.** Peak velocity ($v_p$) and speed ratios (S) of the atomic carbon (C; $^3P$), dicarbon ($C_2$; $X^1\Sigma_g^+/a^3\Pi_u$), ammonia ($NH_3$; $X^1A_1$), and D3-ammonia ($ND_3$; $X^1A_1$) beams along with the corresponding collision energy ($E_C$) and center-of-mass angle ($\Theta_{CM}$) for each reactive scattering experiment.

| beam | $v_p$ (m $s^{-1}$) | S | $E_c$ (kJ $mol^{-1}$) | $\Theta_{CM}$ (deg) |
|---|---|---|---|---|
| C | 2512 ± 49 | 2.9 ± 0.3 | | |
| $ND_3$ | 1091 ± 25 | 10.1 ± 1.3 | 28.1 ± 0.9 | 35.9 ± 0.5 |
| $C_2$ | 1451 ± 19 | 3.2 ± 0.1 | | |
| $NH_3$ | 1143 ± 34 | 10.3 ± 1.4 | 17.0 ± 0.3 | 29.2 ± 0.3 |

**Table S2**. Statistical branching ratios (%) for the reaction of carbon with D3-ammonia ($ND_3$) at the collision energies ($E_c$, kJ $mol^{-1}$) of 0 and 28.1 kJ $mol^{-1}$.

| $E_c$ | **p1** | **p2** | **p3** | **p4** |
|---|---|---|---|---|
| 0 | 8 | 41 | 33 | 18 |
| 28.1 | 7 | 33 | 29 | 31 |

**Table S3.** The RRKM rate constants ($s^{-1}$) for the reaction of carbon with D3-ammonia ($ND_3$) computed at collision energies of 0.0 and 28.1 kJ/mol.

| | **0.0** | **28.1** |
|---|---|---|
| (i1$\rightarrow$i2) | $5.27\times10^{10}$ | $4.13\times10^{11}$ |
| (i2$\rightarrow$i1) | $3.70\times10^{7}$ | $5.36\times10^{8}$ |
| (i1$\rightarrow$p4) | $8.09\times10^{9}$ | $1.37\times10^{11}$ |
| (i2$\rightarrow$i3) | $1.56\times10^{11}$ | $2.73\times10^{11}$ |
| (i3$\rightarrow$i2) | $1.13\times10^{11}$ | $2.01\times10^{11}$ |
| (i2$\rightarrow$p2) | $5.35\times10^{11}$ | $9.16\times10^{11}$ |
| (i2$\rightarrow$p3) | $4.36\times10^{11}$ | $8.12\times10^{11}$ |
| (i2$\rightarrow$p4) | $6.41\times10^{10}$ | $1.64\times10^{11}$ |
| (i3$\rightarrow$i4) | $5.84\times10^{11}$ | $8.16\times10^{11}$ |
| (i4$\rightarrow$i3) | $1.64\times10^{12}$ | $2.48\times10^{12}$ |
| (i3$\rightarrow$p1) | $3.24\times10^{12}$ | $4.45\times10^{12}$ |
| (i3$\rightarrow$p2) | $7.71\times10^{11}$ | $1.26\times10^{12}$ |
| (i3$\rightarrow$p3) | $4.60\times10^{11}$ | $8.50\times10^{11}$ |
| (i4$\rightarrow$p1) | $2.12\times10^{13}$ | $2.98\times10^{13}$ |

**Table S4**. Key bimolecular reactions and photodissociation processes associated with the $C/C_2 – NH_3$ systems newly incorporated into the astrochemical model.

| Reactant 1 | Reactant 2 | Products | α | β | γ | No. |
|---|---|---|---|---|---|---|
| $NH_3$ | C | $H_2CN$+H | 4.00E-10 | 0 | 0 | R1 |
| $NH_3$ | $C_2$ | $CH_2CN$+H | 4.00E-10 | 0 | 0 | R2 |
| $CH_3$ | N | $H_2CN$+H | 4.00E-10 | 0 | 0 | R3 |
| $C_2H_3$ | N | $CH_2CN$+H | 4.00E-10 | 0 | 0 | R4 |
| $NH_3$ | CH | $CH_2NH$+H | 4.00E-10 | 0 | 0 | R5 |
| $CH_2NH$ | hν | $H_2CN$+H | 3.00E-09 | 0 | 3.1 | R6 |
| $CH_2NH$ | C | $CH_2CN$+H | 4.00E-10 | 0 | 0 | R7 |
| $C_2H_3$ | CN | $C_2H_2CN$+H | 4.00E-10 | 0 | 0 | R8 |
| $CH_2CN$ | CH | $C_2H_2CN$+H | 4.00E-10 | 0 | 0 | R9 |
| $C_2H_3CN$ | hν | $C_2H_2CN$+H | 3.00E-09 | 0 | 3.1 | R10 |
| $CH_2NH$ | CH | $CH_2CNH$+H | 4.00E-10 | 0 | 0 | R11 |
| $CH_2CNH$ | C | $C_2H_2CN$+H | 4.00E-10 | 0 | 0 | R12 |
| $CH_2CN$ | hν | $C_2HN$+H | 3.00E-09 | 0 | 3.1 | R13 |
| $C_2HN$ | $CH_2$ | $C_2H_2CN$+H | 4.00E-10 | 0 | 0 | R14 |
| $C_3H_3$ | N | $C_2H_2CN$+H | 4.00E-10 | 0 | 0 | R15 |
| $C_3H_3$ | hν | $C_3H_2$+H | 3.00E-09 | 0 | 3.1 | R16 |
| $C_3H_2$ | NH | $C_2H_2CN$+H | 4.00E-10 | 0 | 0 | R17 |
| $C_3H_2$ | hν | $C_3H$+H | 3.00E-09 | 0 | 3.1 | R18 |
| $C_3H$ | $NH_2$ | $C_2H_2CN$+H | 4.00E-10 | 0 | 0 | R19 |
| $C_2H_4$ | hν | $C_2H_3$+H | 3.00E-09 | 0 | 3.1 | R20 |
| $CH_3$ | CH | $C_2H_3$+H | 4.00E-10 | 0 | 0 | R21 |
| $C_2H_4$ | CN | $C_2H_3CN$+H | 4.00E-10 | 0 | 0 | R22 |
| $C_2H_4$ | CH | $C_3H_4$+H | 4.00E-10 | 0 | 0 | R23 |
| $C_3H_4$ | hν | $C_3H_3$+H | 3.00E-09 | 0 | 3.1 | R24 |
| $C_2H_6$ | hν | $C_2H_4$+$H_2$ | 3.00E-09 | 0 | 3.1 | R25 |
| $C_2H_6$ | CH | $C_3H_6$+H | 4.00E-10 | 0 | 0 | R26 |
| $C_2H_3$ | CH | $C_3H_3$+H | 4.00E-10 | 0 | 0 | R27 |
| $C_2H_4$ | C | $C_3H_3$+H | 4.00E-10 | 0 | 0 | R28 |
| $C_3H_6$ | CH | $C_4H_6$+H | 4.00E-10 | 0 | 0 | R29 |
| $H_2CN$ | $C_4H_3$ | $C_5H_4N$+H | 4.00E-10 | 0 | 0 | R30 |
| $CH_2CN$ | $C_3H_3$ | $C_5H_4N$+H | 4.00E-10 | 0 | 0 | R31 |
| $C_2H_2CN$ | $C_2H_3CN$ | $C_5H_5N$+CN | 4.00E-10 | 0 | 0 | R32 |
| $C_4H_6$ | CN | $C_5H_5N$+H | 4.00E-12 | 0 | 0 | R33 |
| $C_5H_5N$ | hν | $C_5H_4N$+H | 3.00E-09 | 0 | 3.1 | R34 |
| $C_5H_4N$ | $C_4H_4$ | $C_9H_7N$+H | 4.00E-10 | 0 | 0 | R35 |
| $CH_3CN$ | hν | $CH_2CN$+H | 3.00E-09 | 0 | 3.1 | R36 |
| $C_2H_3$ | NH | $CH_3CN$+H | 4.00E-10 | 0 | 0 | R37 |
| CN | $CH_3$ | $CH_2CN$+H | 4.00E-10 | 0 | 0 | R38 |
| $C_2H$ | $NH_2$ | $CH_2CN$+H | 4.00E-10 | 0 | 0 | R39 |
| $C_2H_3$ | hν | $C_2H_2$+H | 3.00E-09 | 0 | 3.1 | R40 |
| $C_2H_4$ | hν | $C_2H_2$+$H_2$ | 3.00E-09 | 0 | 3.1 | R41 |
| $CH_3$ | C | $C_2H_2$+H | 4.00E-10 | 0 | 0 | R42 |
| $CH_2$ | CH | $C_2H_2$+H | 4.00E-10 | 0 | 0 | R43 |
| $CH_2$ | NH | $H_2CN$+H | 4.00E-10 | 0 | 0 | R44 |
| CH | $NH_2$ | $H_2CN$+H | 4.00E-10 | 0 | 0 | R45 |
| $CH_3$ | NH | $CH_2NH$+H | 4.00E-10 | 0 | 0 | R46 |
| $CH_2$ | $NH_2$ | $CH_2NH$+H | 4.00E-10 | 0 | 0 | R47 |
| $C_4H_5N$ | CH | $C_5H_5N$+H | 4.00E-10 | 0 | 0 | R48 |
| $C_3H_6$ | CN | $C_4H_5N$ +H | 4.00E-10 | 0 | 0 | R49 |
| $CH_3CN$ | C | $C_2H_2CN$+H | 4.00E-10 | 0 | 0 | R50 |

**Table S5**. Optimized Cartesian coordinates (Angstrom) and vibrational frequencies ($cm^{-1}$) for the intermediates, transition states, reactants, and products involved in the reactions of carbon (C) with ammonia ($NH_3$) at the CCSD(T)/aug-cc-pVQZ level.

**Reagents**

**$NH_3$**

| | | | |
|---|---|---|---|
| H | 0.2920763332 | 0.6431570428 | -0.0000000704 |
| N | 0.2426306353 | -0.3684334478 | -0.0000000879 |
| H | 0.7606464234 | -0.6823617387 | -0.8116631239 |
| H | 0.7606466082 | -0.6823618562 | 0.8116632825 |

Frequencies

| | | |
|---|---|---|
| 1059.43 | 1673.76 | 1673.79 |
| 3476.22 | 3607.15 | 3607.31 |

**Products**

$CNH_2$

| | | | |
|---|---|---|---|
| C | -0.9933203864 | -0.9214443079 | 0.0019111645 |
| N | 0.1124136467 | -0.2955814902 | 0.2602648600 |
| H | 0.2975499258 | 0.6506602979 | -0.0669802735 |
| H | 0.8643903439 | -0.7113712499 | 0.8065141390 |

Frequencies

| | | |
|---|---|---|
| 729.14 | 1026.37 | 1431.05 |
| 1627.18 | 3380.82 | 3452.96 |

$H_2CN$

| | | | |
|---|---|---|---|
| H | -2.6544106164 | 0.4788743740 | -0.8724769137 |
| H | -2.6544073940 | -1.0629837221 | 0.1942219438 |
| C | -2.1083030395 | -0.2077878248 | -0.2173258026 |
| N | -0.9021919501 | -0.0214978272 | 0.0519437726 |

**Trans-HCNH**

| | | | |
|---|---|---|---|
| C | -1.2403610158 | 0.8976528810 | 0.1235983255 |
| N | -0.0966015804 | 0.4604591028 | -0.0630484877 |
| H | -1.8674903239 | 1.3560037585 | -0.6456500726 |
| H | 0.3919166204 | 0.0540003576 | 0.7329070547 |

Frequencies

| | | |
|---|---|---|
| 915.27 | 975.80 | 1198.91 |
| 1747.93 | 3057.20 | 3463.99 |

**Cis-HCNH**

| | | | |
|---|---|---|---|
| H | -1.9207551690 | 1.3703203969 | -0.6286635164 |
| C | -1.2152373401 | 0.9038689002 | 0.0673580619 |
| N | -0.0764022072 | 0.4558729561 | -0.0482882515 |
| H | 0.3992817164 | 0.4798667468 | -0.9545312940 |

Frequencies

| | | |
|---|---|---|
| 856.12 | 895.91 | 1026.32 |
| 1796.26 | 3008.66 | 3349.23 |

**Intermediates**

**i1**

| | | | |
|---|---|---|---|
| C | -1.1894678183 | -0.9542707245 | 0.0851101292 |
| N | 0.3105105939 | -0.3613889756 | 0.0178554606 |
| H | 0.2937207108 | 0.6557867370 | 0.0036902535 |
| H | 0.8508218659 | -0.6604236666 | 0.8263574396 |
| H | 0.7864481175 | -0.6844401806 | -0.8213033347 |

Frequencies

| | | |
|---|---|---|
| 564.20 | 774.84 | 775.14 |
| 1311.16 | 1651.71 | 1651.94 |
| 3408.61 | 3522.57 | 3522.89 |

**i2**

| | | | |
|---|---|---|---|
| H | -1.6922382002 | -1.5923622445 | 0.3507131705 |
| C | -1.2046714483 | -0.8190124445 | -0.2381403996 |
| H | 0.2700264633 | 0.4934781122 | 0.2219812975 |
| N | 0.1158154341 | -0.4496963491 | -0.1096604394 |
| H | 0.6980534112 | -0.6416152941 | -0.9143410390 |

Frequencies

| | | |
|---|---|---|
| 467.14 | 642.25 | 1086.37 |
| 1111.59 | 1261.55 | 1635.63 |
| 3116.57 | 3503.61 | 3585.55 |

**i3**

| | | | |
|---|---|---|---|
| H | -1.9526769089 | 1.3839273143 | 0.0312517729 |
| H | -1.9522667175 | -0.4066837897 | 0.5091839033 |
| C | -1.4082796893 | 0.5307877328 | 0.4277311559 |
| N | -0.0762237674 | 0.4380610522 | 0.0791584036 |
| H | 0.5575364800 | 0.6455537103 | 0.8561111650 |

Frequencies

| | | |
|---|---|---|
| 597.96 | 725.46 | 1022.97 |
| 1033.04 | 1208.50 | 1407.81 |
| 3081.94 | 3181.15 | 3388.57 |

**i4**

| | | | |
|---|---|---|---|
| H | -2.5271326402 | 0.5275089738 | 0.8455204920 |
| H | -2.5271282332 | 0.3823144949 | -0.9293346016 |
| H | -2.5271189379 | -1.0821206285 | 0.0838181791 |
| C | -2.1449804709 | -0.0574401193 | -0.0000055465 |
| N | -0.7243634224 | -0.0574442745 | -0.0000000289 |

Frequencies

| | | |
|---|---|---|
| 970.68 | 970.70 | 1052.49 |
| 1388.10 | 1439.90 | 1440.07 |
| 2966.17 | 3043.50 | 3043.74 |

**Transition states**

**i1-i2**

| | | | |
|---|---|---|---|
| C | -1.2904643133 | -0.1622041734 | 0.0528994377 |
| N | 0.2574410299 | -0.0943573664 | -0.0086854701 |
| H | -0.4716490707 | 0.8827577429 | 0.0054427490 |
| H | 0.7858269331 | -0.3004802937 | 0.8309508648 |
| H | 0.7188444210 | -0.3257159094 | -0.8806075814 |

Frequencies

| | | |
|---|---|---|
| 1682.01i | 669.38 | 828.24 |
| 837.16 | 1135.33 | 1516.26 |
| 2214.69 | 3458.50 | 3595.22 |

**i2-i3**

| | | | |
|---|---|---|---|
| H | -1.2978772894 | -0.6995688413 | 0.5912701261 |
| C | -0.6602691565 | -0.1125924912 | -0.0618714970 |
| H | 0.0653807966 | 0.8378001215 | 0.3924678722 |
| N | 0.7399349670 | -0.1331852620 | -0.0102571205 |
| H | 1.1528306823 | 0.1075474730 | -0.9116093807 |

Frequencies

| | | |
|---|---|---|
| 2020.90i | 568.49 | 885.60 |
| 1145.06 | 1155.40 | 1252.39 |
| 2362.67 | 3156.27 | 3455.42 |

**i3-i4**

| | | | |
|---|---|---|---|
| H | -0.6199616799 | 0.4011258166 | -1.0020966152 |
| H | 1.0814113318 | -0.0349079901 | 0.0564017817 |
| H | -0.7299868098 | -0.9753435963 | 0.2554958557 |
| C | -0.2111511007 | -0.1011742743 | -0.1292105004 |
| N | 0.4796892585 | 0.7103000441 | 0.8194094782 |

Frequencies

| | | |
|---|---|---|
| 2007.75i | 808.34 | 863.01 |
| 1086.54 | 1108.77 | 1370.37 |
| 2338.18 | 3070.11 | 3186.26 |

TS4: **i4**-$H_2CN$+H

| | | | |
|---|---|---|---|
| H | -2.4961864208 | 0.5206942498 | 0.8919184391 |
| H | -2.4962626561 | 0.3678856178 | -0.9739658543 |
| H | -2.7882606466 | -1.4991596464 | 0.1179012034 |
| C | -1.9580041580 | 0.2681002147 | -0.0266157351 |
| N | -0.7120101186 | 0.0552975640 | -0.0092390531 |

Frequencies

| | | |
|---|---|---|
| 737.72i | 377.59 | 535.50 |
| 940.38 | 986.62 | 1365.80 |
| 1577.32 | 3003.40 | 3080.00 |

TS5: **i1-**$CNH_2$+H

| | | | |
|---|---|---|---|
| C | -1.0709041950 | -1.0215745615 | 0.0815068427 |
| N | 0.1992995094 | -0.5514812678 | 0.0249796829 |
| H | 0.4075914569 | 0.9178384447 | -0.0046321396 |
| H | 0.7910227493 | -0.6624576001 | 0.8488020870 |
| H | 0.7250244794 | -0.6870620152 | -0.8389464730 |

Frequencies

| | | |
|---|---|---|
| 1677.25i | 693.77 | 908.04 |
| 1093.86 | 1107.31 | 1236.49 |
| 1597.59 | 3373.76 | 3438.58 |

TS6: **i2**-HNCH-trans+H

| | | | |
|---|---|---|---|
| H | -1.7934092894 | -1.4433736508 | 0.4134862479 |
| C | -1.1054579840 | -1.0667244438 | -0.3437882844 |
| H | 0.3372112649 | 0.7926245217 | 0.3620798321 |
| N | 0.1039854473 | -0.7542380077 | -0.1412403940 |
| H | 0.6446555612 | -0.5374964194 | -0.9799844017 |

Frequencies

| | | |
|---|---|---|
| 1288.62i | 567.37 | 614.87 |
| 892.88 | 1034.33 | 1210.47 |
| 1576.15 | 3094.54 | 3436.63 |

TS7: **i3**-HCNH-trans +H

| | | | |
|---|---|---|---|
| H | -2.0469895379 | 1.2516261050 | -0.0816561852 |
| H | -2.1449398866 | -0.6628754098 | 0.4781219017 |
| C | -1.2234029792 | 0.9056857999 | 0.5507866290 |
| N | -0.0756399919 | 0.6235325772 | 0.1335310565 |
| H | 0.6590613956 | 0.4736759277 | 0.8226535980 |

Frequencies

| | | |
|---|---|---|
| 974.09i | 533.61 | 706.01 |
| 920.00 | 996.70 | 1210.30 |
| 1641.00 | 3051.46 | 3477.31 |

TS8: **i3**-$H_2CN$+H

| | | | |
|---|---|---|---|
| H | -2.0104593049 | 1.4105545666 | 0.0854255674 |
| H | -2.0100006747 | -0.4028402326 | 0.5695027994 |
| C | -1.4669167972 | 0.4700470050 | 0.2002930435 |
| N | -0.2294788357 | 0.4022699271 | -0.0547783161 |
| H | 0.8849456124 | 0.7116157339 | 1.1029929059 |

Frequencies

| | | |
|---|---|---|
| 1196.87i | 469.21 | 605.36 |
| 901.92 | 952.89 | 1363.78 |
| 1568.83 | 3022.32 | 3106.34 |

TS9: **i3**-HCHN-cis +H

| | | | |
|---|---|---|---|
| N | -0.0626492823 | -0.0578199324 | -0.6073326617 |
| C | 0.1398712320 | -0.0646430046 | 0.6169361531 |
| H | -0.2603333692 | 0.8121467167 | -1.1067808513 |
| H | 0.3959790624 | 0.7181600998 | 1.3408320025 |
| H | -1.5188517119 | 0.0147509759 | 1.4235729461 |

Frequencies

| | | |
|---|---|---|
| 932.16i | 414.52 | 631.96 |
| 852.59 | 897.59 | 975.79 |
| 1705.45 | 2992.14 | 3371.73 |

TS10: **i2**-HCHN-cis +H

| | | | |
|---|---|---|---|
| C | 0.0496204860 | -0.1001076326 | -0.6874776856 |
| N | 0.0878062752 | 0.0911919989 | 0.5499932305 |
| H | -0.6805889055 | 0.1387471724 | -1.4641976588 |
| H | -0.6197544000 | 0.7120633340 | 0.9556393538 |
| H | -0.4015921914 | -1.1739958577 | 1.5316373109 |

Frequencies

| | | |
|---|---|---|
| 1146.69i | 483.01 | 604.33 |
| 838.40 | 963.20 | 1087.21 |
| 1631.65 | 3046.70 | 3349.79 |

**Table S6**. Optimized Cartesian coordinates (Angstrom) and vibrational frequencies ($cm^{-1}$) for the intermediates, transition states, reactants, and products involved in the reactions of dicarbon ($C_2$) with ammonia ($NH_3$) at the CCSD(T)/aug-cc-pVTZ level.

**Reagents**

**CC-singlet**

| | | | |
|---|---|---|---|
| C | 0.9952159254 | 2.4895524659 | 0.0000000000 |
| C | 1.8541340746 | 1.5802375341 | 0.0000000000 |

Frequencies

1839.74

**CC-triplet**

| | | | |
|---|---|---|---|
| C | 0.9714406887 | 2.5147227080 | 0.0000000000 |
| C | 1.8779093113 | 1.5550672920 | 0.0000000000 |

Frequencies

1626.64

**$NH_3$**

| | | | |
|---|---|---|---|
| N | 0.9032657888 | 1.7990987104 | -0.0283246089 |
| H | 1.9168543871 | 1.7731051376 | 0.0161169887 |
| H | 0.5896810578 | 2.2379536100 | 0.8313585748 |
| H | 0.5896587665 | 0.8346425420 | 0.0110690454 |

Frequencies

| | | |
|---|---|---|
| 1062.66 | 1672.48 | 1672.58 |
| 3463.43 | 3592.09 | 3592.37 |

**Products – Hydrogen loss**

**p1'-$H_2$CCN**

| | | | |
|---|---|---|---|
| C | -3.5113218329 | 0.6957951603 | 0.0000000000 |
| C | -2.2758036638 | 1.3417953683 | 0.0000000000 |
| N | -1.2344952172 | 1.8863012203 | 0.0000000000 |
| H | -3.5525242695 | -0.3831820446 | 0.0000000000 |
| H | -4.4208950166 | 1.2776602956 | 0.0000000000 |

Frequencies

| | | |
|---|---|---|
| 361.82 | 410.91 | 639.50 |
| 1028.14 | 1032.94 | 1453.51 |
| 2106.03 | 3176.35 | 3288.10 |

**p2'-CNC$H_2$**

| | | | |
|---|---|---|---|
| C | 0.2302786477 | 0.0388504735 | -1.1737183331 |
| N | 0.0765014478 | -0.0300441832 | 0.1628446127 |
| C | -0.0590572679 | -0.0846127020 | 1.3440269914 |
| H | 0.6338526302 | -0.8175478564 | -1.6908521866 |
| H | -0.0633999503 | 0.9450738400 | -1.6799461854 |

Frequencies

| | | |
|---|---|---|
| 267.65 | 351.17 | 516.29 |
| 1082.88 | 1130.14 | 1476.20 |
| 2014.43 | 3172.00 | 3302.50 |

**p3'-HNCCH**

| | | | |
|---|---|---|---|
| C | -0.0000744816 | -0.0329406061 | 0.0594034496 |
| C | -0.0000823576 | 0.0061864152 | 1.3036509529 |
| N | 0.0000552715 | 0.0838882631 | -1.2183936925 |
| H | 0.0000833841 | -0.8287540002 | -1.6778857414 |
| H | 0.0010174992 | -0.0181745353 | 2.3664470687 |

Frequencies

| | | |
|---|---|---|
| 282.32 | 409.51 | 450.54 |
| 460.30 | 1094.69 | 1167.82 |
| 1878.30 | 3444.31 | 3452.74 |

**p4'-CNCH2-cic-ccsd.out**

| | | | |
|---|---|---|---|
| C | -1.0315761990 | 1.0726409199 | 0.4815831706 |
| C | 0.4044794514 | 1.0816451737 | 0.0562025756 |
| N | -0.9511133655 | 0.3419132509 | -0.5012599035 |
| H | 0.7739125694 | 1.8785380463 | -0.5738577767 |
| H | 1.1121975438 | 0.4212326091 | 0.5373319340 |

Frequencies

| | | |
|---|---|---|
| 451.03 | 698.74 | 914.41 |
| 972.08 | 1053.07 | 1477.66 |
| 1799.62 | 3147.81 | 3264.77 |

**p5'-HCNCH-cic**

| | | | |
|---|---|---|---|
| C | -2.7315575846 | 0.7532997051 | 0.3325871697 |
| C | -1.3693002226 | 0.7562223352 | 0.0316040394 |
| N | -1.8744114271 | -0.4253195170 | -0.0797057126 |
| H | -0.4075840205 | 1.2343059149 | -0.0899684272 |
| H | -3.6091267453 | 1.1091715618 | -0.1945170694 |

Frequencies

| | | |
|---|---|---|
| 676.60 | 806.02 | 900.98 |
| 931.99 | 1017.71 | 1263.96 |
| 1595.50 | 3164.78 | 3228.10 |

**p6'-HCNCH**

| | | | |
|---|---|---|---|
| C | 0.1436927101 | -0.0459816405 | -1.1364480880 |
| N | 0.0277999840 | -0.1807549654 | 0.0932931191 |
| C | -0.0452783852 | -0.2416461749 | 1.3355246718 |
| H | 0.7674068264 | -0.6490795594 | -1.7859582321 |
| H | -0.9799064978 | -0.1798303414 | 1.8842323676 |

Frequencies

| | | |
|---|---|---|
| 418.61 | 431.84 | 548.08 |
| 645.35 | 930.91 | 1268.72 |
| 1793.80 | 3165.51 | 3187.08 |

**p7'-$H_2$NCC**

| | | | |
|---|---|---|---|
| C | -3.3943594394 | 1.4021634927 | -0.0330769642 |
| C | -2.1955504586 | 0.9231724138 | -0.0141800666 |
| N | -4.6095313447 | 1.8872458411 | -0.0524795501 |
| H | -4.7745758274 | 2.8794911116 | -0.1298858597 |
| H | -5.4143929298 | 1.2822671408 | 0.0106224407 |

Frequencies

| | | |
|---|---|---|
| 250.95 | 277.15 | 457.39 |
| 1089.32 | 1136.16 | 1643.87 |
| 1992.71 | 3545.24 | 3655.93 |

**Products – Diatomic loss**

**$CCH_2$**

| | | | |
|---|---|---|---|
| C | -4.4936456595 | 1.0604997146 | 0.0000000000 |
| C | -5.7991757611 | 1.0026148420 | 0.0000000000 |
| H | -3.9909460566 | 2.0244443888 | 0.0000000000 |
| H | -3.9076025228 | 0.1448310545 | 0.0000000000 |

Frequencies

| | | |
|---|---|---|
| 334.50 | 735.53 | 1218.48 |
| 1658.20 | 3122.94 | 3212.72 |

**$CH_3$**

| | | | |
|---|---|---|---|
| C | -3.7439913569 | 0.9044748851 | 0.0000000000 |
| H | -3.6827293956 | 1.9822470432 | 0.0000000000 |
| H | -2.8412402496 | 0.3125198490 | 0.0000000000 |
| H | -4.7080389979 | 0.4186282227 | 0.0000000000 |

Frequencies

| | | |
|---|---|---|
| 496.26 | 1419.04 | 1419.32 |
| 3114.51 | 3294.20 | 3294.52 |

**CN**

| | | | |
|---|---|---|---|
| C | -3.9235798168 | 0.0526859896 | -0.0505174148 |
| N | -2.7505461210 | -0.0086756109 | 0.0511838882 |

Frequencies

2047.02

**HCCH**

| | | | |
|---|---|---|---|
| C | -2.2659004487 | 0.7915210348 | -0.5489139075 |
| C | -1.4048754979 | 1.1017278813 | 0.2430052110 |
| H | -3.0229375347 | 0.5189936474 | -1.2450661960 |
| H | -0.6478374687 | 1.3743001563 | 0.9391389024 |

Frequencies

| | | |
|---|---|---|
| 592.63 | 744.05 | 748.25 |
| 1994.29 | 3394.15 | 3502.03 |

**NH**

| | | | |
|---|---|---|---|
| N | -5.0314562212 | 0.8270822024 | 0.0000000000 |
| H | -6.0681537788 | 0.8168677976 | 0.0000000000 |

Frequencies

3316.66

**Products – triatomic loss**

**$C_2H$**

| | | | |
|---|---|---|---|
| C | -2.1251664279 | 1.4561036200 | 0.0000000000 |
| C | -0.9116097073 | 1.3878936144 | 0.0000000000 |
| H | 0.1513161352 | 1.3281227656 | 0.0000000000 |

Frequencies

| | | |
|---|---|---|
| 277.84 | 277.93 | 2004.03 |
| 3444.00 | | |

**$CH_2$**

| | | | |
|---|---|---|---|
| C | 0.5643229808 | 2.6925678760 | 0.0000000000 |
| H | 1.5382405669 | 2.1585901950 | 0.0000000000 |
| H | 0.8863651323 | 3.7555525990 | 0.0000000000 |

Frequencies

| | | |
|---|---|---|
| 1395.37 | 2911.72 | 2982.21 |

**HCN**

| | | | |
|---|---|---|---|
| C | -2.9341609294 | 1.4371232641 | 0.0000000000 |
| H | -4.0012095834 | 1.4232707382 | 0.0000000000 |
| N | -1.7740994872 | 1.4521659977 | 0.0000000000 |

Frequencies

| | | |
|---|---|---|
| 717.02 | 717.04 | 2107.00 |
| 3432.82 | | |

**HNC**

| | | | |
|---|---|---|---|
| H | -3.5816631508 | 0.0896920456 | 0.0556058680 |
| N | -2.5926335666 | -0.0099767083 | -0.0321272858 |
| C | -1.4270769626 | -0.1272434672 | -0.1351145323 |

Frequencies

| | | |
|---|---|---|
| 453.55 | 2036.73 | 3805.94 |

**$NH_2$**

| | | | |
|---|---|---|---|
| N | -2.2893248185 | 0.0194787741 | 0.0000000000 |
| H | -1.9068507422 | 0.9731738754 | 0.0000000000 |
| H | -3.3042044392 | 0.1802273504 | 0.0000000000 |

Frequencies

| | | |
|---|---|---|
| 1539.75 | 3359.56 | 3453.36 |

**Intermediates**

**i1'**

| | | | |
|---|---|---|---|
| C | -4.9343582319 | 0.5184036386 | -0.1030368723 |
| C | -3.7076291179 | 0.7389389301 | -0.2407041336 |
| N | -6.3357239293 | 0.2660792143 | 0.0551349170 |
| H | -6.7458168567 | 0.8773959426 | 0.7659701927 |
| H | -6.8433217313 | 0.4250964786 | -0.8191006153 |
| H | -6.5095501328 | -0.7008142042 | 0.3417365115 |

Frequencies

| | | |
|---|---|---|
| 868.10 | 1057.09 | 1057.19 |
| 1480.33 | 1661.85 | 1662.13 |
| 2018.74 | 3371.28 | 3442.49 |
| 3442.70 | | |

**i2'**

| | | | |
|---|---|---|---|
| C | 0.9349002884 | -0.0001857275 | 0.4189000342 |
| C | -0.3637498793 | 0.0000851182 | 0.6439974543 |
| N | -0.3176538573 | -0.0001051324 | -0.8046253922 |
| H | -1.2138540846 | 0.0004960622 | 1.3023129801 |
| H | -0.4366503735 | 0.8551862482 | -1.3228665887 |
| H | -0.4370151917 | -0.8554804727 | -1.3226433891 |

Frequencies

| | | |
|---|---|---|
| 414.40 | 597.27 | 645.83 |
| 803.38 | 1036.24 | 1110.82 |
| 1135.85 | 1596.01 | 1674.95 |
| 3287.21 | 3553.84 | 3670.72 |

**i3'**

| | | | |
|---|---|---|---|
| C | -3.0316242732 | 0.1390802675 | 0.0355302879 |
| C | -1.7491437648 | 0.1074562676 | -0.0128452674 |
| N | -2.4410981845 | -1.1663409900 | -0.4898229342 |
| H | -0.7535942490 | 0.5055079132 | 0.0235402942 |
| H | -4.0008000538 | 0.5856085706 | 0.1459827921 |
| H | -2.4276194747 | -1.8219920289 | 0.2976148273 |

Frequencies

| | | |
|---|---|---|
| 537.43 | 556.91 | 716.09 |
| 879.65 | 973.32 | 1059.64 |
| 1147.21 | 1365.30 | 1723.66 |
| 3288.42 | 3342.30 | 3350.49 |

**i4'**

| | | | |
|---|---|---|---|
| C | -2.9969018464 | 0.6442965597 | 0.0734523904 |
| C | -1.7898594514 | 0.6788815786 | -0.0258221061 |
| N | -4.3502018494 | 0.6193080518 | 0.2609644739 |
| H | -0.7317082326 | 0.7090581293 | -0.1139163400 |
| H | -4.8295219218 | 1.3519598506 | -0.2448121841 |
| H | -4.7585066984 | -0.2813441699 | 0.0501337660 |

Frequencies

| | | |
|---|---|---|
| 335.66 | 381.20 | 495.32 |
| 651.93 | 696.26 | 1053.69 |
| 1215.43 | 1649.41 | 2195.72 |
| 3468.12 | 3535.29 | 3624.70 |

**i5'**

| | | | |
|---|---|---|---|
| C | -1.5915491794 | 0.0743675121 | 0.5100854444 |
| C | -0.0594291041 | -0.0042577375 | 0.0875463765 |
| N | -1.1225259530 | -0.9377432845 | -0.1395598083 |
| H | 0.2763211526 | 0.5856735264 | -0.7567187504 |
| H | 0.6743481260 | -0.2568823608 | 0.8433776770 |
| H | -1.3753350423 | -1.8263976556 | -0.5447309394 |

Frequencies

| | | |
|---|---|---|
| 428.94 | 711.07 | 960.97 |
| 965.30 | 1030.83 | 1145.56 |
| 1206.01 | 1485.16 | 1587.05 |
| 3116.20 | 3216.67 | 3589.33 |

**i6'**

| | | | |
|---|---|---|---|
| C | -3.5336584338 | 0.9452242240 | 0.1434747168 |
| C | -2.2282978370 | 1.0377713288 | -0.0069751689 |
| N | -4.7341358669 | 0.8686798763 | 0.4143163364 |
| H | -5.3444076380 | 0.7680235655 | -0.3960908261 |
| H | -1.7677965926 | 2.0065409788 | -0.1369635323 |
| H | -1.6219536317 | 0.1435100266 | -0.0177615257 |

Frequencies

| | | |
|---|---|---|
| 402.99 | 457.40 | 705.10 |
| 903.17 | 996.93 | 1044.59 |
| 1132.94 | 1437.75 | 2072.91 |
| 3173.21 | 3268.21 | 3478.17 |

**i7'**

| | | | |
|---|---|---|---|
| C | -1.3448702716 | 0.5221913550 | -0.1555211309 |
| C | 0.1003924279 | 0.4534726637 | 0.0799687810 |
| N | -2.4906355389 | 0.5766588663 | -0.3422119556 |
| H | 0.4254170455 | 1.3310641561 | 0.6387662219 |
| H | 0.6284265979 | 0.4206070286 | -0.8729854907 |
| H | 0.3388797392 | -0.4431640697 | 0.6519835744 |

Frequencies

| | | |
|---|---|---|
| 359.03 | 359.12 | 920.87 |
| 1063.21 | 1063.22 | 1416.36 |
| 1490.92 | 1491.09 | 2291.43 |
| 3060.88 | 3144.47 | 3144.55 |

**i8'**

| | | | |
|---|---|---|---|
| C | 0.1312397935 | -0.0273422675 | 0.8936355675 |
| N | -0.7270848325 | -0.2576981029 | -0.3841334780 |
| H | 0.0732889768 | 0.9641139028 | 1.3271505201 |
| H | 0.1787200430 | -0.8601831883 | 1.5852918077 |
| H | 1.2827312425 | -0.2652989951 | -1.2663092735 |
| C | 0.5284456091 | -0.2009850074 | -0.4954186625 |

Frequencies

| | | |
|---|---|---|
| 700.41 | 772.30 | 987.40 |
| 996.28 | 1039.35 | 1106.96 |
| 1264.68 | 1510.77 | 1684.47 |
| 3115.68 | 3208.15 | 3226.58 |

**i9'**

| | | | |
|---|---|---|---|
| C | 0.2188237836 | 0.0350389259 | -1.1584936983 |
| N | 0.0069347177 | -0.0494125908 | 0.1073631491 |
| C | -0.0294278283 | -0.0629584557 | 1.3335387238 |
| H | 0.6493078956 | -0.8101672039 | -1.6764675568 |
| H | -0.0348452538 | 0.9508938049 | -1.6731689524 |
| H | -0.9471891349 | -0.4190442406 | 1.8082663247 |

Frequencies

| | | |
|---|---|---|
| 332.28 | 465.08 | 690.84 |
| 903.79 | 946.16 | 1140.10 |
| 1195.07 | 1492.92 | 1952.44 |
| 3085.95 | 3152.93 | 3267.13 |

**Transition states**

**i1'-i2'**

| | | | |
|---|---|---|---|
| C | -0.0000224478 | 0.7315569642 | -0.6999477987 |
| N | 0.0000078738 | -0.2131191257 | 0.8006015074 |
| H | -0.0000032176 | 1.0446573757 | 0.5807381277 |
| H | -0.8609997040 | -0.5015151593 | 1.2458768708 |
| H | 0.8610342858 | -0.5015046720 | 1.2458473465 |
| C | -0.0000170286 | -0.5976761089 | -0.7801970036 |

Frequencies

| | | |
|---|---|---|
| 1510.65i | 513.43 | 654.46 |
| 800.15 | 847.88 | 963.18 |
| 1134.38 | 1526.01 | 1579.27 |
| 2149.09 | 3511.20 | 3643.54 |

**i1'-i4'**

| | | | |
|---|---|---|---|
| C | -0.2529780174 | -0.0008754995 | 0.1981897162 |
| C | 0.1785624366 | -0.0003380137 | 1.3920151241 |
| N | 0.0615389166 | 0.0001468492 | -1.2358851854 |
| H | -0.2009444208 | 0.8507839909 | -1.7169130755 |
| H | 0.9308727346 | 0.0017262380 | -0.3319439119 |
| H | -0.1975490456 | -0.8512597811 | -1.7173722504 |

Frequencies

| | | |
|---|---|---|
| 1759.06i | 195.48 | 274.13 |
| 819.44 | 893.05 | 922.61 |
| 1128.04 | 1582.34 | 1893.70 |
| 2328.51 | 3502.90 | 3621.04 |

**i2'-i3'**

| | | | |
|---|---|---|---|
| C | 0.0448363805 | 0.8898131938 | -0.5404035000 |
| N | 0.0872635416 | -0.3366935268 | 0.8994138445 |
| H | 0.2089632179 | 1.0173338244 | 0.7083085280 |
| H | -0.0357540472 | -1.2899873136 | -1.1967322541 |
| H | -0.8823848033 | -0.3946520806 | 1.2379218943 |
| C | 0.0280379802 | -0.4297992173 | -0.5592476326 |

Frequencies

| | | |
|---|---|---|
| 1416.76i | 650.51 | 689.32 |
| 921.68 | 937.33 | 1083.79 |
| 1138.29 | 1282.63 | 1602.68 |
| 2281.15 | 3327.76 | 3338.81 |

**i2'-i4'**

| | | | |
|---|---|---|---|
| C | 0.7460718136 | -0.0008484134 | 0.2875963435 |
| C | -0.3196236681 | 0.0010365390 | 0.9766882194 |
| N | -0.2478966855 | -0.0004215887 | -0.9775218373 |
| H | -1.1888487821 | 0.0012048268 | 1.6040810095 |
| H | 0.0369609750 | 0.8229815062 | -1.4941677273 |
| H | 0.0371980701 | -0.8232698933 | -1.4948207961 |

Frequencies

| | | |
|---|---|---|
| 756.30i | 320.00 | 489.03 |
| 626.36 | 893.82 | 989.82 |
| 1127.04 | 1567.13 | 1704.97 |
| 3365.06 | 3500.56 | 3608.21 |

**i3'-i5'**

| | | | |
|---|---|---|---|
| C | -0.3216049178 | 0.1710877088 | -0.6577835066 |
| C | 0.8956883618 | -0.1455110473 | -0.1164197254 |
| N | -0.4989883855 | -0.1104094931 | 0.7458525875 |
| H | 0.2134383539 | -0.9357774491 | -0.9649628872 |
| H | -0.9008108393 | 0.5536032240 | -1.4866576269 |
| H | -0.4021987404 | 0.7840517503 | 1.2324938414 |

Frequencies

| | | |
|---|---|---|
| 876.61i | 508.19 | 878.75 |
| 990.77 | 1005.94 | 1086.16 |
| 1161.90 | 1388.33 | 1481.90 |
| 2188.82 | 3226.55 | 3417.58 |

**i3'-i6'**

| | | | |
|---|---|---|---|
| C | -0.3735348985 | -0.0338544063 | 0.0080311260 |
| C | 0.1948256843 | 0.1321644801 | -1.3068605764 |
| N | 0.2716163410 | 0.0687364067 | 1.1209335899 |
| H | 0.7706378488 | -0.7649202927 | -1.5962517444 |
| H | -1.4720391525 | -0.1259082484 | -0.0657321539 |
| H | -0.3856521880 | 0.0593289481 | 1.9001504240 |

Frequencies

| | | |
|---|---|---|
| 391.57i | 456.39 | 640.99 |
| 948.38 | 1022.60 | 1055.92 |
| 1171.83 | 1367.03 | 1587.79 |
| 2950.23 | 2996.74 | 3471.31 |

**i3'-i8'**

| | | | |
|---|---|---|---|
| C | 0.3441624285 | -0.1048274843 | 0.7596440336 |
| N | -0.6513021268 | -0.1457778301 | -0.1255944657 |
| H | -0.6678028412 | 0.7499824985 | 0.8026652307 |
| H | 0.3110423673 | -0.5852570560 | 1.7372110931 |
| H | 0.9108965991 | -0.6554080270 | -1.2674923173 |
| C | 0.6619026486 | 0.2890835998 | -0.7475152404 |

Frequencies

| | | |
|---|---|---|
| 2241.42i | 317.10 | 800.63 |
| 873.77 | 1013.42 | 1056.44 |
| 1165.14 | 1254.33 | 1374.70 |
| 1996.15 | 2943.23 | 3112.76 |

**i5'-i8'**

| | | | |
|---|---|---|---|
| C | -0.8151416221 | -0.0010334925 | 0.1955680092 |
| N | 0.5094102063 | 0.0051654081 | 0.5842569311 |
| H | -0.3260317015 | -0.8154763461 | 1.0029810384 |
| H | 0.4882612246 | -0.9019780951 | -1.4454982314 |
| H | 0.4154263761 | 0.9424893992 | -1.3311765600 |
| C | 0.2400444065 | -0.0183163736 | -0.8715841674 |

Frequencies

| | | |
|---|---|---|
| 1412.19i | 762.82 | 811.62 |
| 989.44 | 1013.18 | 1069.26 |
| 1177.60 | 1314.72 | 1516.85 |
| 2485.80 | 3143.12 | 3251.84 |

**i6'-i7'**

| | | | |
|---|---|---|---|
| C | 0.0422766629 | -0.0585159724 | -0.0443649745 |
| C | -0.0048674323 | 0.1636868456 | 1.3592232958 |
| N | -0.0444476319 | 0.0055481870 | -1.2698351051 |
| H | 0.0076805630 | -1.1954458879 | -0.2095180442 |
| H | 0.9297972361 | -0.0872839651 | 1.8689138675 |
| H | -0.9124443554 | -0.0992440420 | 1.8237209925 |

Frequencies

| | | |
|---|---|---|
| 1083.22i | 327.97 | 329.58 |
| 473.90 | 591.61 | 998.85 |
| 1018.06 | 1430.86 | 1892.96 |
| 2734.56 | 3155.23 | 3259.54 |

**i8'-i9'**

| | | | |
|---|---|---|---|
| C | 0.0499750424 | 0.1949768750 | -0.9930802840 |
| N | 0.2894284082 | -0.5415881485 | 0.1538498343 |
| C | -0.1118500962 | 0.2326341600 | 1.1166681652 |
| H | 0.8707706579 | 0.2854362009 | -1.6976648404 |
| H | -0.7619545692 | 0.9242909147 | -1.0237810089 |
| H | -1.0315171604 | 0.8476462491 | 1.0177935396 |

Frequencies

| | | |
|---|---|---|
| 482.75i | 496.84 | 693.06 |
| 827.26 | 1027.89 | 1102.36 |
| 1183.70 | 1383.06 | 1497.18 |
| 2883.13 | 3057.47 | 3198.96 |

**Triplet reactants-$C_2H+NH_2$**

| | | | |
|---|---|---|---|
| C | 0.21406831 | -0.55043530 | 1.21402335 |
| C | -0.13445067 | 0.29974809 | 2.05287814 |
| H | 0.28214970 | -0.69848108 | -0.08832047 |
| N | 0.03155981 | -0.06023769 | -1.10680139 |
| H | -0.95380658 | 0.19947821 | -1.04096675 |
| H | 0.56047946 | 0.80992776 | -1.03081286 |

Frequencies

| | | |
|---|---|---|
| 1479.449i | 164.414 | 185.358 |
| 444.698 | 571.514 | 827.791 |
| 1290.143 | 1443.355 | 1600.149 |
| 1857.841 | 3455.361 | 3554.673 |

**Van der Waals Minimum**

| | | | |
|---|---|---|---|
| C | -0.4406937533 | -0.3234263552 | -0.5941362302 |
| C | 0.4705218199 | 0.3564968030 | -1.2407949600 |
| N | 0.0202666611 | -0.0432962812 | 1.3608243596 |
| H | 0.8208029174 | 0.5774970992 | 1.3966044248 |
| H | -0.7836959985 | 0.3718426209 | 1.8157068741 |
| H | 0.2415157759 | -0.9464499038 | 1.7620211911 |

Frequencies

| | | |
|---|---|---|
| 72.15 | 238.98 | 291.73 |
| 472.85 | 489.16 | 1005.41 |
| 1587.60 | 1639.27 | 1644.47 |
| 3466.06 | 3625.12 | 3634.49 |

**Table S7**. Optimized Cartesian coordinates (Angstrom) and vibrational frequencies ($cm^{-1}$) for the intermediates, transition states, reactants, and products involved in the pathways from $H_2CN^{\bullet}$, cis-HCNH, and $H_2CCN^{\bullet}$ radicals to pyridine and pyridinyl radicals.

**Intermediates**

**i1"**

| | | | |
|---|---|---|---|
| N | -0.385437 | 1.299259 | 0.167033 |
| C | 0.787331 | 1.773708 | 0.079994 |
| C | -1.863082 | -0.540348 | 0.158601 |
| C | -0.603813 | -0.095142 | 0.059028 |
| C | 0.496031 | -0.989878 | -0.150056 |
| C | 1.451701 | -1.697261 | -0.325122 |
| H | 0.913723 | 2.848672 | 0.168024 |
| H | 1.681512 | 1.166729 | -0.078485 |
| H | -2.095289 | -1.591704 | 0.084763 |
| H | -2.663680 | 0.167381 | 0.316134 |
| H | 2.281003 | -2.341415 | -0.479913 |

Frequencies

| | | |
|---|---|---|
| 131.5830 | 165.9401 | 264.8221 |
| 336.0832 | 487.7798 | 555.2548 |
| 632.9441 | 655.7867 | 694.0683 |
| 729.1735 | 771.4176 | 819.6866 |
| 935.5021 | 958.2878 | 1066.3564 |
| 1205.9243 | 1268.3081 | 1414.7085 |
| 1489.4531 | 1660.3799 | 1693.8677 |
| 2199.5097 | 3035.5270 | 3149.6127 |
| 3162.8817 | 3261.2002 | 3467.0127 |

**i2"**

| | | | |
|---|---|---|---|
| N | 0.4438576679 | 1.1505584071 | -0.4309098078 |
| C | 1.3596449225 | 0.2077985965 | 0.0086988525 |
| C | -0.7608831099 | 1.1416076566 | -0.0493850848 |
| C | -0.6875494153 | -2.4916374783 | -1.2059713972 |
| C | 0.1987293616 | -1.8394121976 | -0.7184024829 |
| C | 1.276856614 | -1.128066558 | -0.1386006873 |
| H | 2.2720901289 | 0.6343490767 | 0.4080753452 |
| H | -1.4520424298 | 1.8672561668 | -0.4694036441 |
| H | -1.146666684 | 0.4428795255 | 0.6959319629 |
| H | -1.4717914353 | -3.0597788639 | -1.6390362887 |
| H | 2.1336763794 | -1.7160243314 | 0.1673072321 |

Frequencies

| | | |
|---|---|---|
| 111.2549 | 155.6359 | 236.0377 |
| 347.0416 | 415.6753 | 579.3046 |
| 610.2701 | 631.4413 | 682.1259 |
| 795.1164 | 840.1914 | 906.5206 |
| 970.9944 | 1055.7787 | 1077.9638 |
| 1216.9261 | 1258.5385 | 1423.4055 |
| 1505.9993 | 1631.4523 | 1724.7337 |
| 2194.7745 | 3036.1360 | 3146.4363 |
| 3151.7320 | 3170.0200 | 3470.4076 |

**i3"**

| | | | |
|---|---|---|---|
| N | 0.3209032145 | -1.5983031008 | 0.0236674737 |
| C | 1.6024032627 | -1.1074968393 | -0.3335764311 |
| C | -1.9438180083 | -0.5456244821 | -0.3252197502 |
| C | -0.7489193153 | -1.0470903198 | -0.1636461813 |
| C | 2.2686961089 | 1.4092046462 | -0.4228595565 |
| C | 1.9337404891 | 0.1549244401 | -0.4021729944 |
| H | 2.3385884111 | -1.8844604064 | -0.4919169681 |
| H | -2.3606993856 | 0.1245493957 | 0.4125978694 |
| H | -2.5268069396 | -0.7853842295 | -1.2025963735 |
| H | 2.2085983158 | 1.9957374816 | -1.3327669586 |
| H | 2.6212788468 | 1.9166014144 | 0.46852487043 |

Frequencies

| | | |
|---|---|---|
| 70.9759 | 125.8867 | 245.8180 |
| 327.8856 | 356.5984 | 504.1446 |
| 529.3447 | 674.3859 | 702.7065 |
| 721.9612 | 892.0032 | 898.4185 |
| 904.8004 | 995.1426 | 1017.6454 |
| 1116.4550 | 1280.3020 | 1366.5653 |
| 1454.8339 | 1473.1939 | 2031.5375 |
| 2112.9385 | 3110.5007 | 3152.2684 |
| 3171.1149 | 3179.9485 | 3234.5683 |

**i4''**

| | | | |
|---|---|---|---|
| N | -0.6988758407 | 1.1267296462 | -0.1774601689 |
| C | 0.5682500143 | 1.154281835 | -0.1619021145 |
| C | -1.4122794543 | -0.1378133206 | -0.4947422254 |
| C | -0.5620833098 | -1.3458232391 | -0.1640845948 |
| C | 0.6994898319 | -1.1377496743 | -0.4947863741 |
| C | 1.3749682009 | -0.0192347218 | -0.6559623015 |
| H | 1.0601625156 | 2.0972593193 | 0.0609501725 |
| H | -1.753908383 | -0.0406292022 | -1.5295642806 |
| H | -0.8951523952 | -2.0719687212 | 0.5659305601 |
| H | 2.2096434071 | 0.1605588106 | -1.3203999596 |
| H | -2.3080365866 | -0.171020732 | 0.1261082866 |

Frequencies

| | | |
|---|---|---|
| 246.8213 | 369.9522 | 478.5709 |
| 567.6994 | 593.9747 | 716.9793 |
| 786.0319 | 840.5193 | 869.5373 |
| 888.6101 | 929.9435 | 969.1541 |
| 991.3862 | 1142.6242 | 1204.4800 |
| 1268.2462 | 1285.1240 | 1321.1815 |
| 1404.3504 | 1450.4074 | 1638.3408 |
| 1819.3129 | 3024.1687 | 3080.1730 |
| 3124.1528 | 3163.8738 | 3174.0067 |

**i5''**

| | | | |
|---|---|---|---|
| N | -0.3810283198 | 1.0165112229 | 0.3024468279 |
| C | 0.9087039471 | 1.1856420968 | 0.1400686544 |
| C | -1.9127503971 | -0.8033176124 | 0.1238249184 |
| C | -0.6769963999 | -0.3091058572 | -0.0094365759 |
| C | 0.6089252133 | -0.9925425453 | -0.4163426467 |
| C | 1.6923780152 | -0.0174441861 | -0.126319208 |
| H | 1.3355448665 | 2.1699258047 | 0.2878534745 |
| H | -2.1449506564 | -1.8211336869 | -0.1565246692 |
| H | -2.7038582424 | -0.1901712525 | 0.5315453209 |
| H | 0.7732134739 | -1.9750929209 | 0.0295433203 |
| H | 0.6558474996 | -1.1603920631 | -1.5044304166 |

Frequencies

| | | |
|---|---|---|
| 132.0457 | 280.4002 | 360.7234 |
| 405.5388 | 692.5977 | 712.5371 |
| 756.2907 | 848.8171 | 869.9720 |
| 940.4539 | 954.4918 | 961.4774 |
| 973.9444 | 1036.8040 | 1129.9303 |
| 1225.9002 | 1252.0573 | 1346.7961 |
| 1368.8297 | 1424.6288 | 1447.2034 |
| 1682.9441 | 2978.2972 | 3072.2903 |
| 3152.3826 | 3176.7638 | 3244.6811 |

**i6''**

| | | | |
|---|---|---|---|
| N | -1.440513 | 0.168703 | -0.292883 |
| C | -0.572181 | 1.172447 | 0.069215 |
| C | -1.264781 | -1.016243 | 0.139034 |
| C | 1.163742 | -1.410815 | 0.050037 |
| C | 1.513044 | -0.159311 | -0.146926 |
| C | 0.773248 | 1.064131 | 0.116716 |
| H | -1.023403 | 2.151383 | 0.174095 |
| H | -1.858029 | -1.834564 | -0.254051 |
| H | -0.604927 | -1.260777 | 0.981913 |
| H | 2.538667 | -0.096021 | -0.522514 |
| H | 1.352854 | 1.957807 | 0.302286 |

Frequencies

| | | |
|---|---|---|
| 86.1636 | 176.4700 | 258.7647 |
| 347.4921 | 423.2845 | 545.3816 |
| 616.0966 | 713.7198 | 822.9208 |
| 844.0340 | 947.9610 | 971.3034 |
| 993.1331 | 1053.1191 | 1079.6650 |
| 1184.1962 | 1248.8240 | 1420.5409 |
| 1479.6947 | 1568.9124 | 1662.0984 |
| 1674.6659 | 2958.0010 | 3060.8211 |
| 3161.0552 | 3163.4516 | 3187.6947 |

**i7''**

| | | | |
|---|---|---|---|
| N | -0.325656 | 1.178499 | 0.148884 |
| C | 0.825181 | 1.674035 | 0.066574 |
| C | -1.702139 | -0.744909 | 0.121362 |
| C | -0.479647 | -0.209490 | 0.034025 |
| C | 0.741180 | -1.015762 | -0.180564 |
| C | 1.991599 | -0.654373 | -0.287177 |
| H | 0.995142 | 2.743063 | 0.148872 |
| H | 1.748304 | 1.041667 | -0.098365 |
| H | -1.857816 | -1.809823 | 0.037566 |
| H | -2.558147 | -0.106873 | 0.278806 |
| H | 0.622001 | -2.096034 | -0.269984 |

Frequencies

| | | |
|---|---|---|
| 195.0156 | 222.3138 | 304.2281 |
| 395.7098 | 493.7103 | 520.6605 |
| 626.1489 | 640.6000 | 766.0342 |
| 794.2101 | 801.2346 | 928.0259 |
| 946.9070 | 990.3141 | 1015.3876 |
| 1149.4864 | 1264.4663 | 1424.0187 |
| 1490.3829 | 1653.8548 | 1701.4364 |
| 1733.0366 | 2518.7911 | 3111.1026 |
| 3148.1583 | 3167.7159 | 3262.1691 |

**i8''**

| | | | |
|---|---|---|---|
| N | -0.7292145011 | 1.1458653784 | -0.6421200114 |
| C | 0.5911327122 | 1.2000919561 | -0.5123567789 |
| C | -1.3503864737 | 0.0178200226 | -0.5872979444 |
| C | -0.6272193001 | -1.1892936419 | -0.0647640312 |
| C | 0.8383711285 | -1.079798316 | -0.3024108535 |
| C | 1.4045805051 | 0.0451595367 | -0.7806193936 |
| H | 1.0183638986 | 2.1187201801 | -0.1236424972 |
| H | -2.3190276717 | -0.0649873219 | -1.0711340439 |
| H | -1.0291569608 | -2.0792618149 | -0.5615428072 |
| H | -0.7912178821 | -1.3706727716 | 1.0028607772 |
| H | 2.4400045451 | 0.1406307923 | -1.0773814159 |

Frequencies

| | | |
|---|---|---|
| 239.1324 | 406.3987 | 442.8284 |
| 524.8076 | 617.7342 | 755.7827 |
| 784.0421 | 819.1163 | 858.8821 |
| 884.0443 | 930.0744 | 993.1080 |
| 1074.6735 | 1163.2201 | 1199.8168 |
| 1248.4787 | 1279.6555 | 1325.8255 |
| 1370.3485 | 1391.7727 | 1526.3940 |
| 1604.2493 | 3009.3741 | 3042.1096 |
| 3133.1478 | 3143.2087 | 3173.1282 |

**i9''**

| | | | |
|---|---|---|---|
| N | 0.112793 | -1.101983 | 0.010418 |
| C | 1.478875 | -0.598472 | 0.029725 |
| C | -2.058297 | -0.182799 | -0.015805 |
| C | -0.687032 | -0.052345 | 0.075816 |
| C | 0.072171 | 1.263454 | 0.010626 |
| C | 1.324579 | 0.875721 | -0.128973 |
| H | 2.084819 | -1.137611 | -0.703298 |
| H | 1.932565 | -0.784653 | 1.007767 |
| H | -2.727263 | 0.557886 | 0.398561 |
| H | -2.479662 | -0.993945 | -0.591413 |
| H | -0.381785 | 2.238858 | -0.012874 |

Frequencies

| | | |
|---|---|---|
| 209.7937 | 301.0197 | 339.3796 |
| 411.1292 | 583.8559 | 656.8328 |
| 693.1627 | 786.2121 | 804.3294 |
| 866.2269 | 882.9250 | 903.5660 |
| 969.7584 | 1049.9006 | 1128.0107 |
| 1192.1379 | 1207.2359 | 1291.1352 |
| 1389.8395 | 1455.1981 | 1550.0055 |
| 1591.3609 | 3027.7601 | 3059.2289 |
| 3154.1627 | 3244.8736 | 3252.8396 |

**i10''**

| | | | |
|---|---|---|---|
| N | -0.3984892466 | 1.0595669939 | 0.2427558971 |
| C | 0.881075444 | 1.1849419039 | 0.1001726806 |
| C | -1.8992440879 | -0.8237742875 | 0.0721508462 |
| C | -0.6654819112 | -0.3128727266 | 0.0095638082 |
| C | 0.58968907 | -1.0079827176 | -0.2901709714 |
| C | 1.5581631984 | -0.0736927105 | -0.2333051205 |
| H | 1.3695444353 | 2.1435233327 | 0.2250157762 |
| H | -2.0822148687 | -1.8735395937 | -0.1091624369 |
| H | -2.7401774234 | -0.1864497137 | 0.3057695905 |
| H | 0.6858786258 | -2.0595926349 | -0.5068632194 |
| H | 2.6166527642 | -0.201119846 | -0.396295850 |

Frequencies

| | | |
|---|---|---|
| 215.8776 | 361.2693 | 524.2419 |
| 656.7713 | 714.7008 | 771.6220 |
| 804.8083 | 855.5774 | 910.1132 |
| 927.0609 | 948.6039 | 960.3147 |
| 982.7142 | 995.3293 | 1095.7845 |
| 1236.0254 | 1317.5072 | 1364.7241 |
| 1438.0710 | 1500.8976 | 1611.1601 |
| 1706.4605 | 3154.7640 | 3172.3051 |
| 3217.6542 | 3238.7427 | 3248.0791 |

**i11''**

| | | | |
|---|---|---|---|
| N | -0.4128560963 | -0.9499703571 | -0.823772183 |
| C | 0.8775819644 | -1.1529291594 | -0.3702104976 |
| C | -1.6169684187 | -0.2530727576 | -0.2440794855 |
| C | -0.569414476 | 0.4122938206 | -1.0057193738 |
| C | 0.6209357895 | 1.0655804857 | -0.7627440414 |
| C | 1.5629422893 | 0.0422027433 | -0.4590315302 |
| H | 1.2468197735 | -2.1434745015 | -0.1606408189 |
| H | -1.6674825309 | -0.0921730301 | 0.8326830741 |
| H | -2.5346880776 | -0.5873117863 | -0.7091868534 |
| H | 0.7953582393 | 2.1282670884 | -0.7613717438 |
| H | 2.6213735434 | 0.1750144541 | -0.2926415465 |

Frequencies

| | | |
|---|---|---|
| 262.8914 | 398.2524 | 596.4878 |
| 682.9826 | 775.3911 | 805.0076 |
| 844.1164 | 860.3412 | 910.7915 |
| 956.2375 | 966.4662 | 1028.8614 |
| 1044.1842 | 1076.6567 | 1133.4898 |
| 1149.5762 | 1232.4524 | 1324.8803 |
| 1401.4021 | 1450.3796 | 1501.4291 |
| 1582.4501 | 3062.2408 | 3179.8226 |
| 3207.1971 | 3234.2922 | 3242.1391 |

**i13''**

| | | | |
|---|---|---|---|
| N | -0.7035404545 | 1.150910699 | -0.6016999379 |
| C | 0.569996411 | 1.1446064701 | -0.4303233198 |
| C | -1.3260999043 | -0.117810772 | -0.2757175434 |
| C | -0.498594353 | -1.3389319302 | -0.2538119554 |
| C | 0.837314346 | -1.2719527284 | -0.5646155919 |
| C | 1.3773223949 | -0.0301150079 | -0.235839485 |
| H | 1.099808584 | 2.0877659626 | -0.5491258399 |
| H | -2.1796956596 | -0.2933024771 | -0.946382539 |
| H | -1.8075426329 | -0.0182932947 | 0.7061168766 |
| H | 1.4413163462 | -2.1383540769 | -0.8051844051 |
| H | 2.4042369222 | 0.0852571556 | 0.0920477409 |

Frequencies

| | | |
|---|---|---|
| 214.7525 | 303.9880 | 442.0225 |
| 551.3136 | 584.7212 | 778.9524 |
| 832.1376 | 883.3665 | 910.9268 |
| 933.1649 | 946.0195 | 989.5555 |
| 1018.0630 | 1162.0432 | 1179.9352 |
| 1260.9375 | 1307.1619 | 1342.9664 |
| 1364.4234 | 1381.8756 | 1495.1358 |
| 1607.0779 | 2988.3159 | 2997.2003 |
| 3108.1051 | 3146.1260 | 3159.8011 |

**i12''**

| | | | |
|---|---|---|---|
| N | -0.7359722139 | 1.2061105275 | 0. |
| C | 0.5981266297 | 1.2073644302 | 0. |
| C | -1.3472272486 | 0.0205355566 | 0. |
| C | -0.669957128 | -1.1933918478 | 0. |
| C | 0.7181483568 | -1.1770787145 | 0. |
| C | 1.3677129828 | 0.0498986353 | 0. |
| H | 1.0742733952 | 2.1815763129 | 0. |
| H | -2.4313680107 | 0.0426687654 | 0. |
| H | -1.2205890411 | -2.1238774727 | 0. |
| H | 1.2816251307 | -2.1007559891 | 0. |
| H | 2.4470424663 | 0.1136463322 | 0. |

Frequencies

| | | |
|---|---|---|
| 385.0580 | 421.1848 | 617.1777 |
| 670.6838 | 720.6418 | 768.8550 |
| 900.1629 | 963.6626 | 1011.1029 |
| 1012.2452 | 1022.8882 | 1051.6637 |
| 1080.1513 | 1095.8266 | 1172.8986 |
| 1243.4233 | 1282.6877 | 1390.4050 |
| 1477.0811 | 1517.4330 | 1620.7286 |
| 1627.0112 | 3145.5433 | 3148.2899 |
| 3171.6967 | 3188.3994 | 3195.9953 |

**i14''**

| | | | |
|---|---|---|---|
| C | -0.155584975 | 1.1258001963 | 0.0972934457 |
| N | 1.0081607036 | 1.6163262798 | -0.0089725479 |
| C | -1.6689545179 | -0.7775443367 | 0.1955595128 |
| C | -0.4091521361 | -0.330601499 | 0.0776780519 |
| C | 0.6967173745 | -1.2194217619 | -0.0665490791 |
| C | 1.6257650094 | -1.9695639599 | -0.1878275458 |
| H | -1.0624442902 | 1.7293118447 | 0.2117290345 |
| H | 0.9787932647 | 2.6339834168 | 0.0266889756 |
| H | -1.9011762181 | -1.8318966054 | 0.18634337 |
| H | -2.4911679911 | -0.0828289749 | 0.303823763 |
| H | 2.4535217761 | -2.6246175996 | -0.2954239807 |

Frequencies

| | | |
|---|---|---|
| 155.3187 | 155.6702 | 260.1579 |
| 311.9809 | 475.1999 | 534.2345 |
| 670.9766 | 674.7578 | 679.8206 |
| 734.4054 | 759.8282 | 863.1626 |
| 960.9739 | 962.4278 | 1109.5932 |
| 1192.6383 | 1332.2631 | 1408.8986 |
| 1450.1202 | 1649.2735 | 1697.2320 |
| 2216.9262 | 3013.8644 | 3149.7411 |
| 3243.3312 | 3462.3234 | 3471.3871 |

**i15"**

| | | | |
|---|---|---|---|
| C | -0.5146406869 | 1.0024750489 | 0.917546642 |
| N | 0.661235018 | 1.4780399394 | 0.8035080412 |
| C | -1.6206042789 | -1.2933411086 | 0.9524892429 |
| C | -0.5537421853 | -0.4792349362 | 0.8641956392 |
| C | 0.8082377258 | -0.9768097089 | 0.6915978855 |
| C | 1.562423995 | 0.3011315002 | 0.6563630381 |
| H | -1.3803530044 | 1.6463090937 | 1.0399334145 |
| H | -1.4791418424 | -2.3639609286 | 0.8948945671 |
| H | -2.6301496803 | -0.9187919145 | 1.0806658659 |
| H | 2.3292484925 | 0.3064835655 | 1.4412005219 |
| H | 2.1360954469 | 0.373273449 | -0.2762118584 |

Frequencies

| | | |
|---|---|---|
| 126.0641 | 278.2498 | 348.6203 |
| 549.3659 | 621.5931 | 678.2593 |
| 803.3156 | 804.0417 | 898.6820 |
| 924.1759 | 933.6198 | 960.0546 |
| 1016.5948 | 1053.9580 | 1158.6038 |
| 1247.2384 | 1270.5116 | 1303.2150 |
| 1361.8508 | 1436.2478 | 1644.6926 |
| 1662.0507 | 3009.7081 | 3024.0987 |
| 3130.0616 | 3140.9697 | 3229.8471 |

**i16”**

| | | | |
|---|---|---|---|
| C | -0.5479631203 | 1.3582705445 | 0.1064556861 |
| N | 0.6508625964 | 1.8363209874 | -0.0102590296 |
| C | -1.4729096331 | -0.9218547931 | 0.1213267252 |
| C | -0.4167836403 | -0.0951764916 | 0.0348529178 |
| C | 1.0089124695 | -0.325771122 | -0.1343139427 |
| C | 1.7426811556 | 0.829485942 | -0.172490844 |
| H | -1.4425141092 | 1.9517590306 | 0.2303174431 |
| H | 0.8595400707 | 2.8237733985 | 0.0045736483 |
| H | -1.3519792129 | -1.9956664968 | 0.0659854637 |
| H | -2.4795354612 | -0.5458306894 | 0.2491024161 |
| H | 1.4338418849 | -1.3139393103 | -0.2205424841 |

Frequencies

| | | |
|---|---|---|
| 132.5455 | 303.5686 | 334.7713 |
| 605.6927 | 676.7597 | 702.8028 |
| 734.2577 | 813.3894 | 820.2104 |
| 824.3547 | 910.5704 | 946.7966 |
| 970.0152 | 990.1267 | 1160.4473 |
| 1188.5396 | 1286.6489 | 1398.2481 |
| 1436.0355 | 1494.0619 | 1552.6028 |
| 1673.0249 | 3140.0503 | 3200.9534 |
| 3208.1957 | 3226.8313 | 3597.2696 |

**i17”**

| | | | |
|---|---|---|---|
| C | -0.2162550272 | 1.0529015843 | 0.1044361807 |
| N | 0.966805281 | 1.5471159254 | -0.0080562418 |
| C | -1.2799285937 | -1.2225485004 | 0.136454034 |
| C | -0.212375471 | -0.4230577089 | 0.0482228905 |
| C | 1.2014504235 | -0.7491532395 | -0.1209631373 |
| C | 1.8477860721 | 0.4306433592 | -0.1480399181 |
| H | -1.097185939 | 1.6704624632 | 0.2255328782 |
| H | -1.1932683273 | -2.3000526209 | 0.0861077685 |
| H | -2.2742816716 | -0.8141290899 | 0.2622130526 |
| H | 1.6195081274 | -1.7382983915 | -0.2047482334 |
| H | 2.9046921258 | 0.614935219 | -0.2583012739 |

Frequencies

| | | |
|---|---|---|
| 208.3867 | 335.1508 | 493.3692 |
| 652.7495 | 691.1667 | 727.8829 |
| 802.2892 | 831.3998 | 900.4739 |
| 948.5291 | 949.8403 | 965.4203 |
| 991.7074 | 999.6659 | 1121.1981 |
| 1261.1888 | 1307.6438 | 1343.5292 |
| 1458.9608 | 1544.0775 | 1605.5293 |
| 1705.4387 | 3138.7221 | 3176.3439 |
| 3219.8309 | 3223.2182 | 3244.3470 |

**i18”**

| | | | |
|---|---|---|---|
| C | -1.2717463681 | 1.4519263728 | -0.735551344 |
| N | -0.0081141748 | 1.5511325065 | -0.9693561761 |
| C | -2.0647036922 | 0.3929190629 | -1.443937182 |
| C | -1.2090137499 | -0.7760333608 | -1.7888368684 |
| C | 0.1263244518 | -0.7628225822 | -1.6102491437 |
| C | 0.6437118775 | 0.5784270591 | -1.5955532682 |
| H | -1.6940287427 | 2.0249411744 | 0.0845558301 |
| H | -2.5495823334 | 0.7472951326 | -2.3598925171 |
| H | -2.8914710261 | 0.0833955969 | -0.7951923047 |
| H | 0.7768660095 | -1.6174710767 | -1.7356290419 |
| H | 1.5322947484 | 0.8513761144 | -2.1553489841 |

Frequencies

| | | |
|---|---|---|
| 237.2123 | 406.0228 | 442.5767 |
| 524.3662 | 617.4667 | 755.5948 |
| 784.2408 | 819.4467 | 859.0195 |
| 883.8848 | 929.5247 | 993.1945 |
| 1074.8721 | 1162.9924 | 1199.5593 |
| 1247.6127 | 1280.0867 | 1325.5719 |
| 1369.7941 | 1391.4365 | 1525.8554 |
| 1603.9923 | 3008.4050 | 3041.9796 |
| 3132.3920 | 3142.6631 | 3171.8031 |

**i19”**

| | | | |
|---|---|---|---|
| N | -1.1615262955 | -0.5324013648 | -0.1944882569 |
| C | -1.2682572799 | 0.7263624778 | -0.0390043289 |
| C | -0.1281879117 | 1.6182274805 | 0.1742874634 |
| C | 1.1773758941 | 1.2967980398 | 0.2426449022 |
| C | 1.7704996895 | 0.0215492889 | 0.125653045 |
| C | 2.3829726262 | -1.0105960172 | 0.0421374688 |
| H | -2.0772051801 | -0.9583059077 | -0.3278899009 |
| H | -2.2359021138 | 1.2401465915 | -0.0541355402 |
| H | -0.3802088611 | 2.6651672344 | 0.2900611935 |
| H | 1.8794195662 | 2.1077666156 | 0.4081568257 |
| H | 2.8764218663 | -1.9463264388 | -0.0387038716 |

Frequencies

| | | |
|---|---|---|
| 70.6927 | 141.9694 | 268.8271 |
| 278.4458 | 467.9024 | 501.6666 |
| 649.1345 | 678.7968 | 745.5298 |
| 818.4114 | 850.6556 | 882.3810 |
| 1006.6841 | 1024.3301 | 1116.6376 |
| 1237.6138 | 1267.2272 | 1416.7194 |
| 1462.3331 | 1632.2349 | 1698.9888 |
| 2196.1259 | 3004.3147 | 3132.1995 |
| 3167.9687 | 3458.7031 | 3471.4914 |

**i20”**

| | | | |
|---|---|---|---|
| N | -0.7602801275 | -0.8984451246 | 0.2250473021 |
| C | -1.1140966009 | 0.3888143201 | 0.0544479831 |
| C | -0.1237597465 | 1.3429198375 | 0.1031259027 |
| C | 1.1973127634 | 0.9362167593 | 0.322904806 |
| C | 1.6065004275 | -0.4031744054 | 0.5081823232 |
| C | 0.5326465619 | -1.2807480742 | 0.4435711609 |
| H | -1.4853131525 | -1.6048928449 | 0.1888958922 |
| H | -2.1594312458 | 0.5973878696 | -0.1120659874 |
| H | -0.3920238676 | 2.3833405344 | -0.0316927456 |
| H | 1.9476429644 | 1.7240069747 | 0.3485532318 |
| H | 0.6280380236 | -2.3568198465 | 0.5620501309 |

Frequencies

| | | |
|---|---|---|
| 129.8156 | 382.1494 | 607.0255 |
| 645.4805 | 672.7497 | 741.5567 |
| 831.7271 | 893.7848 | 945.3990 |
| 970.3572 | 1013.3146 | 1044.9947 |
| 1060.0701 | 1087.7355 | 1195.7398 |
| 1262.2073 | 1315.1955 | 1358.5382 |
| 1452.1637 | 1482.0709 | 1601.6751 |
| 1621.1859 | 3090.4651 | 3113.3922 |
| 3163.0351 | 3216.9526 | 3516.5337 |

**i21”**

| | | | |
|---|---|---|---|
| N | -1.0313652824 | -0.9688044562 | 0.1126587621 |
| C | -1.2607656365 | 0.2489442163 | -0.2276044545 |
| C | -0.2663782189 | 1.2820489217 | -0.3421885533 |
| C | 0.9787909829 | 1.0777380588 | 0.2498352056 |
| C | 1.3801843649 | -0.2305176693 | 0.1385080381 |
| C | 0.3774027527 | -1.3112246758 | 0.0798713378 |
| H | -2.2995337049 | 0.5558940597 | -0.3336674101 |
| H | -0.5315285249 | 2.2174813991 | -0.8220054928 |
| H | 1.6398119219 | 1.885810472 | 0.5392242628 |
| H | 0.6010249812 | -2.0306881951 | 0.8808559446 |
| H | 0.5645976395 | -1.8967424279 | -0.8298630337 |

Frequencies

| | | |
|---|---|---|
| 215.0134 | 303.7726 | 441.7982 |
| 551.2716 | 584.8037 | 779.3563 |
| 832.1040 | 883.7259 | 911.1397 |
| 933.2786 | 945.3735 | 989.8643 |
| 1018.3801 | 1161.9735 | 1180.1615 |
| 1261.2461 | 1307.3500 | 1343.2197 |
| 1364.0781 | 1381.8298 | 1495.3549 |
| 1607.2588 | 2988.3178 | 2996.6725 |
| 3106.6045 | 3143.8585 | 3157.7417 |

**i22”**

| | | | |
|---|---|---|---|
| N | -0.6170445952 | -1.1824065585 | -0.0823793066 |
| C | -1.0013441891 | 0.1110564795 | -0.2584714891 |
| C | -0.058389694 | 1.0968665664 | -0.218623403 |
| C | 1.2749130543 | 0.7145778872 | 0.00524343 |
| C | 1.5908272994 | -0.6158605717 | 0.1751885052 |
| C | 0.637438308 | -1.6727628624 | 0.140564373 |
| H | -1.355495545 | -1.8713411485 | -0.1214403727 |
| H | -2.053933769 | 0.2950508364 | -0.4241011862 |
| H | -0.3413183093 | 2.1296239805 | -0.3558748228 |
| H | 2.0455269011 | 1.476392115 | 0.0419553942 |
| H | 2.6252415388 | -0.887820724 | 0.3465308779 |

Frequencies

| | | |
|---|---|---|
| 328.7826 | 399.6054 | 620.8401 |
| 646.0188 | 677.9736 | 763.2289 |
| 863.2379 | 898.6014 | 971.3153 |
| 977.1490 | 1033.3269 | 1044.6121 |
| 1057.5776 | 1084.4232 | 1179.3570 |
| 1230.2539 | 1242.5224 | 1402.1942 |
| 1455.5157 | 1517.2436 | 1572.2467 |
| 1653.9083 | 3142.7609 | 3160.8356 |
| 3182.7943 | 3208.9729 | 3545.9573 |

**Transition states**

**i13''-i12''**

| | | | |
|---|---|---|---|
| N | -0.7351698553 | 1.1576358409 | -0.1588559638 |
| C | 0.5524844988 | 1.1693709263 | -0.0164820846 |
| C | -1.3381647923 | -0.114510272 | -0.0811487193 |
| C | -0.6675863986 | -1.4054543088 | -0.1117556207 |
| C | 0.7651877367 | -1.2372140998 | -0.0315679543 |
| C | 1.3507389747 | -0.0124658128 | 0.0992107974 |
| H | 1.0457027999 | 2.1373082699 | -0.0401356413 |
| H | -2.4071939219 | -0.0876993898 | -0.2676404689 |
| H | -1.2684287235 | -0.5536955815 | 1.0005813187 |
| H | 1.3879260562 | -2.1226515052 | -0.1005670387 |
| H | 2.4223666253 | 0.0957169328 | 0.2225433756 |

Frequencies

550.0235i

| | | |
|---|---|---|
| 201.3572 | 371.4614 | 597.3076 |
| 625.2420 | 701.4467 | 778.6399 |
| 877.1027 | 918.3343 | 997.5089 |
| 1006.0660 | 1026.7763 | 1041.4261 |
| 1062.6254 | 1145.2094 | 1214.2870 |
| 1243.7459 | 1372.0576 | 1397.8448 |
| 1448.4657 | 1527.0841 | 1630.8172 |
| 2479.0553 | 3122.3929 | 3135.5134 |
| 3136.2927 | 3160.4337 | |

**i8"-i12"**

| | | | |
|---|---|---|---|
| N | -0.7933788696 | 1.2728390424 | -0.1095229789 |
| C | 0.5765891892 | 1.2742962329 | -0.0910645812 |
| C | -1.3794083908 | 0.1185945903 | -0.0110150796 |
| C | -0.6406870491 | -1.1193315665 | 0.1322088907 |
| C | 0.8068064978 | -1.1755922855 | 0.0943515016 |
| C | 1.3513355828 | 0.15504288 | 0.0392205792 |
| H | 1.0250230618 | 2.2576501578 | -0.189122245 |
| H | -2.466501421 | 0.107018112 | -0.0350047712 |
| H | -1.2035535085 | -2.0471851642 | 0.0688075557 |
| H | -0.1063820078 | -1.235071221 | 1.2028234877 |
| H | 2.4292679151 | 0.2751832219 | 0.0381876409 |

Frequencies

| | | |
|---|---|---|
| 822.0571i | 137.3827 | 378.0750 |
| 589.4335 | 659.9247 | 690.0920 |
| 707.8510 | 849.9117 | 932.0338 |
| 962.0004 | 1000.3362 | 1032.9380 |
| 1035.8846 | 1072.1104 | 1204.8966 |
| 1220.0697 | 1251.4997 | 1378.0573 |
| 1409.4799 | 1462.4062 | 1495.6375 |
| 1624.3412 | 2291.2158 | 3106.4321 |
| 3120.3603 | 3131.4924 | 3150.7284 |

**i4"-i13"**

| | | | |
|---|---|---|---|
| N | -0.7618752425 | 1.0458503489 | -0.4159834251 |
| C | 0.5225210754 | 1.100550771 | -0.2926214624 |
| C | -1.5958604465 | -0.1120368561 | -0.1633311928 |
| C | -0.5751075438 | -1.2386516867 | -0.313489258 |
| C | 0.6888964836 | -1.1429755824 | -0.1696632745 |
| C | 1.4588556525 | 0.0271727865 | -0.134667549 |
| H | 0.953530751 | 2.0992073436 | -0.3446962089 |
| H | -2.3928960151 | -0.1967821961 | -0.9006397252 |
| H | -2.0526967828 | -0.1000063198 | 0.8336106559 |
| H | 0.1717877153 | -1.7970662497 | 0.7374646259 |
| H | 2.521732353 | 0.1522536407 | -0.0598141859 |

Frequencies

1327.7355i

| | | |
|---|---|---|
| 113.9325 | 226.7294 | 426.7299 |
| 487.6655 | 581.0323 | 621.8820 |
| 695.1350 | 754.4839 | 810.1230 |
| 917.4382 | 953.1687 | 1005.4572 |
| 1033.8544 | 1129.7973 | 1163.5573 |
| 1241.8821 | 1280.1604 | 1315.8303 |
| 1510.7312 | 1576.5807 | 1880.2880 |
| 2250.4975 | 2969.6929 | 3069.5312 |
| 3109.3720 | 3277.8840 | |

**i4"-i8"**

| | | | |
|---|---|---|---|
| N | -0.7986857649 | 1.2254916837 | -0.609325027 |
| C | 0.5254314677 | 1.212212221 | -0.4941077512 |
| C | -1.4003167419 | 0.0226514896 | -0.5625973649 |
| C | -0.6351464366 | -1.2291915044 | -0.4495030857 |
| C | 0.7785519143 | -1.2436258602 | -0.5107822209 |
| C | 1.3122679899 | 0.0496957506 | -0.4919824403 |
| H | 0.9971377795 | 2.1917718039 | -0.4864392469 |
| H | -2.4790682067 | -0.0030904016 | -0.6932072796 |
| H | -1.2293880157 | -2.1326818315 | -0.3394748871 |
| H | -1.1428295646 | -0.5293178589 | 0.5527199626 |
| H | 2.386463579 | 0.1938955076 | -0.573371659 |

Frequencies

| | | |
|---|---|---|
| 984.4739i | 265.9603 | 375.4741 |
| 571.6704 | 663.6181 | 693.4994 |
| 794.8957 | 907.3305 | 925.5698 |
| 978.8283 | 1022.0935 | 1043.5917 |
| 1091.2923 | 1110.5681 | 1192.2137 |
| 1206.4869 | 1297.0351 | 1336.0182 |
| 1388.8011 | 1439.5591 | 1521.4565 |
| 1542.1443 | 2127.4152 | 3104.2441 |
| 3110.2876 | 3119.4962 | 3124.6703 |

**i2"-i4"**

| | | | |
|---|---|---|---|
| N | 0.447705 | 1.268685 | -0.097589 |
| C | 1.318133 | 0.328559 | 0.147543 |
| C | -0.879491 | 1.185405 | -0.021674 |
| C | -1.468721 | -0.790466 | -0.054879 |
| C | -0.326253 | -1.308930 | -0.036847 |
| C | 1.008127 | -1.054571 | 0.000558 |
| H | 2.359587 | 0.630364 | 0.199084 |
| H | -1.430536 | 1.798085 | -0.730169 |
| H | -1.330611 | 1.124103 | 0.962949 |
| H | -2.414591 | -0.885348 | 0.440463 |
| H | 1.771443 | -1.707985 | -0.397406 |

Frequencies

483.9000i

| | | |
|---|---|---|
| 231.6835 | 374.8444 | 449.4055 |
| 543.3739 | 580.9336 | 618.0123 |
| 708.9049 | 783.6927 | 829.0423 |
| 909.9062 | 939.2935 | 1035.2837 |
| 1070.2840 | 1133.6327 | 1164.7793 |
| 1185.4140 | 1379.4966 | 1414.7602 |
| 1517.7165 | 1558.4816 | 1933.6978 |
| 3076.5727 | 3137.3913 | 3150.1708 |
| 3184.3964 | 3303.3200 | |

**i2"-i6"**

| | | | |
|---|---|---|---|
| N | -0.7541836711 | 0.9890895214 | -0.507972708 |
| C | 0.6022291704 | 0.9939240544 | -0.2225501258 |
| C | -1.5991613344 | 0.4044279674 | 0.2275606569 |
| C | 0.0557365408 | -2.2996578152 | -0.6580911179 |
| C | 0.9792412558 | -1.4210863491 | -0.4953744998 |
| C | 1.4222060188 | -0.0696518451 | -0.2475579502 |
| H | 1.0254065748 | 1.983008726 | -0.0917032418 |
| H | -2.6429619236 | 0.3974958752 | -0.0730054516 |
| H | -1.337386265 | -0.0912024403 | 1.1652806573 |
| H | 1.7196395267 | -2.2592935695 | -0.6486581563 |
| H | 2.4814831068 | 0.0945908748 | -0.1199250629 |

Frequencies

| | | |
|---|---|---|
| 127.0088i | | |
| 64.8564 | 162.3611 | 242.3091 |
| 298.3166 | 436.1732 | 546.6085 |
| 620.9449 | 761.7136 | 795.3058 |
| 897.5876 | 945.0907 | 958.6717 |
| 1035.6192 | 1089.4206 | 1216.2248 |
| 1251.5165 | 1418.3948 | 1508.8004 |
| 1642.5657 | 1729.8938 | 1848.4707 |
| 2803.0816 | 3033.3127 | 3150.8521 |
| 3160.3521 | 3203.9121 | |

**i10"-i11"**

| | | | |
|---|---|---|---|
| N | -0.444685 | -0.867740 | -0.321104 |
| C | 0.834555 | -1.051887 | 0.112151 |
| C | -1.776126 | 0.109972 | 0.293374 |
| C | -0.610590 | 0.517316 | -0.373715 |
| C | 0.567706 | 1.186996 | -0.086393 |
| C | 1.509443 | 0.165200 | 0.148696 |
| H | 1.225669 | -2.047826 | 0.246317 |
| H | -1.961729 | 0.392996 | 1.328565 |
| H | -2.581395 | -0.385763 | -0.225241 |
| H | 0.716542 | 2.250176 | -0.006463 |
| H | 2.563786 | 0.299013 | 0.339869 |

Frequencies

| | | |
|---|---|---|
| 510.0623i | | |
| 276.2802 | 561.5614 | 635.0294 |
| 739.8253 | 776.4670 | 805.6227 |
| 827.4376 | 868.1204 | 921.7999 |
| 993.3516 | 1012.8552 | 1055.1752 |
| 1071.8298 | 1120.4874 | 1127.7848 |
| 1219.5432 | 1303.9035 | 1428.9981 |
| 1432.2758 | 1488.7863 | 1618.9560 |
| 3082.7307 | 3203.4112 | 3223.9921 |
| 3226.3824 | 3244.6997 | |

**i5"-i10"**

| | | | |
|---|---|---|---|
| N | -0.3281884203 | 1.0022300262 | 0.3386926469 |
| C | 0.9492721532 | 1.1342290372 | 0.1813712608 |
| C | -1.8532734018 | -0.8650108212 | 0.1640959777 |
| C | -0.6264119276 | -0.345721007 | 0.0753319442 |
| C | 0.650556238 | -1.0236631995 | -0.2682703303 |
| C | 1.7353302345 | -0.1143207551 | -0.1242035523 |
| H | 1.4161451936 | 2.1020795433 | 0.3155808302 |
| H | -2.0368305342 | -1.9067179139 | -0.058465936 |
| H | -2.6883116304 | -0.2500767159 | 0.468736374 |
| H | 0.756428008 | -2.098327109 | -0.3539890357 |
| H | 1.1603800871 | -0.6237950851 | -1.3055021795 |

Frequencies

| | | |
|---|---|---|
| 823.6594i | | |
| 206.4253 | 357.9444 | 451.3212 |
| 615.5658 | 706.6333 | 765.6990 |
| 832.2671 | 868.9730 | 923.3707 |
| 937.8860 | 961.7153 | 982.5912 |
| 986.8373 | 1128.9701 | 1250.4627 |
| 1272.3529 | 1325.6540 | 1350.6624 |
| 1437.2171 | 1530.0035 | 1702.8866 |
| 2232.2943 | 3153.9825 | 3170.5719 |
| 3177.7586 | 3246.5108 | |

**i11"-i13"**

| | | | |
|---|---|---|---|
| N | 0.804389 | 0.942175 | -0.242806 |
| C | -0.476865 | 1.185473 | 0.030847 |
| C | 1.534532 | -0.234598 | 0.160068 |
| C | 0.414415 | -1.146191 | -0.051484 |
| C | -0.889111 | -1.121377 | -0.116120 |
| C | -1.443001 | 0.219432 | 0.109431 |
| H | -0.753873 | 2.235674 | -0.007950 |
| H | 1.875904 | -0.269505 | 1.206533 |
| H | 2.397325 | -0.361857 | -0.491395 |
| H | -1.513253 | -2.002622 | -0.174604 |
| H | -2.476647 | 0.386657 | 0.370605 |

Frequencies

609.9614i

| | | |
|---|---|---|
| 152.8741 | 430.7562 | 511.3330 |
| 549.3905 | 716.7450 | 787.8045 |
| 824.1657 | 858.9552 | 894.7317 |
| 972.5747 | 985.5300 | 1028.1509 |
| 1102.4835 | 1177.2104 | 1184.2307 |
| 1243.2821 | 1322.4975 | 1399.3943 |
| 1482.2799 | 1539.9813 | 1727.5092 |
| 2914.6931 | 3074.4913 | 3129.1524 |
| 3184.1367 | 3212.4858 | |

**i3"-i5"**

| | | | |
|---|---|---|---|
| N | 0.128095 | -1.190535 | 0.191607 |
| C | 1.374042 | -0.698548 | -0.050657 |
| C | -2.048294 | -0.046466 | -0.140871 |
| C | -0.767044 | -0.252393 | 0.131438 |
| C | 0.337592 | 1.350720 | 0.148631 |
| C | 1.417188 | 0.644632 | -0.351435 |
| H | 2.210719 | -1.381935 | -0.030989 |
| H | -2.556026 | 0.850752 | 0.179251 |
| H | -2.594527 | -0.750697 | -0.756582 |
| H | -0.104376 | 2.180273 | -0.391465 |
| H | 0.266642 | 1.447685 | 1.235898 |

Frequencies

620.2095i

| | | |
|---|---|---|
| 215.8097 | 280.0213 | 385.6445 |
| 493.2943 | 545.7355 | 655.7528 |
| 708.6902 | 831.9449 | 849.0103 |
| 902.3089 | 946.9508 | 985.4724 |
| 1018.0446 | 1073.2442 | 1116.6834 |
| 1241.9815 | 1344.4161 | 1413.4911 |
| 1455.0751 | 1493.2451 | 1756.4769 |
| 3047.7192 | 3137.2244 | 3160.3225 |
| 3204.1172 | 3231.1628 | |

**i1"-i7"**

| | | | |
|---|---|---|---|
| N | 1.299595 | 0.083652 | -0.033481 |
| C | 1.611732 | -1.146235 | -0.014653 |
| C | -0.331664 | 1.797774 | -0.032381 |
| C | -0.050377 | 0.490400 | -0.042975 |
| C | -1.069929 | -0.533077 | -0.055827 |
| C | -1.823915 | -1.538248 | -0.061569 |
| H | 2.666081 | -1.404399 | -0.012551 |
| H | 0.892154 | -1.968111 | 0.004005 |
| H | -1.346976 | 2.163663 | -0.054981 |
| H | 0.480227 | 2.508016 | 0.005797 |
| H | -2.326924 | -0.453438 | 0.298615 |

Frequencies
624.0383i

| | | |
|---|---|---|
| 80.8279 | 138.9801 | |
| 183.6088 | 260.4162 | 343.9896 |
| 491.7323 | 539.0984 | 656.2537 |
| 728.8609 | 757.8705 | 812.8737 |
| 925.6964 | 946.9996 | 1071.9757 |
| 1205.0256 | 1261.4851 | 1419.3536 |
| 1500.6093 | 1663.2089 | 1695.0736 |
| 1976.2048 | 2400.5993 | 3036.6145 |
| 3156.3335 | 3166.4134 | 3263.4356 |

**i7''-i9''**

| | | | |
|---|---|---|---|
| N | 0.159230 | -1.085745 | -0.150579 |
| C | 1.418469 | -0.785867 | 0.121612 |
| C | -2.033610 | -0.199974 | 0.026385 |
| C | -0.694278 | -0.012904 | 0.051494 |
| C | 0.054219 | 1.248544 | 0.019838 |
| C | 1.354922 | 1.037245 | -0.181517 |
| H | 2.201440 | -1.416951 | -0.282216 |
| H | 1.640833 | -0.369585 | 1.105946 |
| H | -2.721748 | 0.604627 | 0.237347 |
| H | -2.430004 | -1.167166 | -0.245101 |
| H | -0.403461 | 2.227026 | 0.011209 |

Frequencies
506.3992i

| | | |
|---|---|---|
| 162.1764 | 284.8884 | |
| 366.9635 | 478.7418 | 608.6932 |
| 642.1583 | 677.7601 | 773.1673 |
| 840.3596 | 883.2086 | 911.9406 |
| 957.6467 | 1006.7173 | 1120.5869 |
| 1159.8832 | 1268.4322 | 1381.9846 |
| 1414.5680 | 1492.7266 | 1559.7357 |
| 1611.4110 | 3035.4619 | 3161.9136 |
| 3170.4109 | 3200.0188 | 3259.0737 |

**i19''-i10''**

| | | | |
|---|---|---|---|
| N | 0.108101 | -1.145932 | 0.057211 |
| C | 1.393474 | -0.607452 | 0.016444 |
| C | -2.041287 | -0.127032 | -0.052805 |
| C | -0.683640 | -0.047723 | 0.126298 |
| C | 0.079186 | 1.215639 | 0.042190 |
| C | 1.387977 | 0.836734 | -0.005274 |
| H | 1.733076 | -0.277583 | -1.095438 |
| H | 2.222314 | -1.208042 | 0.372627 |
| H | -2.688902 | 0.702817 | 0.195559 |
| H | -2.479371 | -1.003081 | -0.510769 |
| H | -0.358081 | 2.186417 | -0.123578 |

Frequencies
1223.9527i

| | | |
|---|---|---|
| 212.2061 | 332.5372 | |
| 486.6764 | 561.9663 | 620.5464 |
| 724.3827 | 737.8804 | 784.1763 |
| 829.6455 | 922.5950 | 977.8802 |
| 991.2115 | 1070.2856 | 1138.4339 |
| 1198.3232 | 1235.5216 | 1264.6527 |
| 1357.1576 | 1390.9710 | 1424.5058 |
| 1566.1711 | 2214.7166 | 3148.0133 |
| 3159.8449 | 3226.4216 | 3248.1124 |

**i2''-i6''**

| | | | |
|---|---|---|---|
| N | -1.402010 | 0.054007 | -0.433495 |
| C | -0.607963 | 1.112078 | -0.019583 |
| C | -1.577486 | -0.978437 | 0.273678 |
| C | 1.662133 | -1.370576 | -0.157882 |
| C | 1.530726 | -0.102189 | 0.000712 |
| C | 0.724602 | 1.086395 | 0.147127 |
| H | -1.129543 | 2.059697 | 0.047197 |
| H | -2.174491 | -1.793833 | -0.124592 |
| H | -1.166451 | -1.094000 | 1.279258 |
| H | 2.658213 | -0.047607 | 0.009059 |
| H | 1.234274 | 2.014070 | 0.359230 |

Frequencies

127.8968i

| | | |
|---|---|---|
| 65.3399 | 161.8648 | |
| 242.0683 | 297.9304 | 435.4351 |
| 546.5696 | 620.8126 | 761.7367 |
| 795.0518 | 897.5472 | 944.9988 |
| 958.9305 | 1035.5865 | 1089.5311 |
| 1216.2505 | 1251.4941 | 1418.2637 |
| 1508.7561 | 1642.4265 | 1730.0720 |
| 1848.6106 | 2803.4078 | 3032.6111 |
| 3149.9418 | 3159.2550 | 3202.9243 |

**i6''-i13''**

| | | | |
|---|---|---|---|
| N | -1.425127 | 0.171991 | -0.292627 |
| C | -0.567226 | 1.181268 | 0.062069 |
| C | -1.212839 | -1.010942 | 0.143086 |
| C | 1.085189 | -1.413302 | 0.039375 |
| C | 1.493097 | -0.172289 | -0.145401 |
| C | 0.779907 | 1.062098 | 0.123249 |
| H | -1.014793 | 2.164059 | 0.143310 |
| H | -1.783178 | -1.849488 | -0.238690 |
| H | -0.583207 | -1.219152 | 1.019103 |
| H | 2.519116 | -0.129859 | -0.518400 |
| H | 1.369189 | 1.949503 | 0.308799 |

Frequencies

100.3633i

| | | |
|---|---|---|
| 188.8648 | 282.1239 | |
| 365.6946 | 454.3269 | 558.3478 |
| 620.5991 | 725.6281 | 827.5343 |
| 860.9662 | 954.4602 | 969.6213 |
| 980.1788 | 1063.0984 | 1093.1024 |
| 1179.5801 | 1241.7563 | 1419.8230 |
| 1468.0994 | 1544.8990 | 1641.5659 |
| 1656.6295 | 2951.4970 | 3073.3248 |
| 3164.0261 | 3171.1129 | 3188.0267 |

**i14”-i15”**

| | | | |
|---|---|---|---|
| C | -0.5227327936 | 1.4416933854 | 0.3256776329 |
| N | 0.459948118 | 2.1713911602 | 0.4920282343 |
| C | -1.4646019172 | -0.8950287856 | 0.3578159668 |
| C | -0.4000074048 | -0.0947348604 | 0.1786418035 |
| C | 0.9547469314 | -0.5011627192 | 0.0802840368 |
| C | 1.8488380325 | 0.4980646055 | 0.1013355964 |
| H | -1.5469705193 | 1.8251682021 | 0.3083947464 |
| H | -1.3392591739 | -1.9634713727 | 0.4507506692 |
| H | -2.4621856154 | -0.4887216617 | 0.4677948342 |
| H | 2.8127930641 | 0.5163095111 | 0.5905451525 |
| H | 1.7300352782 | 1.2101045353 | -0.721024673 |

Frequencies

| | | |
|---|---|---|
| 542.7095*i* | 183.1602 | 329.1198 |
| 354.0124 | 469.4557 | 548.5387 |
| 585.5548 | 631.3610 | 699.5581 |
| 758.6110 | 815.6566 | 880.2329 |
| 926.8376 | 955.8403 | 1003.0608 |
| 1121.4964 | 1202.7870 | 1327.9573 |
| 1430.6953 | 1520.3686 | 1645.5225 |
| 1721.8899 | 3039.5087 | 3050.5593 |
| 3145.9408 | 3206.2362 | 3245.0780 |

**i14”-i16”**

| | | | |
|---|---|---|---|
| C | -0.1516478096 | 1.1813518377 | 0.0861964656 |
| N | 0.7422073939 | 2.0667764641 | -0.0533892499 |
| C | -0.8041270658 | -1.1706571343 | 0.1787567741 |
| C | 0.1669842457 | -0.2599826738 | 0.0268624365 |
| C | 1.5432961357 | -0.5979408145 | -0.1976616067 |
| C | 2.7684650825 | -0.7966607408 | -0.3977027114 |
| H | -1.2117102432 | 1.3921548477 | 0.2585329625 |
| H | 0.3519905428 | 3.0048546127 | 0.0144227473 |
| H | -0.6145761297 | -2.2331855768 | 0.1423225833 |
| H | -1.8238302005 | -0.8535312471 | 0.3451870565 |
| H | 2.0345820482 | -1.8046075749 | -0.285599458 |

Frequencies

| | | |
|---|---|---|
| 528.4077*i* | 158.6675 | 180.6082 |
| 224.0290 | 249.4671 | 330.7306 |
| 490.0488 | 514.7200 | 699.0250 |
| 733.9271 | 744.3921 | 860.0363 |
| 943.9258 | 954.6506 | 1107.0266 |
| 1180.1954 | 1325.4396 | 1404.3688 |
| 1454.1944 | 1652.7855 | 1696.6834 |
| 1989.0840 | 2394.4376 | 3026.2291 |
| 3156.1486 | 3244.5552 | 3472.4704 |

**i15”-i17”**

| | | | |
|---|---|---|---|
| C | -0.4077111759 | 1.3258789691 | 0.3712711844 |
| N | 0.7860547157 | 1.7924282893 | 0.3282077467 |
| C | -1.5312309674 | -0.932486732 | 0.3319088841 |
| C | -0.448447023 | -0.1465889081 | 0.306526473 |
| C | 0.9548073351 | -0.6435897836 | 0.2924879956 |
| C | 1.6332598613 | 0.6066314721 | 0.2332184303 |
| H | -1.2701610801 | 1.978852972 | 0.4459742258 |
| H | -1.422005582 | -2.0077531889 | 0.2936155434 |
| H | -2.534031468 | -0.5271026423 | 0.3986413571 |
| H | 2.6974054006 | 0.7493056632 | 0.3757201634 |
| H | 1.5769229837 | 0.0432948894 | -0.8337370036 |

Frequencies

| | | |
|---|---|---|
| 778.5130*i* | 174.8821 | 348.8342 |
| 497.3210 | 625.3309 | 689.6693 |
| 743.9506 | 820.8871 | 903.0989 |
| 916.2146 | 937.5223 | 966.2530 |
| 986.3258 | 1040.7438 | 1151.5230 |
| 1247.9383 | 1283.0094 | 1299.9517 |
| 1345.3133 | 1440.4076 | 1607.1233 |
| 1689.1518 | 2286.1880 | 3134.6856 |
| 3160.2462 | 3176.4009 | 3229.5280 |

**i16”-i17”**

| | | | |
|---|---|---|---|
| C | -0.5340479111 | 1.3657262918 | 0.086686808 |
| N | 0.660088614 | 1.8131410311 | -0.0394297857 |
| C | -1.5832513487 | -0.8804428804 | 0.1425765879 |
| C | -0.4954996831 | -0.1046452682 | 0.0386014142 |
| C | 0.9135236328 | -0.4481434729 | -0.1374160414 |
| C | 1.6752443328 | 0.6673072867 | -0.1940313856 |
| H | -1.412234361 | 1.9875189605 | 0.2076679387 |
| H | -1.5154928253 | -1.9594223186 | 0.1027155537 |
| H | -2.5696812251 | -0.4545491293 | 0.2701738072 |
| H | 1.2755238836 | -1.4617867001 | -0.2110136949 |
| H | 1.7726958913 | 2.0712331994 | -0.165188202 |

Frequencies

| | | |
|---|---|---|
| 1747.5421*i* | 203.2439 | 292.5521 |
| 338.8882 | 500.7453 | 502.9330 |
| 691.5440 | 701.6129 | 723.6302 |
| 816.1373 | 838.0993 | 896.1263 |
| 938.3769 | 964.8599 | 988.4943 |
| 1174.6297 | 1229.5614 | 1303.7706 |
| 1450.3940 | 1505.5322 | 1554.8849 |
| 1687.5697 | 2340.3095 | 3141.8278 |
| 3176.1959 | 3214.1982 | 3227.1200 |

**i17”-i18”**

| | | | |
|---|---|---|---|
| C | -0.5533879297 | 0.9731757766 | 0.3627605873 |
| N | 0.6265676738 | 1.4771251491 | -0.164679678 |
| C | -1.5784600551 | 0.2353825743 | -0.336283154 |
| C | -0.5326337629 | -0.8360206406 | -0.1451161193 |
| C | 0.7794116579 | -0.8354885137 | -0.3946630564 |
| C | 1.3951160106 | 0.5079837449 | -0.5058040835 |
| H | -0.6495943936 | 1.0382747845 | 1.4411746654 |
| H | -1.7422078752 | 0.3907970982 | -1.4010403665 |
| H | -2.5108433038 | 0.1173279055 | 0.2016229461 |
| H | 1.3295501428 | -1.7252626484 | -0.6769404541 |
| H | 2.3606738353 | 0.6886937697 | -0.971706287 |

Frequencies

| | | |
|---|---|---|
| 357.4599*i* | 302.4777 | 364.4092 |
| 540.3283 | 617.5787 | 728.8100 |
| 796.3015 | 863.5285 | 898.6433 |
| 922.8999 | 943.0836 | 980.3104 |
| 1029.6752 | 1130.0441 | 1167.6498 |
| 1213.1153 | 1288.1895 | 1329.0774 |
| 1418.2280 | 1470.3824 | 1571.1092 |
| 1630.9910 | 3079.9162 | 3127.9383 |
| 3142.5133 | 3157.2658 | 3164.6470 |

**i18”-i12”**

| | | | |
|---|---|---|---|
| C | -0.5422846877 | 1.3240639182 | 0.1917898833 |
| N | 0.7536567187 | 1.4019201134 | 0.1835143141 |
| C | -1.2535808293 | 0.1436960673 | -0.2540835041 |
| C | -0.5804758336 | -1.0083274556 | -0.8193083461 |
| C | 0.8456512415 | -0.8405617981 | -0.7294204149 |
| C | 1.4356381778 | 0.3116365449 | -0.2888290385 |
| H | -1.0919508343 | 2.1889371903 | 0.555569303 |
| H | -2.3345941818 | 0.2153916853 | -0.3433631944 |
| H | -1.1358486946 | -0.7942516994 | 0.488719585 |
| H | 1.486095014 | -1.6404162087 | -1.0849538497 |
| H | 2.5149699091 | 0.4245926425 | -0.2905917379 |

Frequencies

| | | |
|---|---|---|
| 825.0431*i* | 139.9434 | 378.1615 |
| 589.2691 | 660.5656 | 689.2255 |
| 708.1031 | 850.3268 | 932.2789 |
| 962.0216 | 1000.2415 | 1033.1316 |
| 1036.1066 | 1072.2329 | 1204.4380 |
| 1220.9683 | 1251.5625 | 1378.1322 |
| 1409.4713 | 1462.2296 | 1495.6016 |
| 1624.1832 | 2290.8575 | 3105.9597 |
| 3119.8209 | 3130.5138 | 3149.7789 |

**i15”-HCN+$H_2CCCH_2$**

| | | | |
|---|---|---|---|
| C | 0.007081362 | 1.4528025668 | 0.0588489816 |
| N | 1.1013558129 | 1.9185963374 | -0.028091573 |
| C | -1.6618568785 | -0.673699202 | 0.1423114169 |
| C | -0.3729523829 | -0.3963789386 | 0.0320441694 |
| C | 0.8417024568 | -0.9531537204 | -0.0995394027 |
| C | 2.0545870523 | -0.5713498804 | -0.2004365054 |
| H | -0.9964670588 | 1.8397381794 | 0.1662454968 |
| H | 1.9729668183 | 0.6341643964 | -0.1520629223 |
| H | -2.0211236302 | -1.6928743182 | 0.1413475936 |
| H | -2.3983010503 | 0.1133121094 | 0.2370518413 |
| H | 3.0140804983 | -1.0434035297 | -0.3066800964 |

Frequencies

| | | |
|---|---|---|
| 696.7617*i* | 195.7327 | 287.6803 |
| 341.5982 | 374.8231 | 492.1994 |
| 558.2464 | 580.0716 | 636.2786 |
| 743.8294 | 785.0965 | 787.6248 |
| 915.2549 | 957.0779 | 1000.1851 |
| 1013.7941 | 1111.2220 | 1222.7913 |
| 1460.8072 | 1703.1057 | 1729.9712 |
| 1937.3021 | 1964.0841 | 3141.6899 |
| 3197.7531 | 3225.8634 | 3281.0164 |

**i19”-i20”**

| | | | |
|---|---|---|---|
| N | -0.9801232382 | -0.9325469824 | 0.1887475933 |
| C | -1.1567458335 | 0.3385719994 | 0.0375967676 |
| C | -0.0961255378 | 1.3038567605 | 0.1043590036 |
| C | 1.2053126358 | 0.9815382047 | 0.333703847 |
| C | 1.6192772609 | -0.365664579 | 0.5286046559 |
| C | 0.9033271464 | -1.401065194 | 0.5210904701 |
| H | -1.8331651923 | -1.4834544107 | 0.1132252601 |
| H | -2.1573151274 | 0.727926731 | -0.1525020577 |
| H | -0.3798853387 | 2.3382020711 | -0.0399028638 |
| H | 1.9382365823 | 1.7819511448 | 0.3668022304 |
| H | 0.8144426424 | -2.4607097453 | 0.6112950935 |

Frequencies

| | | |
|---|---|---|
| 490.5713*i* | 128.6913 | 278.8504 |
| 465.8367 | 480.9345 | 529.4874 |
| 682.5611 | 754.6480 | 863.1571 |
| 865.0361 | 963.6902 | 1000.1250 |
| 1011.2771 | 1055.2631 | 1088.9702 |
| 1220.6487 | 1231.8653 | 1394.2724 |
| 1457.8980 | 1540.0757 | 1641.0578 |
| 1835.4571 | 3066.8451 | 3120.2639 |
| 3172.7505 | 3373.1448 | 3451.4963 |

**i20”-i21”**

| | | | |
|---|---|---|---|
| N | 1.165125 | 0.759353 | -0.062444 |
| C | -0.039077 | 1.339556 | -0.003682 |
| C | -1.215073 | 0.597450 | -0.018973 |
| C | -1.167638 | -0.792813 | 0.012075 |
| C | 0.033719 | -1.542344 | 0.008087 |
| C | 1.157350 | -0.687025 | -0.058128 |
| H | -0.045416 | 2.421133 | 0.049163 |
| H | -2.162868 | 1.124500 | -0.026206 |
| H | -2.117262 | -1.321955 | 0.031080 |
| H | 2.162051 | -1.065734 | -0.222593 |
| H | 1.391927 | 0.037647 | 0.969389 |

Frequencies

| | | |
|---|---|---|
| 1376.2607*i* | 206.5639 | 302.3420 |
| 595.6909 | 639.4869 | 678.7745 |
| 742.0762 | 808.3860 | 867.6760 |
| 966.5842 | 1007.0874 | 1031.2324 |
| 1045.3807 | 1078.6972 | 1120.7959 |
| 1185.4206 | 1275.8015 | 1330.6463 |
| 1375.7140 | 1435.9655 | 1536.7964 |
| 1567.7601 | 2207.5112 | 3103.9866 |
| 3128.3889 | 3146.7702 | 3169.5060 |

**i20”-i22”**

| | | | |
|---|---|---|---|
| N | -0.6478737264 | -0.9362165864 | 0.0335144308 |
| C | -0.9774899602 | 0.3580048357 | -0.182205058 |
| C | 0.0036786339 | 1.3239816252 | -0.1775349726 |
| C | 1.3246784757 | 0.9572606443 | 0.1194962274 |
| C | 1.6812375897 | -0.3905748365 | 0.120134031 |
| C | 0.6486610889 | -1.3640838135 | 0.0519435046 |
| H | -1.3913589649 | -1.6142492139 | 0.1324963667 |
| H | -2.0237621704 | 0.5813484766 | -0.3344502976 |
| H | -0.2823440101 | 2.3547062327 | -0.3417349259 |
| H | 2.0480214866 | 1.7359207599 | 0.3404862334 |
| H | 1.766378557 | -1.6955951241 | 0.5435994602 |

Frequencies

| | | |
|---|---|---|
| 1840.4806*i* | 185.7124 | 345.6135 |
| 539.0098 | 615.3002 | 651.6884 |
| 721.4421 | 767.6630 | 834.0252 |
| 940.8491 | 987.7462 | 991.7127 |
| 1035.6979 | 1065.0289 | 1087.8393 |
| 1171.3313 | 1225.6197 | 1288.4484 |
| 1426.1310 | 1481.0634 | 1557.6322 |
| 1609.1425 | 2206.0680 | 3128.9077 |
| 3168.3820 | 3195.7131 | 3550.9804 |

**i21”-i12**

| | | | |
|---|---|---|---|
| N | -0.5966716891 | -0.9935343628 | -0.3501708129 |
| C | -0.9376234098 | 0.2562706444 | -0.3450135871 |
| C | -0.0226860314 | 1.3294715269 | -0.1034076215 |
| C | 1.3062922478 | 1.0619223772 | 0.0458469568 |
| C | 1.8178781781 | -0.285563968 | 0.1450701924 |
| C | 0.7400995779 | -1.2546531332 | 0.0150104246 |
| H | -1.9697368999 | 0.4901185561 | -0.5915625419 |
| H | -0.4013007561 | 2.3452753621 | -0.1121000854 |
| H | 2.0168259114 | 1.8811195357 | 0.0708764482 |
| H | 0.9267919678 | -1.0167882606 | 1.1446747729 |
| H | 1.0021759033 | -2.3051872777 | -0.062202146 |

Frequencies

| | | |
|---|---|---|
| 549.0483*i* | 197.9336 | 371.0080 |
| 597.1815 | 624.1241 | 700.5998 |
| 777.7706 | 873.7003 | 918.1706 |
| 996.7244 | 1005.5194 | 1026.4455 |
| 1041.0760 | 1062.1776 | 1144.8915 |
| 1213.6423 | 1241.8732 | 1371.0963 |
| 1397.7505 | 1448.2743 | 1527.0615 |
| 1630.7215 | 2479.0396 | 3121.3574 |
| 3134.8728 | 3135.4084 | 3159.4758 |

**i22”-i12**

| | | | |
|---|---|---|---|
| N | -0.6416910639 | -1.1907736718 | -0.0507401067 |
| C | -1.0188492551 | 0.0847203285 | -0.2422182174 |
| C | -0.0535309706 | 1.0676127657 | -0.2185455554 |
| C | 1.2810447696 | 0.7050463064 | 0.0023532786 |
| C | 1.640506914 | -0.618740669 | 0.1956261562 |
| C | 0.6532278703 | -1.6147150693 | 0.1710962825 |
| H | -0.482660909 | -2.3525108584 | 0.0960769645 |
| H | -2.0687487005 | 0.2850330466 | -0.4071146268 |
| H | -0.3282424294 | 2.1020764834 | -0.3690232347 |
| H | 2.0381097308 | 1.48014171 | 0.0204911084 |
| H | 2.6752350437 | -0.885179372 | 0.3646379508 |

Frequencies

| | | |
|---|---|---|
| 2021.0910*i* | 157.9261 | 401.0930 |
| 462.2959 | 617.0285 | 653.9794 |
| 734.4975 | 775.1464 | 915.1157 |
| 974.5024 | 996.3221 | 1028.6738 |
| 1045.0948 | 1061.1920 | 1110.3685 |
| 1180.3987 | 1187.6122 | 1240.9620 |
| 1437.8159 | 1458.0739 | 1565.0642 |
| 1614.5544 | 2405.9157 | 3151.7609 |
| 3177.6916 | 3180.2558 | 3196.3608 |

**Products**

ortho-$C_5H_4N$

| | | | |
|---|---|---|---|
| N | -0.8052586645 | 1.3487655046 | 0. |
| C | 0.4738599752 | 1.3391442365 | 0. |
| C | -1.4658591521 | 0.175798094 | 0. |
| C | -0.7747361548 | -1.02423758 | 0. |
| C | 0.6207982594 | -1.002375505 | 0. |
| C | 1.2899162394 | 0.2161618929 | 0. |
| H | -2.5475007783 | 0.2145866708 | 0. |
| H | -1.3151371905 | -1.9605094773 | 0. |
| H | 1.1801732349 | -1.9297520623 | 0. |
| H | 2.3678182313 | 0.2796332259 | 0. |

Frequencies

| | | |
|---|---|---|
| 386.4512 | 427.6511 | 578.2906 |
| 663.7073 | 716.6818 | 755.9484 |
| 894.0024 | 956.8675 | 974.8163 |
| 1011.9991 | 1037.4618 | 1074.9040 |
| 1110.6484 | 1167.9487 | 1261.9472 |
| 1332.1845 | 1427.1609 | 1503.2466 |
| 1571.3300 | 1664.1018 | 3163.0009 |
| 3172.6225 | 3195.2661 | 3202.6968 |

meta-$C_5H_4N$

| | | | |
|---|---|---|---|
| N | -0.8285418085 | 1.363456887 | 0. |
| C | 0.5136657578 | 1.3906806876 | 0. |
| C | -1.4393610315 | 0.1797228983 | 0. |
| C | -0.771496371 | -1.0404845127 | 0. |
| C | 0.6249629432 | -1.0280134795 | 0. |
| C | 1.2132000748 | 0.2083194404 | 0. |
| H | 0.9884640476 | 2.3631723026 | 0. |
| H | -2.523118012 | 0.2070303206 | 0. |
| H | -1.323226523 | -1.9715969041 | 0. |
| H | 1.1967559225 | -1.9471426402 | 0. |

Frequencies

| | | |
|---|---|---|
| 390.7281 | 424.5636 | 580.7812 |
| 662.5875 | 692.6503 | 788.5788 |
| 927.0698 | 950.9567 | 991.2707 |
| 998.8739 | 1047.7995 | 1066.2773 |
| 1109.8340 | 1203.9497 | 1260.5414 |
| 1331.6211 | 1440.5374 | 1471.8003 |
| 1542.3851 | 1612.8164 | 3150.1283 |
| 3166.5417 | 3171.4924 | 3183.6070 |

para-$C_5H_4N$

| | | | |
|---|---|---|---|
| N | -0.828993529 | 1.3586896001 | 0. |
| C | 0.5011375049 | 1.3699871214 | 0. |
| C | -1.447473975 | 0.181108184 | 0. |
| C | -0.7753871027 | -1.0486822913 | 0. |
| C | 0.5898289438 | -0.9667174251 | 0. |
| C | 1.2872584859 | 0.2097682587 | 0. |
| H | 0.979766716 | 2.3438109947 | 0. |
| H | -2.532388077 | 0.2010170088 | 0. |
| H | -1.3238796952 | -1.9805952863 | 0. |
| H | 2.3668537283 | 0.2711618349 | 0. |

Frequencies

| | | |
|---|---|---|
| 380.1911 | 446.7267 | 618.8660 |
| 645.2008 | 717.2548 | 772.4021 |
| 832.1109 | 968.6583 | 982.5181 |
| 991.0188 | 1041.5333 | 1075.5029 |
| 1080.5928 | 1230.0602 | 1265.5323 |
| 1322.6611 | 1409.1523 | 1476.3033 |
| 1539.0335 | 1617.6045 | 3139.1489 |
| 3141.6232 | 3183.9297 | 3185.6157 |

**Reactants**

$C_3H_3$

| | | | |
|---|---|---|---|
| C | 1.335257 | 0.000000 | 0. |
| C | -1.248944 | 0.000000 | 0. |
| C | 0.115295 | 0.000000 | 0. |
| H | 2.396116 | 0.000000 | 0. |
| H | -1.802881 | 0.928035 | 0. |
| H | -1.802881 | -0.928035 | 0. |

Frequencies

| | | |
|---|---|---|
| 354.6849 | 407.3892 | 463.8420 |
| 632.8663 | 691.1253 | 1033.7510 |
| 1090.2007 | 1459.7879 | 2011.1814 |
| 3142.9117 | 3231.8145 | 3462.5430 |

$H_2CCN$

| | | | |
|---|---|---|---|
| N | -1.353539 | 0.000000 | 0.0 |
| C | 1.189313 | 0.000000 | 0.0 |
| C | -0.187231 | -0.000000 | -0.0 |
| H | 1.731142 | -0.9333625 | 0.0 |

| | | | | |
|---|---|---|---|---|
| H | | 1.731142 | 0.9333625 | 0.0 |

Frequencies

| | | |
|---|---|---|
| 384.6883 | 435.8502 | 677.7500 |
| 1035.1990 | 1058.8252 | 1447.2572 |
| 2130.2879 | 3162.0467 | 3264.3476 |

*n*-$C_4H_3$

| | | | | |
|---|---|---|---|---|
| C | 0 | 0.000000 | 0.653495 | 0.000000 |
| C | 0 | -0.501531 | 1.745923 | 0.000000 |
| C | 0 | 0.614632 | -0.632781 | -0.000000 |
| C | 0 | -0.057593 | -1.762773 | 0.000000 |
| H | 0 | -0.947673 | 2.708670 | 0.000000 |
| H | 0 | 1.702598 | -0.652455 | -0.000000 |
| H | 0 | -1.087974 | -2.079399 | 0.000000 |

Frequencies

| | | |
|---|---|---|
| 214.9481 | 344.5614 | 512.5788 |
| 648.1910 | 680.2621 | 707.9643 |
| 855.3546 | 875.0441 | 991.7645 |
| 1269.5155 | 1624.6148 | 2198.7835 |
| 3096.4447 | 3240.8951 | 3470.1950 |

*i*-$C_4H_3$

| | | | |
|---|---|---|---|
| C | -0.000060 | -0.752647 | 0.000000 |
| C | -0.000192 | -1.981606 | 0.000000 |
| C | 0.000000 | 0.562640 | 0.000000 |
| C | 0.000193 | 1.865203 | 0.000000 |
| H | -0.000204 | -3.042414 | 0.000000 |
| H | -0.922684 | 2.440573 | 0.000000 |
| H | 0.923245 | 2.440295 | 0.000000 |

Frequencies

| | | |
|---|---|---|
| 84.5647 | 230.5300 | 272.9127 |
| 379.8248 | 544.4460 | 636.5348 |
| 891.6670 | 901.8948 | 980.5452 |
| 1438.3278 | 1800.7049 | 2014.6555 |
| 3067.5758 | 3126.4778 | 3461.1248 |

$H_2CN$

| | | | |
|---|---|---|---|
| H | 0.000005 | -1.075995 | -0.934650 |
| H | 0.000005 | -1.075995 | 0.934650 |
| C | -0.000001 | -0.500874 | -0.000000 |
| N | -0.000001 | 0.736747 | 0.000000 |

Frequencies

| | | |
|---|---|---|
| 934.6859 | 994.8119 | 1378.3835 |
| 1724.6088 | 2959.8511 | 3012.2841 |